\documentclass{article}

\usepackage{arxiv}

\usepackage{graphicx}
\usepackage{newtxtext}
\usepackage{newtxmath}
\usepackage{amsmath}
\usepackage{natbib}
\numberwithin{equation}{section}
\usepackage[hidelinks]{hyperref}       % hyperlinks
\hypersetup{
    colorlinks = true,
    urlcolor   = blue,
    citecolor  = black,
}

\newcommand{\RomanNumeralCaps}[1]
\linenumbers

\usepackage{floatrow}
\usepackage[label font=bf,labelformat=parens]{subfig}% <-- changed
\def\be{\begin{equation}}
\def\ee{\end{equation}}
\usepackage{multirow}
\usepackage{authblk}

\title{Experimental investigations of linear and nonlinear periodic traveling waves in a viscous fluid conduit}

\author[1]{Yifeng Mao\thanks{yifeng.mao@colorado.edu}}
\author[1]{Mark A. Hoefer}
\affil[1]{Department of Applied Mathematics, University of Colorado, Boulder, CO 80309, USA}

\begin{document}

\maketitle

\begin{abstract}
Conduits generated by the buoyant dynamics between two miscible, Stokes fluids with high viscosity contrast exhibit rich nonlinear wave dynamics. However, little is known about the fundamental wave dispersion properties of the medium. In the present work, a pump is used to inject a time-periodic flow that results in the excitation of propagating small and large amplitude periodic traveling waves along the conduit interface. This wavemaker problem is used as a means to measure the linear and nonlinear dispersion relations and corresponding periodic traveling wave profiles. Measurements are favorably compared with predictions from a fully nonlinear, long-wave model (the conduit equation) and the analytically computed linear dispersion relation for two-Stokes flow. A critical frequency is observed, marking the threshold between propagating and non-propagating (spatially decaying) waves. Measurements of wave profiles and the wavenumber-frequency dispersion relation quantitatively agree with wave solutions of the conduit equation. An upshift from the conduit equation's predicted critical frequency is observed and is hypothesized to be explained by incorporating a weak recirculating flow into the full two-Stokes flow model which we observe to be operable in the experiment. When the boundary condition corresponds to the temporal profile of a nonlinear periodic traveling wave solution of the conduit equation, weakly nonlinear and strongly nonlinear, cnoidal-type waves are observed that quantitatively agree with the conduit nonlinear dispersion relation and wave profiles. This wavemaker problem is an important precursor to the experimental investigation of more general boundary value problems in viscous fluid conduit nonlinear wave dynamics.
\end{abstract}

% keywords can be removed

% {\bf MSC Codes }  {\it(Optional)} Please enter your MSC Codes here

\section{Introduction} \label{sec:introduction}

The wavemaker problem consists of waves launched from a boundary into a medium at rest. It is a fundamental problem for inviscid flows with physical applications such as surface waves in reservoirs or lakes produced by the movement of a solid body (\cite{chwang1983porous}). The Sommerfeld radiation condition (\cite{sommerfeld1912greensche}; \cite{sommerfeld1949partial}) can be used to select a particular solution that requires only outgoing waves at infinity. This boundary condition has been applied to many boundary value problems in fluid dynamics and elsewhere (see, e.g. \cite{buhler2014waves}).
The classical theory considers surface water waves forced by a wavemaker, represented by a vertical impermeable plate which oscillates horizontally with a small displacement (\cite{havelock1929lix}; \cite{kennard1949generation}; \cite{chwang1983nonlinear}). The wavemaker problem has also been studied experimentally, primarily in water wave tanks exhibiting various wave structures. For example,  \cite{goring1980generation} successfully produced clean cnoidal waves in shallow water with a piston wavemaker. Other examples include the features and stability of two-dimensional periodic surface wave patterns propagating on deep water with nearly persistent form (\cite{kimmoun1999short}; \cite{hammack2005progressive}; \cite{henderson2006laboratory}). An isolated rogue wave was successfully created in a water tank by \cite{chabchoub2011rogue}.

This paper presents an alternative fluid medium in which to investigate the wavemaker problem. The buoyancy driven, cylindrical free interface between two miscible, Stokes fluids with high viscosity contrast results in fluid flow through a pipe-like conduit. This scenario models viscously deformable media in a variety of physical environments including magma in the earth's mantle (\cite{scott1984magma}) and channelized water flow beneath glaciers (\cite{stubblefield2020solitary}). Beyond applications, this medium has been shown to exhibit a rich variety of nonlinear wave dynamics. Experiments in established conduits have been extensively studied in consideration of solitons (\cite{olson1986solitary}; \cite{scott1986observations}; \cite{helfrich1990solitary}; \cite{lowman2014interactions}), periodic waves (\cite{kumagai1998interaction}; \cite{olson1986solitary}),  solitary wave fission (\cite{anderson2019controlling}; \cite{maiden2020solitary}), dispersive shock waves (DSWs) (\cite{maiden2016observation}), and the interactions between solitons and DSWs (\cite{maiden2018solitonic}). Quantitative observational agreement with the conduit equation has been achieved in certain regimes. But such agreement has largely been confined to the long-wave, albeit strongly nonlinear, regime. Herein, we experimentally investigate long and short periodic traveling waves in a viscous fluid conduit. 

Periodic wavetrains have already been observed and studied in viscous fluid conduits. Some of the earliest apparently involved periodic waves by either quickly increasing the injection rate from one constant value to another constant value (\cite{scott1986observations}; \cite{kumagai1998interaction}) or, at the onset of injection, the generation of periodic waves trailing a rising diapir on plume of injected fluid (\cite{olson1986solitary}). However, since the injection rate was held constant after the initial change, both scenarios involve different dynamics than what we consider here. In the first case, we now understand that a step up in injection rate from a nonzero previous rate results in a dispersive shock wave consisting of a slowly varying periodic traveling wave that ultimately returns to a constant injection rate, i.e. it is transient (\cite{maiden2016observation}). In the latter case, it is likely that an oscillatory (absolute) instability developed due to sufficiently large injection rate rather than the convective instability regime in which we operate (\cite{d'olce_martin_rakotomalala_salin_talon_2009}). In this work, we controllably create high quality linear and nonlinear periodic traveling waves by use of a time-periodic injection rate. The time-series are specifically designed by evaluating exact periodic traveling wave solutions of the model conduit equation.

The conduit equation
\be 
    A_t + (A^2)_z - (A^2(A^{-1}A_t)_z)_z = 0, \label{eq:conduit}
\ee 
modeling the dynamics of the normalized cross-sectional area $A$ of a conduit at vertical height $z$ and time $t$, is an accurate partial differential equation model of conduit interfacial waves that has been studied since the 1980s (\cite{scott1986observations}; \cite{olson1986solitary}; \cite{lowman2013dispersive}; \cite{maiden2016modulations}; \cite{johnson2020modulational}; \cite{simpson2006degenerate}; \cite{harris2006painleve}). Derivation of the conduit equation follows from the Stokes equations for two viscous fluids under a long-wave, high-viscosity contrast assumption (\cite{lowman2013dispersive}). The resulting nonlinear, dispersive equation is an expression of mass conservation that realizes a maximal balance between buoyancy and viscous stress effects.

% Periodic wavetrains play a fundamental role in mathematical models appearing as a special solution to many dynamical equations. Modulational instability of a uniform train of plane waves was discovered and modeled by \cite{benjamin1967instability} which is well known as the Benjamin-Feir instability. In their work, a train of surface waves of finite amplitude in deep water was shown under two circumstances: one close to the wavemaker exhibited uniform structures while the other further downstream pronounced amplitude modulations. Happening around the same time, other significant theoretic studies were obtained. \cite{whitham1965non} and \cite{whitham1967non} introduced a general asymptotic approach which averages over the local oscillations for studying modulated periodic nonlinear dispersive waves. The nonlinear evolution of modulated waves was found to be given by structural properties of the Whitham equation. Using the theory developed by Whitham, \cite{lighthill1965contributions} considered the averaged Lagrangian for a weakly nonlinear Stokes wave on deep water. Other than fluids, the phenomena of modulational instability was also investigated in various physical system including plasma (\cite{hasegawa2012plasma}) and nonlinear optics (\cite{ostrovskii1967propagation}; \cite{krolikowski2001modulational}).

The major objective of this work is the reliable generation of both linear and nonlinear, cnoidal-type waves from a wavemaker. Laboratory measurement constrained by boundary control is realized by injection of an interior fluid of lower viscosity and lower density into a more viscous exterior fluid using a piston pump.  Limitations are investigated and the experiment is compared with the theory of the boundary value problem for the conduit equation. We find that, though the exact periodic traveling wave solutions to the conduit equation exhibit quantitative agreement with our experimental observations, a deviation in the linear dispersion relation occurs for sufficiently short waves. In order to describe the linear waves in the shorter-wavelength regime, we obtain the exact solution for two-Stokes flow linear interfacial wave propagation. Previous work by \cite{yih1963stability,yih1967instability} and \cite{hickox1971instability} studied the linear instability of the flow in a core-annular flow for the Navier-Stokes equations. Within a parameter regime consistent with our experimental conditions, the instability has been shown to be convective (\cite{d'olce_martin_rakotomalala_salin_talon_2009}; \cite{selvam_talon_lesshafft_meiburg_2009}). However, none of these works obtained an analytical solution to the problem.  Herein, we incorporate a model for recirculating core-annular flow and obtain the exact solution to the Stokes equations for azimuthally symmetric, small-amplitude perturbations. Asymptotic analysis is also conducted in the limit of high viscosity contrast, long waves, and small Reynolds number.

\begin{center}
    \textbf{Outline of this work}
\end{center}

The paper is structured as follows. First we provide some background material on periodic traveling waves. Then we derive the exact linear dispersion relation for two-Stokes flow in section \ref{sec:two_Stokes_flow}. Analytical results for an idealized recirculating flow with azimuthally symmetric linear interfacial waves and their asymptotic expansions are presented.  Section \ref{sec:linear_theory} describes linear modulation theory applied to the initial-boundary value problem for both the conduit equation and two-Stokes flow.  In section \ref{sec:experiment}, experimental methods and the analysis of periodic waves in a viscous fluid conduit are provided. Finally, we conclude in section \ref{sec:conclusion}.

\begin{center}
    \textbf{The wavemaker problem and periodic traveling wave solutions}
\end{center}

The conduit equation (\ref{eq:conduit}) obeys the scaling invariance 
\be 
    A^* = \frac{A}{\bar{A}}, \quad z^*=\bar{A}^{-1/2}z, \quad t^* = \bar{A}^{1/2}t,
\ee
where $\bar{A}>0$. Owing to this scaling invariance,  we can always assume the periodic traveling wave has unit mean. In this work, we study the wavemaker problem for a viscous fluid conduit so that a time-dependent periodic wave with amplitude $a$ and injection frequency $\omega_0$ is created at the boundary and propagates into a quiescent medium. The wavemaker initial-boundary value problem considers
\begin{subequations}
\begin{align}
    A(z,0) &= 1,  \quad z \geq 0, \\
    A(0,t) &= 1+f(t), \quad t \geq 0, 
    \quad f(t+T_0)=f(t), \quad t>0, 
    \quad T_0 = \frac{2\pi}{\omega_0},
\end{align}
\end{subequations} 
where $f(t)$ is the periodic profile with frequency $\omega_0>0$, $f(0)=f'(0)=0$, $\int_0^{T_0} f(t) dt =0$.

Conduit periodic traveling wave solutions satisfy $A(z,t) = \phi(\theta)$, $\theta=kz-\omega t$, $\phi(\theta+2\pi)=\phi(\theta)$ for $\theta\in\mathbb{R}$ and a nonlinear ODE (\cite{olson1986solitary})
\be 
    -\omega \phi' + k (\phi^2)' + \omega k^2 (\phi^2 (\phi^{-1}\phi')')'=0. \label{eq:intro_ODE}
\ee
Integrating (\ref{eq:intro_ODE}) twice results in
\be 
    (\phi')^2 = g(\phi) \equiv -\frac{2}{k^2}\phi - \frac{2}{\omega k}\phi^2\ln\phi + c_1 +c_2\phi^2, \label{eq:intro_nonlinearODE}
\ee
where $c_{1,2}$ are integration constants. 

Upon linearization of (\ref{eq:conduit}) as $A(z,t)-1 \propto \cos(k z-\omega t)$, the conduit equation exhibits the bounded dispersion relation 
\be 
    \omega = \frac{2k}{1+k^2} \label{eq:conduit_linear_disp_1}
\ee 
on a unit mean background. For the wavemaker problem, it is helpful to invert the linear dispersion relation (\ref{eq:conduit_linear_disp_1}) and solve for $k(\omega)$. In order to select the correct branch, we apply the radiation condition (positive group velocity) to obtain (\cite{buhler2014waves})
\be 
    k(\omega) = 
    \begin{cases}
       \frac{1-\sqrt{1-\omega^2}}{\omega}, & 0<\omega\leqslant1  \\
       \frac{1+i\sqrt{\omega^2-1}}{\omega}, & \omega>1
    \end{cases}, \label{eq:conduit_linear_disp_2}
\ee 
where wavenumbers $k$ for frequencies $0<\omega\leqslant1$ are real, implying steady wave propagation. For $\omega>1$, $k(\omega)$ exhibits a positive imaginary part so that these waves spatially decay. The frequency $\omega_{cr}=1$ is therefore the critical frequency, above which waves will not propagate. The corresponding critical wavenumber $k_{cr}=1$ satisfies $k_{cr}=k(\omega_{cr})$. 

In the weakly nonlinear regime, we can obtain approximate periodic traveling wave solutions when the amplitude is small and the wavenumber is sufficiently far away from 0. The weakly nonlinear approximation with three harmonics takes the form
\be 
    \phi(\theta; k, a) = 1 + \dfrac{a}{2} \cos(\theta)  + \dfrac{1+k^2}{48k^2} a^2\cos(2\theta) + \dfrac{1+k^2}{1536k^4}a^3\cos(3\theta) + O(a^4),
\ee 
where $\theta=k z-\omega t$, the amplitude $0<a \ll 1$, the wavenumber $k\gg a^2$, and the frequency $\omega$ is
\be 
    \omega(k,a) = \frac{2k}{1+k^2} + a^2 \frac{1-8k^2}{48k(1+k^2)} + O(a^4). \label{eq:intro_weakly_nonlinear_3}
\ee 
Alternatively, we can express the weakly nonlinear expansions in terms of $\omega$ and get
\begin{subequations}
\begin{align}
    \phi(\theta,\omega,a) & = 1 +\frac{a}{2}\cos(\theta)+\frac{a^2}{24\left(1-\sqrt{1-\omega^2}\right)}\cos(2\theta) \nonumber \\ 
    & \hphantom{= 1} +\frac{\left(-1-\sqrt{1-\omega^2}\right) a^3 }{768\left(-2+\omega^2+2\sqrt{1-\omega^2}\right)}\cos(3\theta) + \cdots, \label{eq:intro_weakly_nonlinear_1} \\
    k(\omega,a) &=\frac{1-\sqrt{1-\omega^2}}{\omega} +\frac{\left(-9+ \frac{7}{\sqrt{1-\omega^2}}\right)}{96\omega} a^2 + \cdots. \label{eq:intro_weakly_nonlinear_2}
\end{align} 
\end{subequations}

\begin{figure}[tb!]
    \centering \hspace{-0.5in}
    \includegraphics[width=0.5\textwidth]{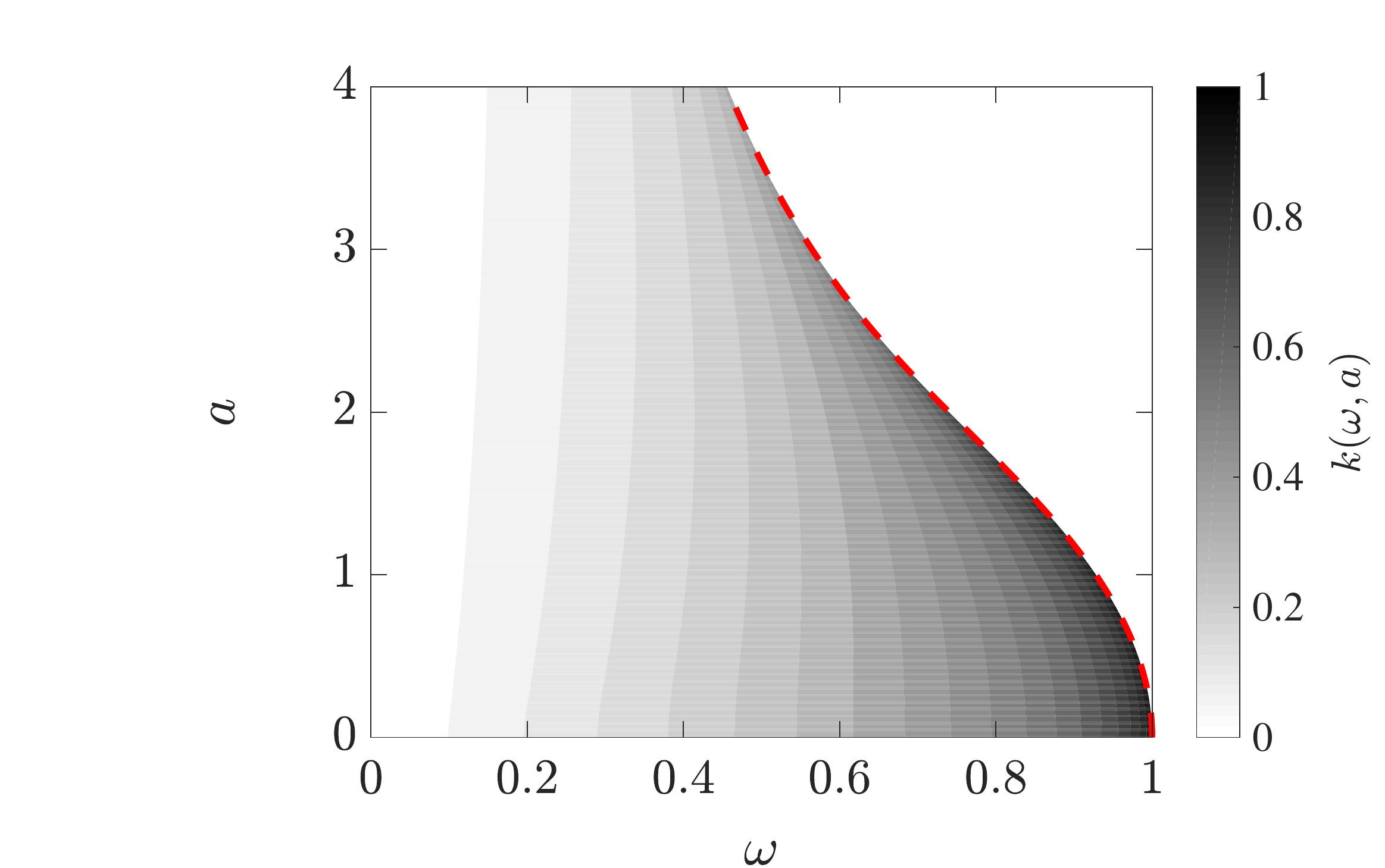}
    \caption{Contour plot of numerically computed nonlinear conduit dispersion relation $k(\omega,a)$. Beyond the red dashed line, no real $k(\omega,a)$ is obtained.}
    \label{fig:intro}
\end{figure}

The nonlinear dispersion relation is explicitly expressed in terms of three roots $0<u_1<u_2<u_3$ of $g(\phi)$ in (\ref{eq:intro_nonlinearODE}) (\cite{lowman2013dispersive}) and phase velocity $U$ so that
\be 
    \exp U = u_1^{\frac{u_1^2(u_2+u_3)}{(u_3-u_1)(u_2-u_1)}} u_2^{-\frac{u_2^2(u_3+u_1)}{(u_3-u_2)(u_2-u_1)}} u_3^{\frac{u_3^2(u_2+u_1)}{(u_3-u_1)(u_3-u_2)}}.
\ee 
The wave amplitude, wavenumber and mean are given by
\be 
    a = u_3-u_2, \quad k=\pi \left( \int_{u_2}^{u_3} \frac{du}{\sqrt{g(u)}} \right)^{-1}, \quad \bar{\phi}=\frac{k}{\pi} \int_{u_2}^{u_3} \frac{u du}{\sqrt{g(u)}}.
\ee 
The numerically computed conduit nonlinear dispersion relation $\omega=kU$ is presented in figure \ref{fig:intro}, where the red dashed line corresponds to the edge of the existence region.

%--------------------------------------------------------------------------------------------------------------%
\section{Two-Stokes flow} \label{sec:two_Stokes_flow}

Since the conduit equation (\ref{eq:conduit}) and its linear dispersion relation (\ref{eq:conduit_linear_disp_1}), (\ref{eq:conduit_linear_disp_2}) are derived under long wavelength assumptions, we now seek a more accurate description of linear waves by analyzing the free interface between two Stokes fluids.

\subsection{Primary recirculating flow}

Interfacial waves generated by the buoyant dynamics between two miscible, Stokes fluids with small viscosity ratio $\epsilon=\mu^{(i)}/\mu^{(e)}$ are described by continuity equations for mass conservation and Stokes equations for momentum conservation
\begin{subequations}
\begin{align}
    \nabla \cdot \mathbf{\Tilde{u}}^{(i,e)} &= 0, \label{eq:twoStokes_mass}\\
    -\nabla \Tilde{p}^{(i,e)} + \nabla \cdot \Tilde{\sigma}^{(i,e)} &= 0, \label{eq:twoStokes_momentum}
\end{align}
\end{subequations}
for the interior and exterior fluids, denoted by a superscript $i$ or $e$, respectively. The velocity fields $\mathbf{\Tilde{u}}^{(i,e)} = (\Tilde{u}_{\Tilde{r}}^{(i,e)}, \Tilde{u}_{\Tilde{z}}^{(i,e)})$ are given in cylindrical coordinates assuming azimuthal symmetry. The pressure deviation from hydrostatic is $\Tilde{p}^{(i,e)}$ and $\Tilde{\sigma}^{(i,e)}=\mu^{(i,e)}[\nabla \mathbf{\Tilde{u}} + (\nabla \mathbf{\Tilde{u}})^T]$ is the deviatoric stress tensor, where $\mu^{(i,e)}$ denotes the fluid visocity. The kinematic boundary condition between the two fluids at $\Tilde{r}=\Tilde{R}(z,t)$ incorporates the time dynamics 
\be 
    \Tilde{u}_{\Tilde{r}}^{(i)} = \frac{\partial \Tilde{R}}{\partial \Tilde{t} } + \Tilde{u}_{\Tilde{z}}^{(i)}\frac{\partial \Tilde{R}}{\partial \Tilde{z}}, \quad \Tilde{r}=\Tilde{R}(\Tilde{z},\Tilde{t}).
\ee 
The remaining boundary conditions include the continuity of normal and tangential velocities 
\be 
    [\mathbf{\Tilde{u}}\cdot \mathbf{\hat{\Tilde{n}}}_c]_j = [\mathbf{\Tilde{u}} \cdot \mathbf{\hat{\Tilde{t}}}_c]_j =0 , \quad \Tilde{r} = \Tilde{R}(\Tilde{z},\Tilde{t}),
\ee 
and continuity of the normal and tangential stresses
\be 
    [\mathbf{\hat{\Tilde{n}}}_c \cdot \mathbf{\Tilde{T}} \cdot \mathbf{\hat{\Tilde{n}}}_c]_j = [\mathbf{\hat{\Tilde{T}}}_c \cdot \mathbf{\Tilde{T}} \cdot \mathbf{\hat{\Tilde{n}}}_c]_j = 0, \quad \Tilde{r} = \Tilde{R}(\Tilde{z},\Tilde{t}),
\ee 
where $[]_j$ represents the difference in the exterior and interior fluid quantities, $\mathbf{\hat{\Tilde{n}}}_c$ is the outward pointing unit normal to the interface and $\mathbf{\hat{\Tilde{t}}}_c$ is the unit tangent to the interface. $\mathbf{\Tilde{T}}^{(i,e)}=\Tilde{\sigma}^{(i,e)} - \mathbf{I} \Tilde{p}^{(i,e)}$ is the stress tensor and $\mathbf{I}$ denotes the identity operator. No-slip and no-penetration boundary conditions at the outer wall $\Tilde{r}=\Tilde{D}$ require $\Tilde{u}_{\Tilde{z}}^{(e)}(\Tilde{D})=0$ and $\Tilde{u}_{\Tilde{r}}^{(e)}(\Tilde{D})=0$. Boundary conditions along the symmetry axis are imposed as 
\be 
    \dfrac{\partial \Tilde{u}_{\Tilde{z}}^{(i)}}{\partial \Tilde{r}}=0, \quad \Tilde{u}_{\Tilde{r}}^{(i)}=0, \quad \dfrac{\partial \Tilde{p}^{(i)}}{\partial \Tilde{r}}=0, \quad \text{at } \Tilde{r}=0. \label{eq:Bndy_sym}
\ee

A nondimensionalization that leads to a maximal balance between buoyant and viscous stress effects taking vertical length scale $L$, velocity scale $U$ and time scale $T$ as $\epsilon\to0$ is (\cite{lowman2013dispersive})
\begin{subequations} \label{eq:nondim}
\begin{align}
    r &= \epsilon^{-1/2} \Tilde{r}/L, \quad z = \Tilde{z}/L,  \quad t = \Tilde{t} / T, \\
    \mathbf{u}^{(i,e)} &= \mathbf{\Tilde{u}}^{(i,e)}/U, \\
    p^{(i,e)} & =  \frac{\Tilde{p}^{(i,e)}-\Tilde{p}_0}{\Pi}, \quad  \Pi=\mu^{(i)}U/L, \\
    \sigma^{(i,e)} &= (L/\mu^{(i,e)}U) \Tilde{\sigma}^{(i,e)},
\end{align}
\end{subequations}
where $\Tilde{p}_0$ is a constant reference pressure. The vertical variation is assumed to be weaker than the radial length scale. With constant conduit radius $R_0$, the scales are defined as
\be 
L= R_0/\sqrt{8\epsilon}, \quad U= \frac{g R_0^2 \Delta}{8\mu^{(i)}}, \quad T=L/U, \quad \Omega = \frac{2\pi}{T}, \label{eq:nondim_scale}
\ee 
where $\Delta=\rho^{(e)}-\rho^{(i)}$ is the difference in the fluid densities.
Substituting the above scalings into the dimensional equations represented by the tilde notation, the nondimensional continuity and linear momentum equations for the interior and exterior fluid are obtained (see Appendix \ref{sec:append1}).

Steady flow requires only that the axial velocity and pressure have initial values different from zero and the vertical velocity is independent of $z$, i.e. $ u_{r,0}^{(i)}=u_{r,0}^{(e)}=0, \frac{\partial u_{z,0}^{(i)}}{\partial z}=\frac{\partial u_{z,0}^{(e)}}{\partial z}=0$. Collecting the governing equations as well as boundary conditions, we obtain the solution for the primary flow 
\begin{subequations}
\begin{align}
    p_0^{(i)}&=(\lambda-1)z + \Lambda,\\
    u_{z,0}^{(i)}&=\dfrac{(\lambda-1) }{4} (r^2-\eta^2) + \dfrac{\epsilon \lambda }{4}(\eta^2-D^2)  + \dfrac{\epsilon \eta^2}{2}\ln \frac{D}{\eta},\\
    p_0^{(e)}&=\lambda z + \Lambda,\\
    u_{z,0}^{(e)}&=\epsilon \left( \dfrac{ \lambda }{4}(r^2-D^2)  + \dfrac{\eta^2}{2}\ln \frac{D}{r} \right),
\end{align}
\end{subequations}
with external fluid pressure gradient $\lambda$, internal fluid pressure gradient $\lambda-1$ and interfacial radius $\eta$. Pressure is determined up to an arbitrary constant, so we take $\Lambda=0$. The vertical volumetric flow rates of the interior and exterior flow through a cross section normal to the $z$ axis are
\begin{subequations}
\begin{align}
      Q^{(i)} &= 2\pi \int^\eta_0 u_{z,0}^{(i)}(r) r dr \nonumber \\
      &= \frac{\pi}{8} \left( 1 - \lambda + 2\epsilon \lambda -  2\epsilon \lambda \frac{D^2}{\eta^2} + 4\epsilon \ln \frac{D}{\eta} \right) \eta^4, \label{eq:twoStokes_Qi}\\
      Q^{(e)} &= 2\pi \int^D_\eta u_{z,0}^{(e)}(r) r dr  \nonumber \\
      &= -\frac{\pi}{8} \epsilon \left( 2 + \lambda - 2(1+\lambda) \frac{D^2}{\eta^2} + \lambda \frac{D^4}{\eta^4} + 4\ln \frac{D}{\eta}  \right) \eta^4. \label{eq:twoStokes_Qe}
\end{align}
\end{subequations}
In the case of a motionless external fluid so that $\epsilon=0$ and $\lambda=0$, the interior flow rate is $Q^{(i)}=\frac{\pi}{8}\eta^4$, which, upon dimensionalization (\ref{eq:nondim_scale}) results in the Poiseuille, pipe flow scaling law $R_0^4=\frac{8\mu^{(i)}}{\pi g \Delta}Q_0$, where $Q_0$ is the interior fluid injection rate.
Note that the externel pressure gradient $\lambda$ is a parameter which is constrained by the direction of fluid flow. Previous studies that derived the conduit equation assumed a high viscosity contrast ($\epsilon\ll1$) and zero external pressure gradient. However, in this work, $\epsilon$ and $\lambda$, as well as $D$ are adapted to model our experiments by introducing an idealized recirculating flow in which the interior fluid rises upward dragging some of the exterior fluid with it, while the rest of the exterior fluid flows downward. This situation dictates that 
\be 
    \left(\frac{\eta}{D}\right)^2 < \lambda < \frac{2\eta^2\ln(D/\eta)}{D^2-\eta^2}. \label{eq:twoStokes_lambda1}
\ee
So far, we have only specified the vertical flow directions for the interior and exterior fluids of an infinitely long conduit. In order to model flow recirculation, we need to appeal to the observed fluid dynamics in our experimental apparatus. The rising interior fluid is observed to pool at the very top of the fluid column, creating a less dense, lower viscosity fluid layer above the exterior fluid (see figure \ref{fig:Fig_setup1} for a schematic). On the other hand, the exterior fluid entrained by the rising interior fluid is observed to recirculate downward and, upon reaching the bottom of the column, recirculates upward. We model these dynamics by invoking the additional requirement of mass conservation for the exterior fluid (i.e. $Q^{(e)}=0$ in (\ref{eq:twoStokes_Qe})), resulting in the external pressure gradient
\be 
    \lambda = \frac{2\eta^2(D^2-\eta^2-2\eta^2\ln(D/\eta))}{(D^2-\eta^2)^2}. \label{eq:twoStokes_lambda2}
\ee
The validity of the requirement (\ref{eq:twoStokes_lambda2}) will be discussed in section \ref{sec:experiment_linear} using experimental measurements.

\subsection{Linear dispersion relation}

\begin{figure}
    \centering
        \subfloat[]{\includegraphics[height=0.3\textwidth]{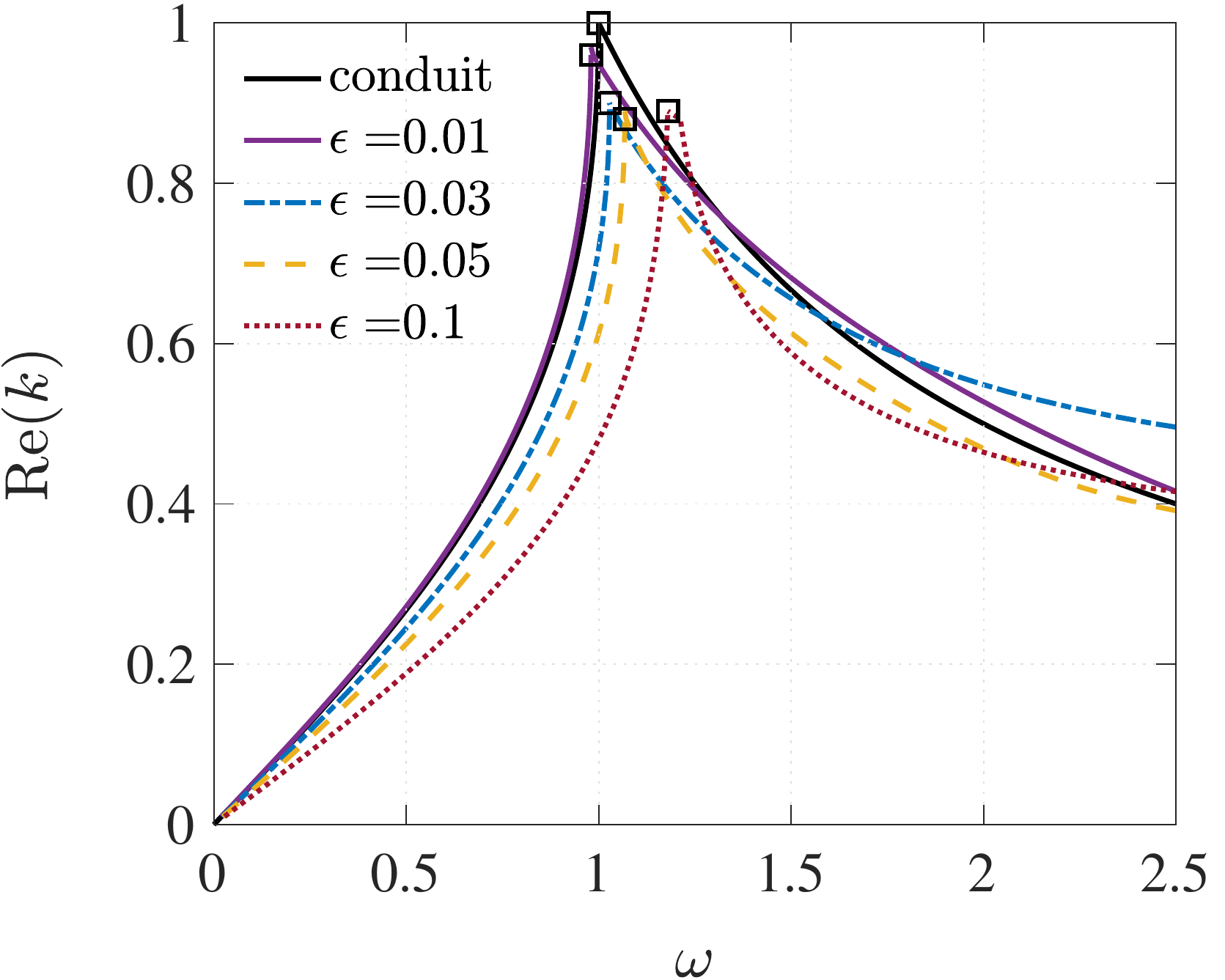}} 
        \subfloat[]{\includegraphics[height=0.3\textwidth]{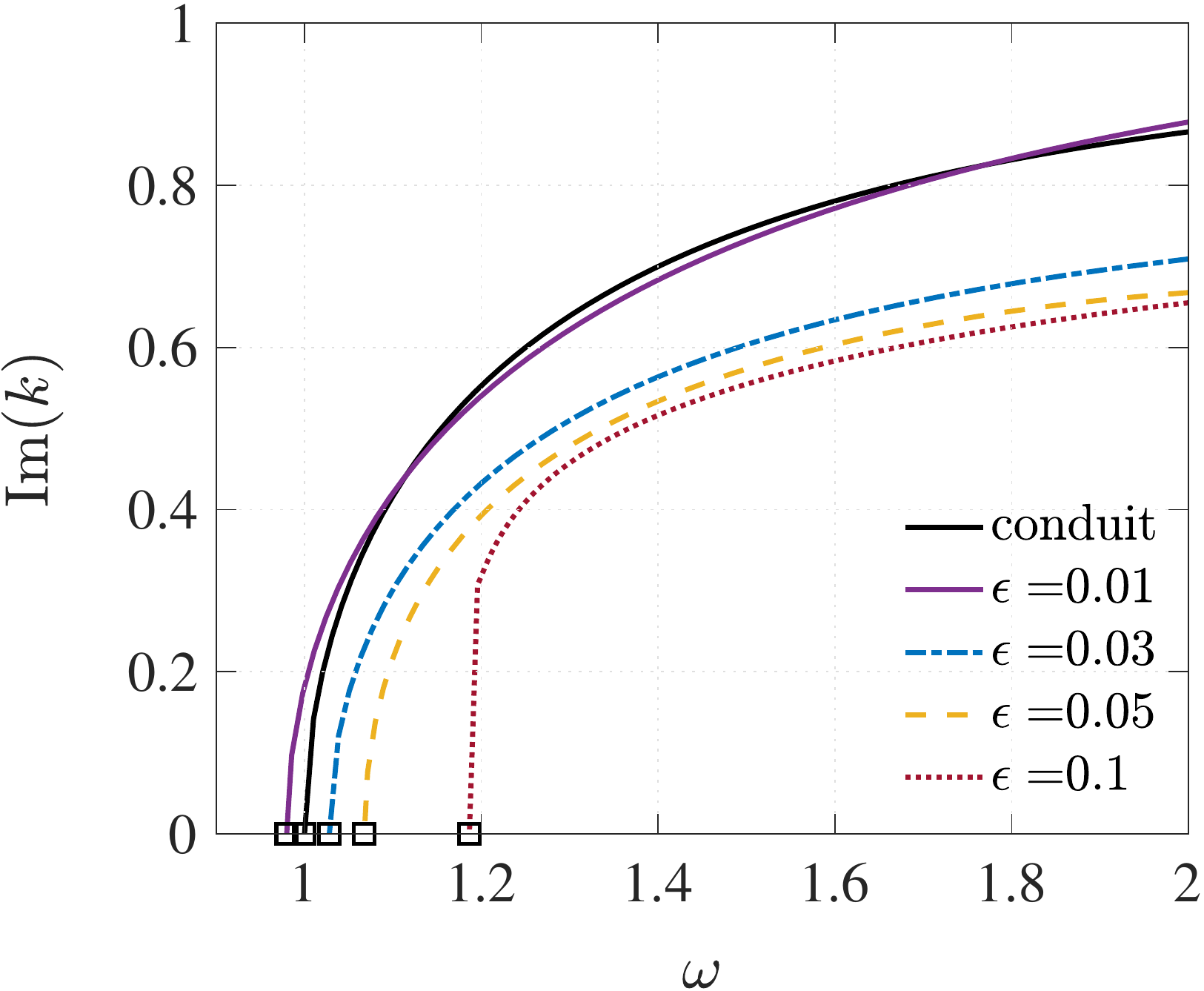}}
    \caption{Two-Stokes recirculating flow dispersion relation with $D/\eta=10$, $\lambda$ in (\ref{eq:twoStokes_lambda2}) and different $\epsilon$. Black squares represent the critical frequency $\omega_{cr}$. (a) Real part of the dispersion relation $\text{Re}(k(\omega))$. (b) Imaginary part $\text{Im}(k(\omega))$.}
    \label{fig:Fig_Stokes_disp1}
\end{figure}

The linear dispersion relation for waves on a straight conduit in two-Stokes flow is investigated by linearizing about the primary flow. Following the procedure as introduced in \cite{batchelor1962analysis} and  \cite{hickox1971instability}, first order velocities and pressures are written as
\begin{align}
    u_1^{(i)}&=[g_z(r),i \epsilon^{1/2} g_r(r)] e^{i(kz-\omega t)}, \label{eq:twoStokes_linearize1}\\
    u_1^{(e)}&=[\epsilon G_z(r),i \epsilon^{1/2} G_r(r)] e^{i(kz-\omega t)}, \label{eq:twoStokes_linearize2}\\
    p_1^{(i)}&=h(r) e^{i(kz-\omega t)}, \label{eq:twoStokes_linearize3}\\
    p_1^{(e)}&=H(r) e^{i(kz-\omega t)}. \label{eq:twoStokes_linearize4}
\end{align}
The $\epsilon$ scaling of the exterior and interior velocity perturbations are chosen to reflect tangential stress balance and the kinematic boundary condition. We assume a constant conduit radius $\eta$ subject to a small disturbance so that 
\be 
    R(z,t)=\eta+ a e^{i(kz-\omega t)}, \quad |a|\ll1.
\ee 
Solving the governing equations directly and applying the boundary conditions, the eigenfunctions $g_r,g_z,h$ and $G_r,G_z,H$ are obtained in terms of a linear algebraic system given in Appendix \ref{sec:append2}. Analytically, inverting the 6 by 6 matrix leads to overly burdensome expressions, so we either evaluate the linear system numerically or consider its asymptotics for small $\epsilon$. The linear dispersion relation is given by the kinematic boundary condition as
\be 
    \omega(k) = k u_{z,0}^{(i)}(\eta) - g_r(\eta,k). \label{eq:two_Stokes_w}
\ee 
In regions where (\ref{eq:two_Stokes_w}) is monotonic in $k$, we can obtain $k(\omega)$ for consideration of the wavemaker problem. In the vicinity of $\omega_k=0$, we use the radiation condition to select the appropriate branch. Selecting $\eta=\sqrt{8}$ for convenience when comparing with the $\epsilon\to0$ dispersion relation (\ref{eq:conduit_linear_disp_2}), example dispersion curves for waves on a unit-mean background are shown in figure \ref{fig:Fig_Stokes_disp1}.  The maximum of  $\text{Re}(k(\omega))$ occurs at the critical frequency $\omega=\omega_{cr}$, defined by the group velocity $c_g(\omega_{cr}) = \frac{d\omega}{dk}(\omega_{cr}) = 0$, above which $k(\omega)$ incorporates a nonzero imaginary part and the wave is spatially damped. The corresponding critical wavenumber is $k_{cr}=k(\omega_{cr})$. In the experiments reported in section \ref{sec:experiment_linear}, $\epsilon>0.01$ and the critical frequency is subject to a deviation from 1.

Consider the limit $\epsilon\to0$. In order for the leading order behavior of the two-Stokes system to be well approximated by the conduit dynamics, we require a maximal balance further incorporating the outer wall's no-slip boundary condition so that
\be 
    D=\epsilon^{-1/2}d, \quad d=O(1). 
\ee
In this case, the corresponding external pressure gradient (\ref{eq:twoStokes_lambda2}) becomes
\be
    \lambda = \frac{16 \epsilon  \left(d^2+8\epsilon  (\ln (8 \epsilon )-2\ln (d))-8 \epsilon \right)}{\left(d^2-8 \epsilon \right)^2} = \epsilon \frac{16}{d^2} + O\left(\epsilon^2 \ln(\epsilon) \right).
\ee
This results in higher order corrections to the conduit unit-mean linear dispersion relation as
\be \label{eq:twoStokes_w(k)_asym}
     \omega(k) = \frac{2 k}{1+k^2} + 4 \epsilon \ln\left( 1/\epsilon \right) k \frac{1+2k^2}{(1+k^2)^2} +  \epsilon \omega_2(k) + \cdots,
\ee
where $\omega_2(k)$ involves a complicated expression that depends on $d$ which can be found in appendix \ref{sec:append2}.
Following the assumptions of high viscosity contrast and long-wavelengths, the conduit linear dispersion relation is obtained at leading order in (\ref{eq:twoStokes_w(k)_asym}). 

\begin{figure}
    \centering
    \includegraphics[width=0.4\linewidth]{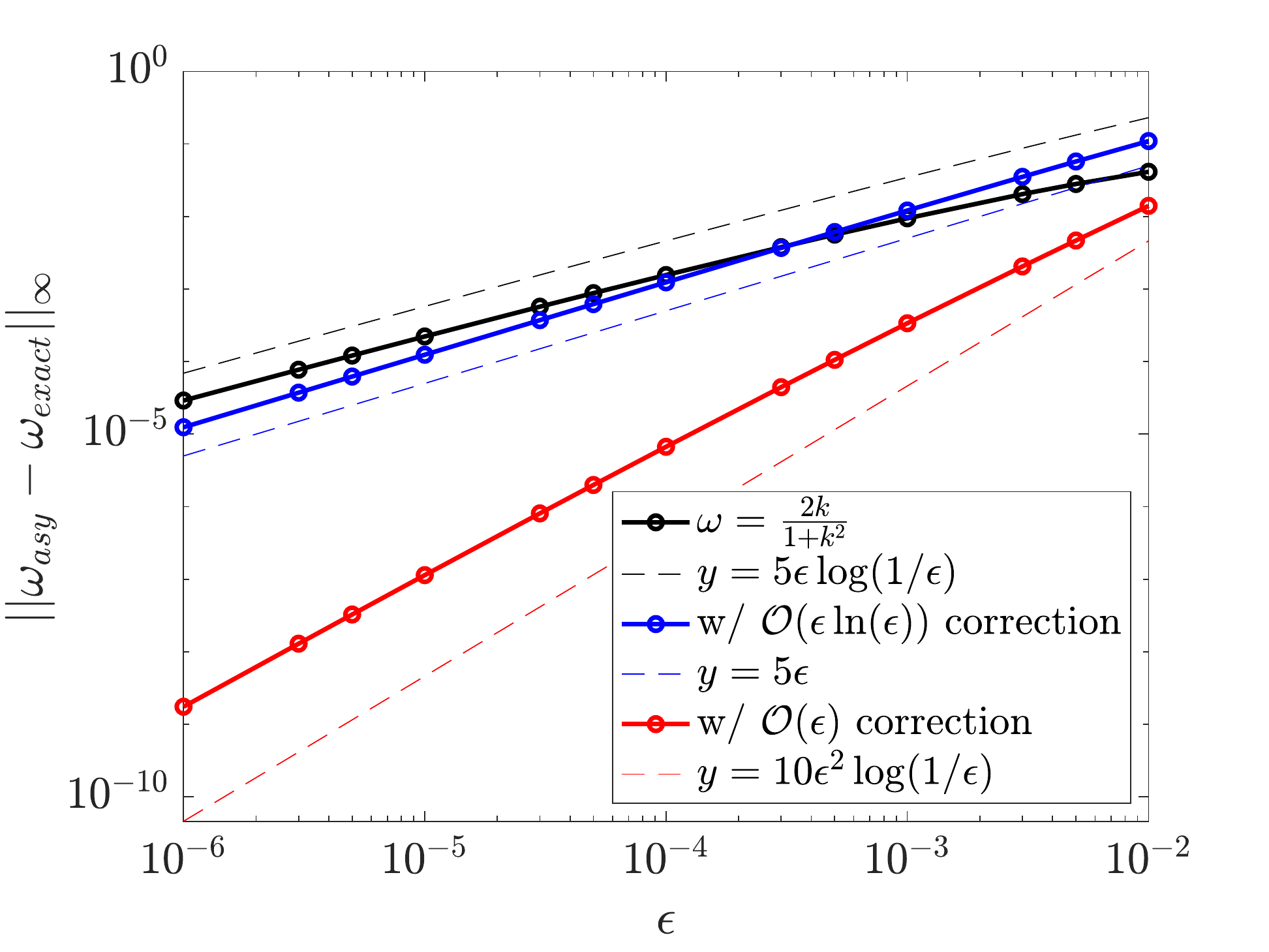}
    \caption{Maximum absolute error in the comparison of exact and asymptotic (\ref{eq:twoStokes_w(k)_asym}) solutions to the two-Stokes linear dispersion relation at $d=5$. }
    \label{fig:append5}
\end{figure}

We provide the 3-term asymptotic expansion (\ref{eq:twoStokes_w(k)_asym}) to the two-Stokes linear dispersion relation because the 2-term expansion has limited applicability. For example, the absolute error in the wave frequency for $0<\omega<\omega_{cr}$ between (\ref{eq:twoStokes_w(k)_asym}) and the full dispersion relation in figure \ref{fig:append5} shows that the second term does not significantly improve upon the leading order term, even for $\epsilon\approx 10^{-6}$ but the 3-term expansion does. Nevertheless, the 2-term expansion predicts an upshift in the critical frequency $\omega_{cr} = 1+3\epsilon\ln(1/\epsilon)$ and the critical wavenumber $k_{cr} = 1+\epsilon\ln(1/\epsilon)$.  While the former is consistent with the numerical calculation of the full dispersion in our experimental regime where $\epsilon\ll1$ (see figure \ref{fig:Fig_Stokes_disp1}),  the latter is not. 

% We note that for the small Reynolds numbers $Re^{(i)}$ achieved in our experiments, inertial terms have negligible effects; see Appendix \ref{sec:append3}. 

\section{Linear modulation theory} \label{sec:linear_theory}

\begin{figure}[tb!]
    \centering
    \includegraphics[height=0.25\textwidth]{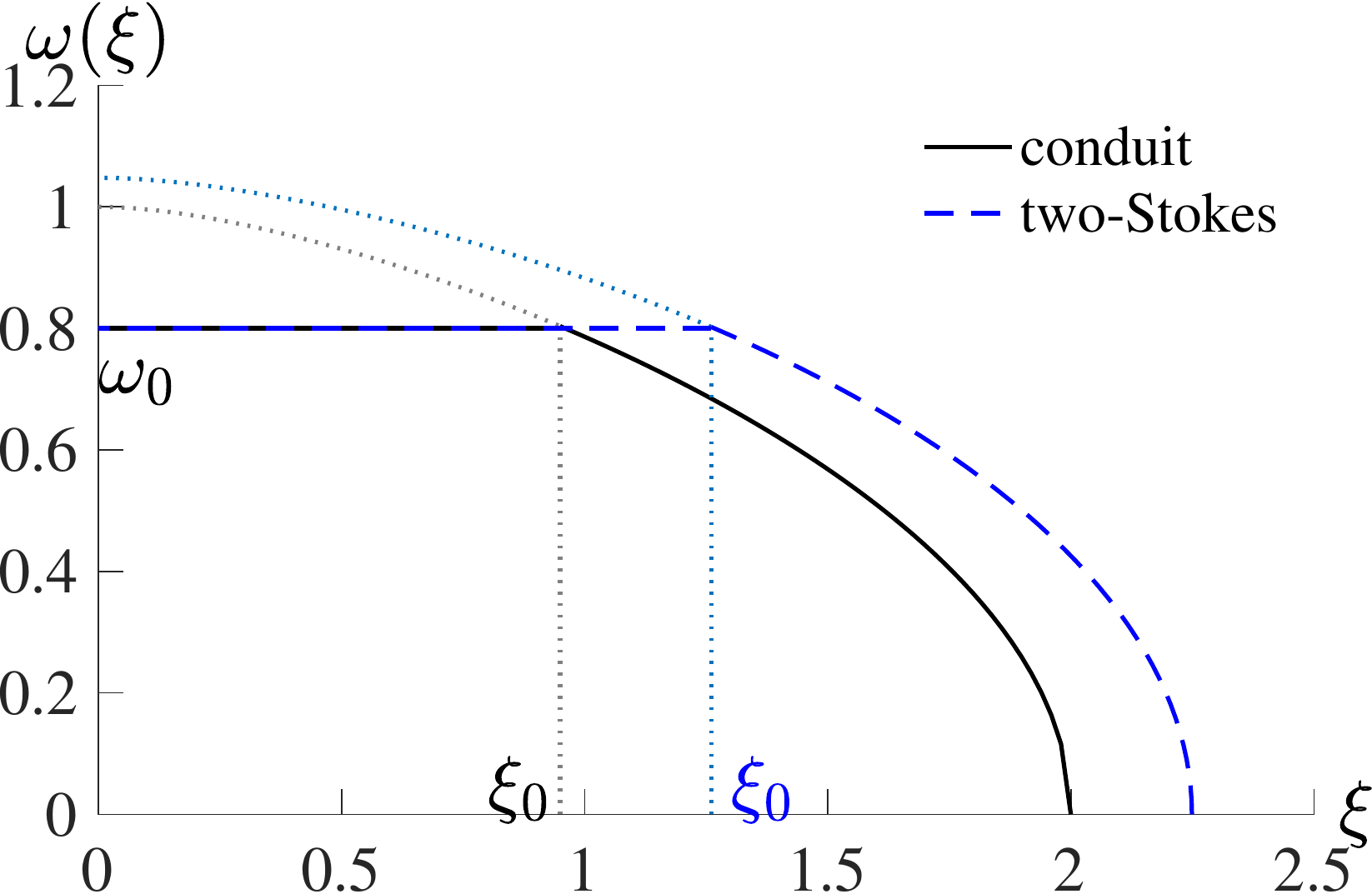}
    \caption{Solution to the linear modulation equation conservation of waves with an input frequency $\omega_0=0.8$ for the conduit equation (black solid) and two-Stokes flow with $\epsilon=0.04,D/\eta=10$ and $\lambda$ at (\ref{eq:twoStokes_lambda2}) (blue dashed).}
    \label{fig:linear_theory}
\end{figure}

To describe slow modulations of the linear periodic wave's amplitude and wavenumber for the wavemaker problem, we use modulation theory. The modulation equation is the conservation of waves (\cite{whitham1974linear})
\be
    k_t + \omega_z =0, \label{eq:linear_Whitham_1}
\ee 
where $\omega$ is the angular frequency and $k(\omega)$ is the local linear dispersion relation. Translating the initial-boundary value problem for $u(z,t)$ to an initial-boundary value problem for $k(z,t)$, we have 
\begin{subequations}
\begin{align}
    \omega(z,0) &= 0, \quad z \geq 0, \\
    \omega(0,t) &= \omega_0, \quad t > 0. \label{eq:linear_Whitham_BC}
\end{align}
\end{subequations}
We seek a self-similar solution such that  $\omega=\omega(\xi)$, $\xi=z/t$, implying $\omega'(\xi)=\xi$ or $k'(\omega)=1/\xi$.
The solution takes the form 
\be 
    \omega(\xi)= 
    \begin{cases}
        \omega_0, & 0\leq\xi\leq c_g(\omega_0) \\
        g(\xi), & c_g(\omega_0) <\xi \leq c_g(0)\\
        0, & \xi> c_g(0)
    \end{cases} \label{eq:modulation_dispersion}
\ee 
where $\omega_0$ is the input frequency and $c_g(\omega_0)$ represents the group velocity for the wavenumber $k_0=k(\omega_0)$. Comparison of the solution (\ref{eq:modulation_dispersion}) for the conduit and two-Stokes dispersion $k(\omega)$ is depicted in figure \ref{fig:linear_theory}. In particular, the conduit equation admits a plane wave solution with constant frequency $\omega_0$ and wavenumber $k_0=\frac{1-\sqrt{1-\omega_0^2}}{\omega_0}$ for $0\leq \xi < \xi_0$, and a decaying oscillatory solution with $\omega(\xi) = \frac{\sqrt{1-2\xi+\sqrt{1+4\xi}}}{\sqrt{2}}$ and $k(\xi)=\frac{\sqrt{-1-\xi+\sqrt{1+4\xi}}}{\sqrt{\xi}}$ for $\xi_0 \leq \xi \leq 2$. The two regions are separated by the line $\xi_0=1-\omega_0^2+\sqrt{1-\omega_0^2}$. For the two-Stokes recirculating flow modulation solution, we have a similar profile for $\omega(\xi)$ but the group velocities $c_g(\omega_0), c_g(0)$ are faster than the corresponding conduit group velocities. These calculations suggest that for an input $\omega_0$ smaller than the critical frequency, we can generate a plane wave that will eventually propagate and fill the entire vertical conduit. We now realize this in experiments described in the next section.

% \begin{figure}[tb!]
%     \centering
%     \subfloat[]{\includegraphics[height=0.3\textwidth]{Fig/Fig_modulation1.pdf} \label{fig:linear_theory_conduit}} 
%     \subfloat[]{\includegraphics[height=0.3\textwidth]{Fig/Fig_modulation2.pdf}  \label{fig:linear_theory_twoStokes}}
%     \caption{Solution (black solid line) to the linear modulation equation conservation of waves for (a) the conduit equation and (b) two-Stokes flow at $\epsilon=0.04,\Lambda=0.2,d=8$.}
% \end{figure}

%--------------------------------------------------------------------------------------------------------------%

\section{Experiment}\label{sec:experiment}

\subsection{Setup}

\begin{figure}
    \centering 
        \subfloat[]{\includegraphics[width=0.52\textwidth]{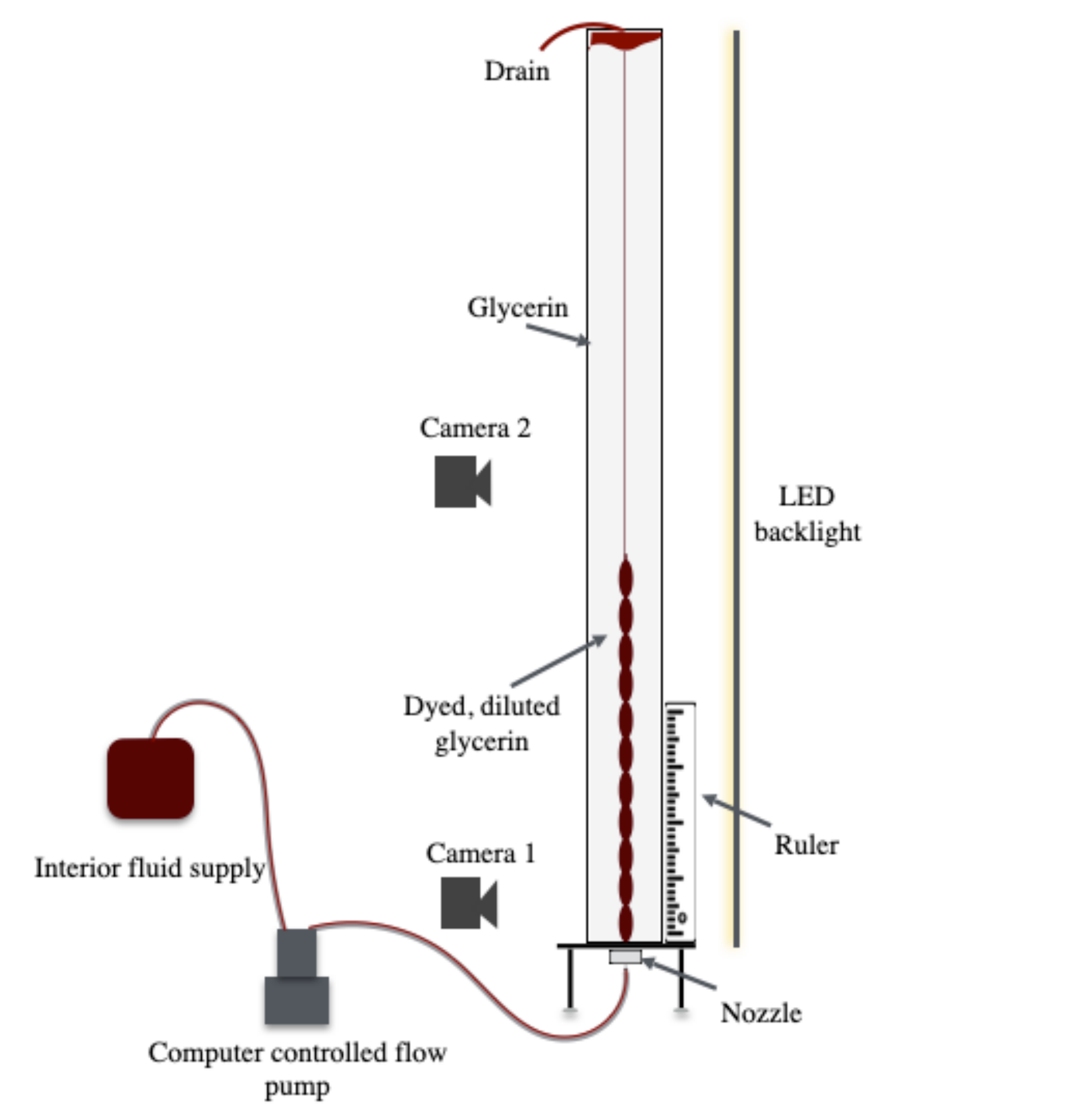} \label{fig:Fig_setup1}} 
        \subfloat[]{\includegraphics[width=0.25\textwidth]{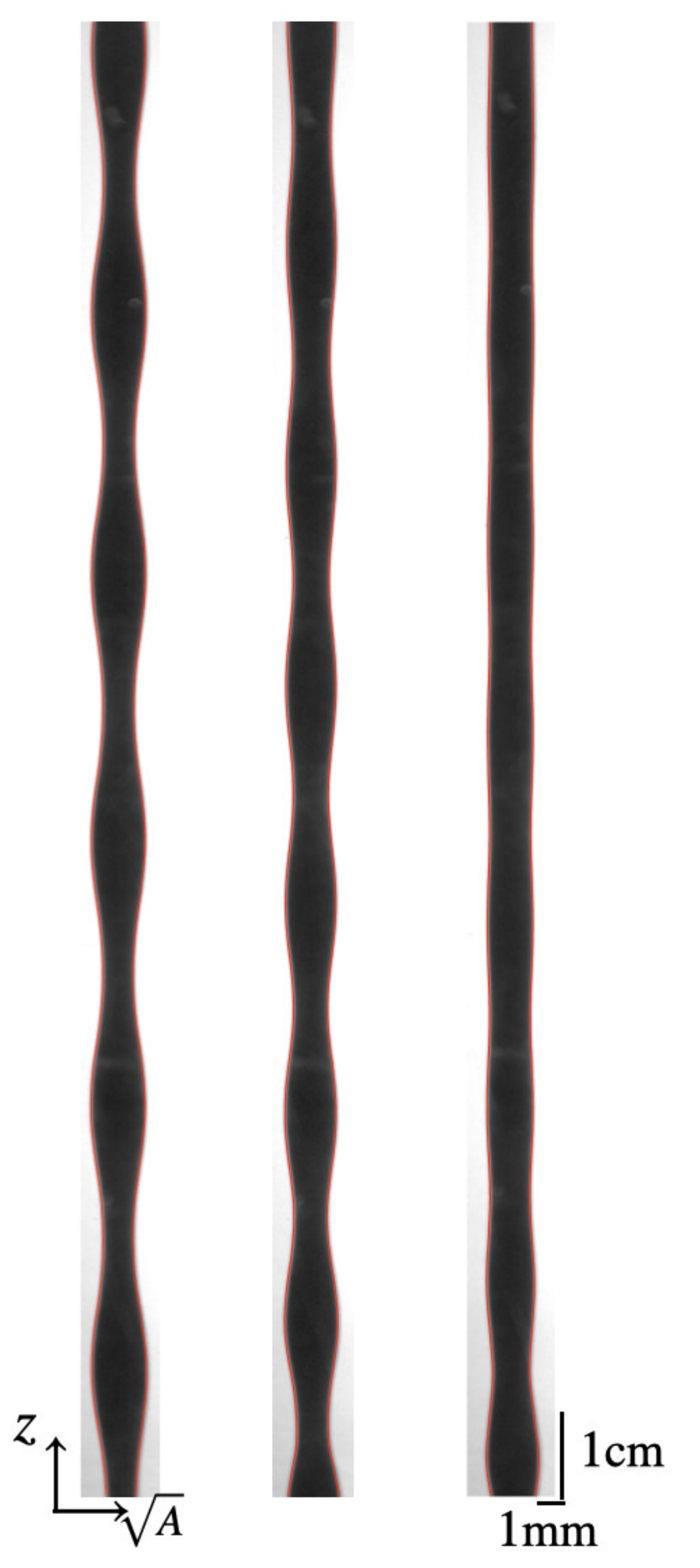} \label{fig:Fig_setup2}} 
    \caption{(a) Schematic of the experimental apparatus. (b) Waves $\omega_0=(0.85,0.95,1.00)/\Omega$ with fixed amplitude $a=0.8$ but increasing frequencies from left to right. The spatial wave is observed to be periodic at the leftmost, but get damped in the middle and arrested at the rightmost. Conduit edges are outlined by the red curves. Measured experimental parameters follows from table \ref{tab:setup}. } 
    
\end{figure}

The experimental setup in figure \ref{fig:Fig_setup1} is identical to those used by \cite{anderson2019controlling} and \cite{maiden2020solitary}. A conduit is formed in a square acrylic column with dimensions $5 \text{ cm} \times 5 \text{ cm} \times 200 \text{ cm} $. The exterior fluid is clean glycerin of high viscosity and the interior fluid consists of a miscible mixture of glycerin ($\sim 70\% $), water ($\sim 20\% $) and food coloring ($\sim 10\% $), which is less dense and less viscous than the exterior fluid. The nominal parameter values used in one of the experiments (dataset 3 in sections \ref{sec:experiment_linear}, \ref{sec:experiment_damping}) are presented in table \ref{tab:setup}. Miscibility of the two fluids implies negligible surface tension at the interface and the free interface is sustained over the timescale of experiment because mass diffusion occurs much more slowly than momentum diffusion.  

A computer-controlled piston pump is used as a wavemaker to inject the interior fluid into the extrusive fluid at the nozzle that then rises buoyantly. Variation in the flow rate results in interfacial waves. Periodic waves are created by a suitable periodic injection rate $Q^{(i)}(t)$:
\be 
    Q^{(i)}(t)=
    q(t), \quad  t \geq 0,
    \quad q(t+T_0)=q(t),
    \quad T_0=\frac{2\pi}{\omega_0}. 
\ee 
Data acquisition is performed using high resolution cameras with one (camera 1 in figure \ref{fig:Fig_setup1}) equipped with macro lenses at the injection nozzle for precise data measurement and one to image the far field (camera 2) and capture fully developed periodic wavetrains. Spatial calibration is achieved with a ruler inside the column within camera view. Example large-amplitude periodic waves with different injection frequencies and the same wave amplitude in time are shown in figure \ref{fig:Fig_setup2}. For a relatively small frequency, the periodic wave maintains its fully developed structure, i.e. it propagates. However, by increasing the input frequency, the spatial decay of the amplitude is observed. The middle wave is weakly, spatially damped, and the rightmost wave is arrested at a larger frequency.

\begin{table}[tb!] \rule{\textwidth}{0.4pt}
\begin{tabular}{cc} 
$\mu^{(i)}$ &  $0.045$ Pa s\\
$\mu^{(e)}$ & $1.11$ Pa s  \\
$\epsilon$ & 0.041 \\
$\rho^{(i)}$ & 1.21 g cm$^{-3}$\\
$\rho^{(e)}$ & 1.26  g cm$^{-3}$\\
$Q_0$ & 3.00 cm$^3$ min$^{-1}$\\
$2R_0$ & 0.36 cm\\
$2D_0$ & 5.00 cm\\
$Re^{(i)}$ & 0.47 \\
$Re^{(e)}$ & 0.020\\
Camera resolution & 375.60 pix cm$^{-1}$\\
\end{tabular} \rule{\textwidth}{0.4pt}
\caption{Example fluid properties measured in experiments: viscosities $\mu^{(i,e)}$, viscosity ratio $\epsilon$, densities $\rho^{(i,e)}$, background flow rate $Q_0$, associated conduit diameter $2R_0$ computed by Poiseuille's law, outer wall diameter $2D_0$, and Reynolds numbers $Re^{(i,e)}$ for interior and exterior fluids.}
\label{tab:setup}
\end{table}

\subsection{Method} \label{sec:experiment_method}

Camera images are processed to extract the conduit edges. A low-pass filter is applied to reduce noise due to pixelation of the photograph and any impurities in the exterior fluid. The conduit diameter is obtained by calculating the number of pixels between the two edges and normalizing by the conduit mean diameter. In each experimental trial, images are taken after the wave has fully developed in the camera view and we identify $t=0$ with the initiation of imaging. A $3$ Hz sampling rate for the close camera and a $1$ Hz sampling rate for the full camera were used. 

We present all dimensional quantities in tilde notations. The circular cross-sectional area $\Tilde{A}$ of the conduit is then modeled by the dimensional conduit equation 
\be 
    \Tilde{A}_{\Tilde{t}} + \frac{g \Delta}{8\pi\mu^{(i)}} \left( \Tilde{A}^2 \right)_{\Tilde{z}} - \frac{\mu^{(e)}}{8\pi \mu^{(i)}} \left( \Tilde{A}^2 \left( \Tilde{A}^{-1} \Tilde{A}_{\Tilde{t}}\right) _{\Tilde{z}} \right)_{\Tilde{z}} =0,
\ee 
where $\Tilde{z},\Tilde{t}$ follow from (\ref{eq:nondim}) and  $\Tilde{A} = \pi R_0^2 A$. 
Profile measurements of linear waves are conducted by scaling the data to unit mean and fitting at every temporal and spatial snapshot a sinusoidal plane wave
\be 
    A(\Tilde{z},\Tilde{t}) = 1+ \frac{a}{2} \cos(\Tilde{k} \Tilde{z}- \Tilde{\omega} \Tilde{t} + \psi), \label{eq:method_fitting}
\ee 
where $\psi$ represents the phase.
The relative $L^2$ norm error in the fit is below $15\%$ in space and $10\%$ in time. To extract the wavenumber for a trial, we compute the mean of $\Tilde{k}$ across a range of times $\Tilde{t}$; similarly, the angular frequency is obtained by averaging $\Tilde{\omega}$ over a range of heights $\Tilde{z}$. The standard error in parameter measurements is roughly as low as $1\%$ for $\Tilde{k}$ and $0.5\%$ for $\Tilde{\omega}$, indicating that the fit (\ref{eq:method_fitting}) to the data results in a reliable and accurate measurement of linear periodic waves. However, this method is not suitable for waves beyond the linear regime. Instead, larger-amplitude waves with macroscopic edge changes are processed by directly splitting the waves into periods at each snapshot and computing the averaged wavelengths and wave periods.  The standard error for $\Tilde{k},\Tilde{\omega}$ measurements with this technique is typically below $5\%$.  

To compare the experimental results with theory, we require the vertical length, speed and time scales $L,U,T=L/U$ in (\ref{eq:nondim_scale}) for nondimensionlization. Upon generation of linear periodic waves, we nondimensionalize the wavenumber and angular frequency data by using two fitting methods: one is to determine $L_c$ and $U_c$ by fitting the dimensional data $\Tilde{\omega},\Tilde{k}$ to the conduit linear dispersion relation
\be 
    f(\Tilde{k}) = \dfrac{2U_c \Tilde{k}}{1+L_c^2 \Tilde{k}^2}, \label{eq:method_conduit_fitting}
\ee 
and the other is by fitting $\Tilde{\omega}, \Tilde{k}$ with the two-Stokes linear dispersion relation using the spatial $L_S$ and velocity $U_S$ scales, as well as $\epsilon,\lambda$ and $D$. Treating $\epsilon,\lambda$ and $D$ as effective fitting parameters, the two-Stokes linear dispersion relation is determined and fitted to experiments. We note that the fitting method is sensitive to the initial guess for the fitting parameters. These two techniques, which construct a direct relationship between experimental data and theory, return reliable approximations of the dimensional scaling coefficients $L$ and $U$. Differences in the two methods will be discussed in section \ref{sec:experiment_linear}. The scales in our experiments from either fitting approach are typically in the range
\be 
    L: 0.3-0.4 \text{ cm}, \quad U: 0.3-0.4 \text{ cm/s},
\ee 
with camera resolutions of $250-400 $ pixel cm$^{-1}$. Although $L$ scales like $R_0/\sqrt{\epsilon}$ for modeling long-wavelength wave dynamics, in our experimental regime, it is comparable with the conduit diameter $2R_0$. Nevertheless, for wavenumbers $0<k \lessapprox 1$, the wavelengths are on the centimeter scale while the conduit diameter is on the scale of millimeters.
Prior to each experiment described below, around 10 trials of linear periodic waves were generated for calibrating $L$ and $U$ according to the method described here.

\subsection{Definition of linear, weakly nonlinear and fully nonlinear regimes } 
\label{sec:Fourier_mode}
\begin{figure}
    \centering
    \includegraphics[width=0.4\linewidth]{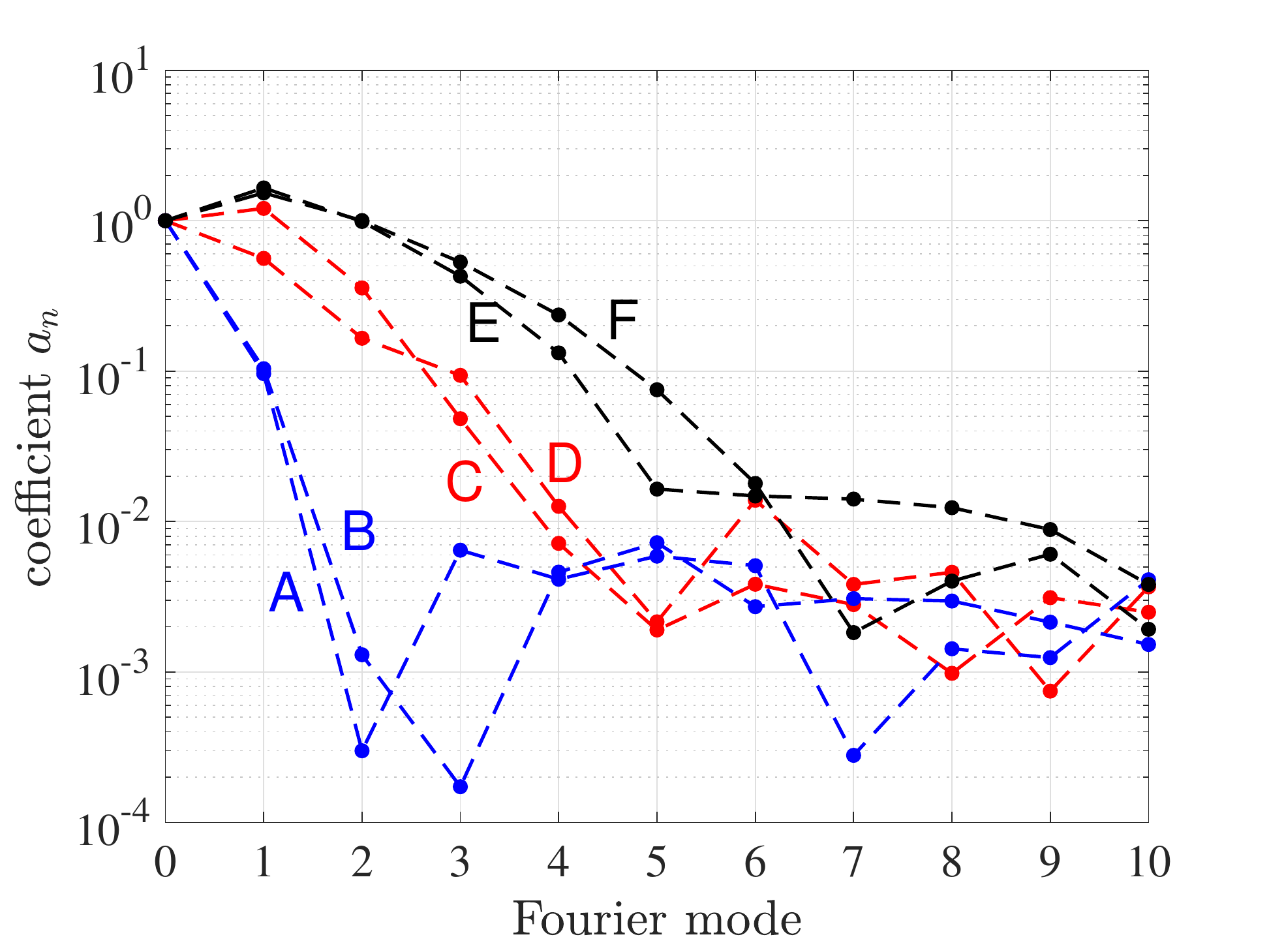}
    \caption{Fourier mode of typical periodic waves in experiments. Waves A,B with dominant $a_1$ can be approximated by a single sinewave. To describe larger amplitude waves C,D, higher harmonic terms are needed. Waves E,F with the largest amplitudes acquire non-negligible Fourier modes even at the fourth or fifth order.}
    \label{fig:Fig_fourier_mode}
\end{figure}

To distinguish the traveling periodic wave structures obtained in experiments, we perform a preliminary analysis by expressing the spatial waves as a Fourier series at a fixed time $t=t_0$:
\be 
    A(z, t_0) = a_0 + \sum^\infty_{n=1} \frac{a_n}{2} \cos (nz + \psi),
\ee 
where $a_0=1$ is the wave mean. Dividing the periodic traveling wave data into periods and applying the Fourier cosine transform, we obtain the Fourier components $a_n$.
Fourier components of some selected periodic waves are plotted in figure \ref{fig:Fig_fourier_mode}.  Waves with negligible second-order and higher modes $|a_n| \lessapprox 0.05$, $n=2,3,\dots$ are identified as lying in the linear regime and can be well described by the linear theory. For larger-amplitude waves $(0.4\lessapprox |a_1| \lessapprox 1.6)$, a single harmonic is insufficient and higher order harmonic terms are needed. These waves are considered to be in the weakly nonlinear regime. A clear distinction between weakly nonlinear and fully nonlinear waves $(|a_1|\gtrapprox 1.6)$ occurs when the fourth-order coefficient is sufficiently large, in which case the weakly nonlinear approximation of three harmonics fails. We now assess each wave class beginning with linear waves.

\subsection{Linear periodic traveling waves} \label{sec:experiment_linear}

\begin{figure}
    \centering
        \subfloat[]{\includegraphics[width=0.45\textwidth]{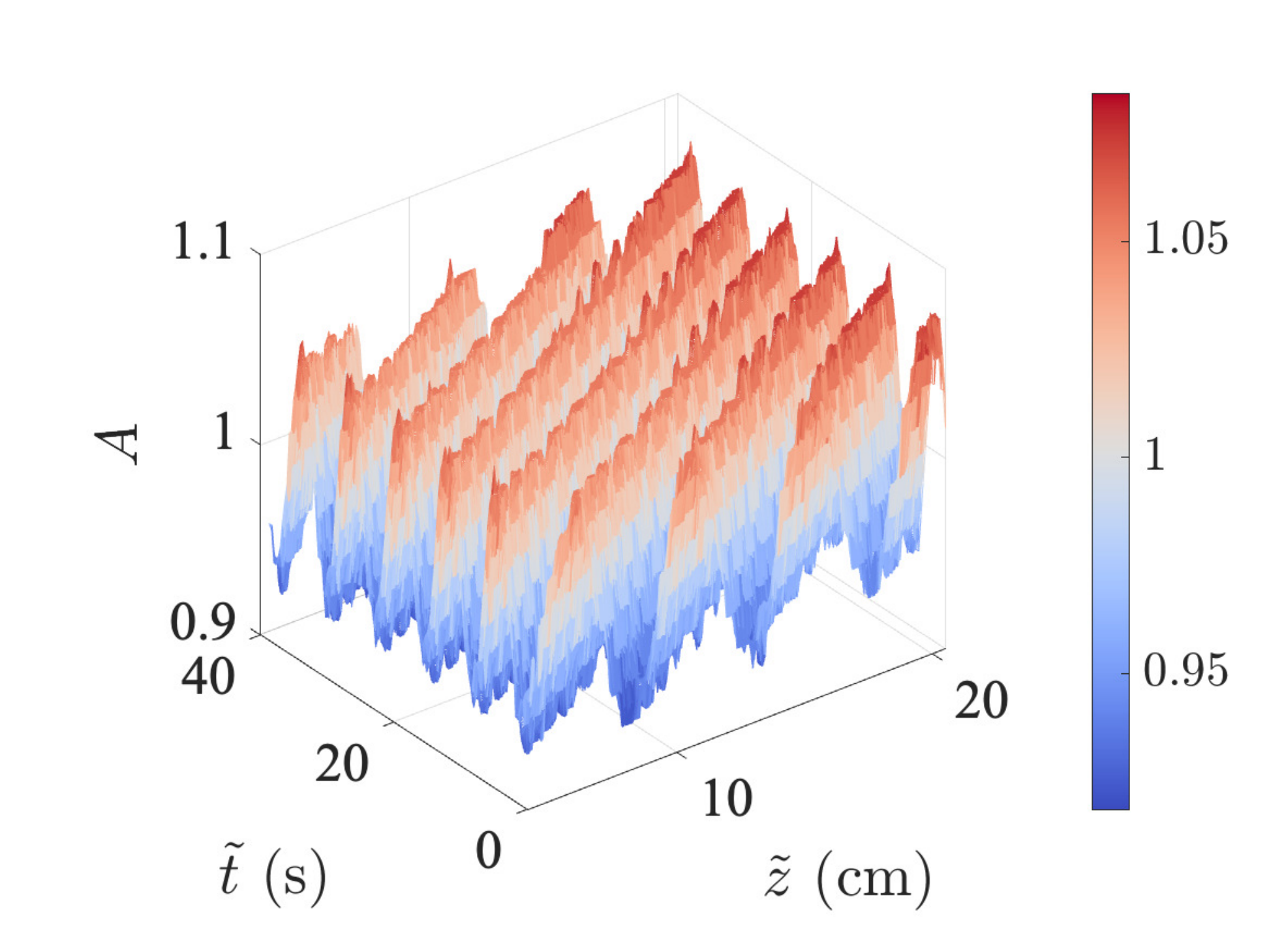} \label{fig:Fig_linear12a} }
        \subfloat[]{\includegraphics[width=0.45\textwidth]{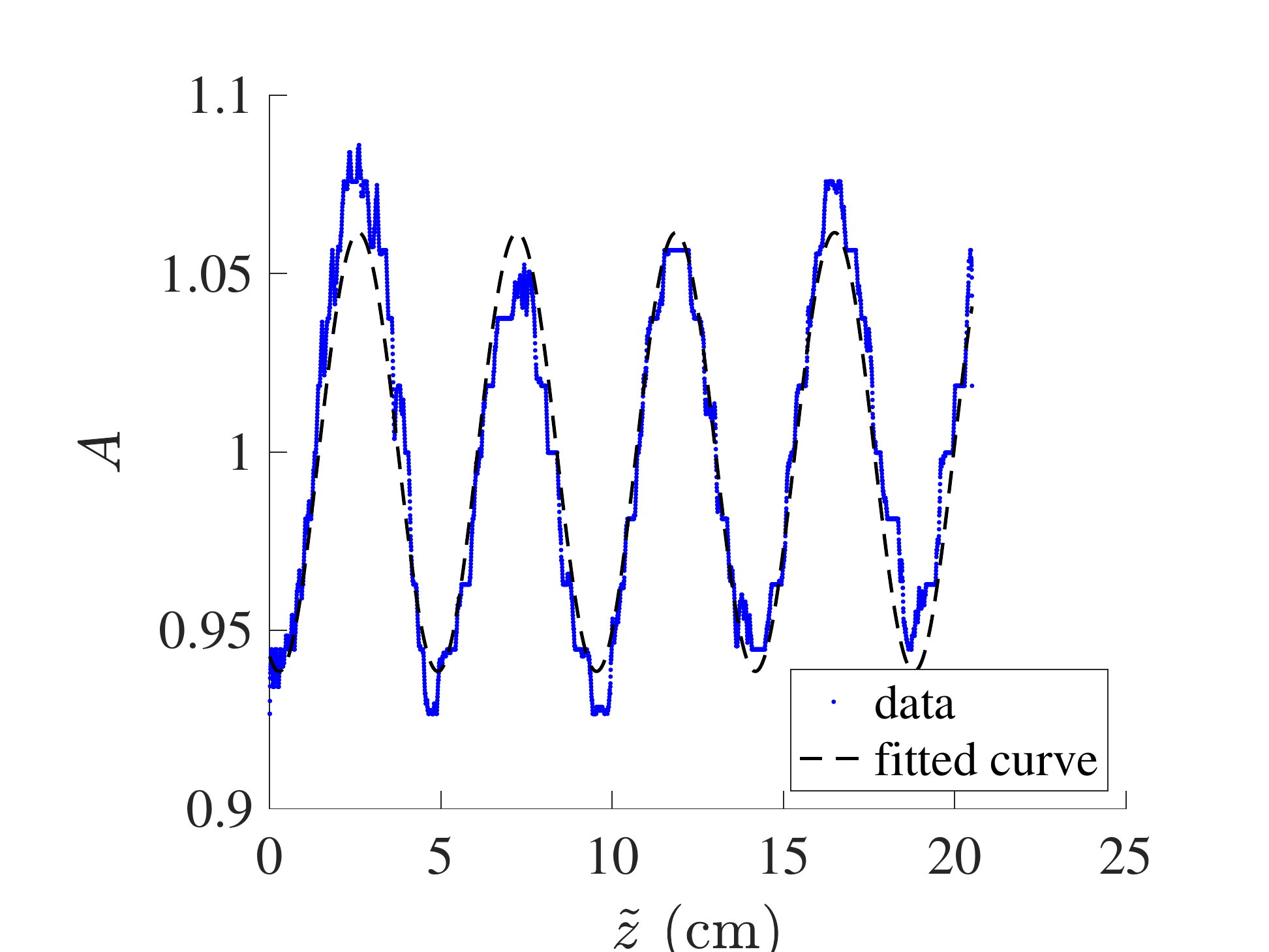} \label{fig:Fig_linear12b}}
    \caption{(a) Example experimental data of linear periodic traveling waves in the viscous fluid conduit. Measured wave parameters are $a=0.12$, $k=1.34\pm0.04$ rad/cm and $\omega=0.860\pm0.002$ rad/s. (b) Linear wave data in spatial domain fitted with sinusoidal waveform. } 
\end{figure}

\begin{figure}
    \centering
    \subfloat[]{\includegraphics[height=0.3\textwidth]{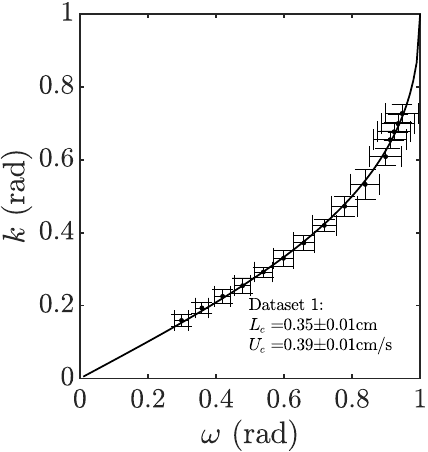}}
    \subfloat[]{\includegraphics[height=0.3\textwidth]{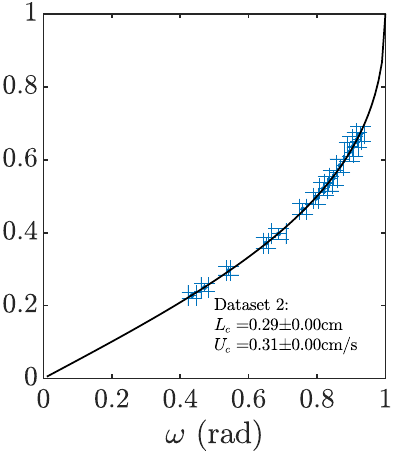}}
    \subfloat[]{\includegraphics[height=0.3\textwidth]{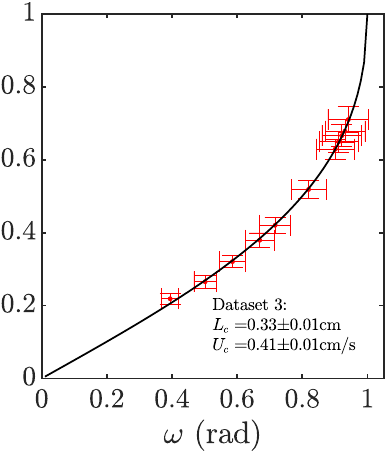} \label{fig:Fig_conduit_vs_expc}}
    \caption{Fit of the experimental measurements $k(\omega)$ (dots) to the conduit linear dispersion relation (black curve). Error bars take account of the errors in measurements using (\ref{eq:method_fitting}) and errors in nondimensionalization. }
    \label{fig:Fig_linear3}
\end{figure}

An example linear periodic traveling wave launched in a viscous fluid conduit by the sinusoidal boundary condition $Q(\Tilde{t})=Q_0(1+\frac{a}{2}\cos(\Tilde{\omega} \Tilde{t} + \pi/2))^2$ for $\Tilde{t}>0$ is depicted in figure \ref{fig:Fig_linear12a}. The wave traveling in the positive $\Tilde{z}$ direction is periodic in both time and space. Extracting the spatial wave at a fixed time, we fit the periodic wave with a cosine function in figure \ref{fig:Fig_linear12b}. This result confirms the prediction from modulation theory that a fully developed periodic wave is generated for input frequency less than the critical value (see equations (\ref{eq:conduit_linear_disp_2}) and (\ref{eq:modulation_dispersion})). Three independent experiments with different fluid properties resulting in three datasets of measured wavenumbers $\Tilde{k}$ and wave frequencies $\Tilde{\omega}$ were collected. In figure \ref{fig:Fig_linear3}, we compare the dimensionless experimental observations of $k(\omega)$ with the conduit linear dispersion relation (\ref{eq:conduit_linear_disp_2}). Across all three datasets, good agreement between the experimental data and the theoretical prediction is achieved for sufficiently long wavelengths $\Tilde{\zeta}=\frac{2\pi}{k}L_c \gtrapprox 2.3$ cm ($k<0.8$). The linear dispersion relation of the conduit equation is quantitatively verified in this regime. However, the experimental critical frequency $\omega_{cr}$ and critical wavenumber $k_{cr}$ corresponding to the values between the last datapoint of each dataset in figure \ref{fig:Fig_linear3} and the first point where the wave is damped are subject to a significant shift from 1, the prediction from the conduit equation. The relative discrepancies in $\omega_{cr}$ and $k_{cr}$ are $(\Delta \omega_{cr}, \Delta k_{cr})=(5\%, 27\%),(7\%, 33\%), (6\%, 29\%)$, respectively. It is notable that the observed critical frequency and wavenumber vary across the three datasets where the fluid properties differ. The interior flow rate is calculated using Poiseuille's law and (\ref{eq:nondim_scale}) for a motionless external fluid so that $Q^{(i)}=\pi U R_0^2$, where $U=U_c$ is the fitted velocity scale and $R_0$ is the conduit radius measured from experiments. Datasets 1-3 have measured radius $R_0=0.19,0.17,0.18$ cm. As expected for the asymptotic validity of the conduit equation, the vertical wavelength is much larger than the horizontal diameter, i.e. $\Tilde{\zeta}\gg 2R_0$. We obtain $Q^{(i)}=2.7,1.7,2.5$ ml/min for each dataset compared to the nominal input flow rates $Q_0=3.0\pm0.1, 2.0\pm0.1, 3.0\pm0.1$ ml/min, respectively.

% \begin{table}[tb!] \rule{\textwidth}{0.4pt}
% \begin{tabular}{cccccccc}
% Dataset & \multicolumn{2}{c}{Measured} & \multicolumn{2}{c}{Conduit} & \multicolumn{2}{c}{Relative error}\\
%  & $\omega_c$ & $k_c$ & $\omega_c$ & $k_c$ & $\Delta \omega_c$ & $\Delta k_c$ \\
% 01 & $0.947\pm 0.060$ & $0.727\pm0.024$ & 1.000 & 1.000 & 5.1\%  & 26.7\%  \\
% 02 & $0.927\pm0.030$ & $0.671\pm0.011$ & 1.000 & 1.000 & 7.3\%  & 32.9\% \\
% 03 & $0.941\pm0.068$ & $0.710\pm0.036$ & 1.000 & 1.000 & 5.9\%  & 29.0\%  
% \end{tabular} \rule{\textwidth}{0.4pt}
% \caption{Comparison of critical frequency $\omega_c$ and critical wavenumber $k_c$ between measurements with $L_c,U_c$ and the conduit linear dispersion relation. $\Delta \omega_c$ and $\Delta k_c$ report the relative differences.}
% \label{tab:linear_1}
% \end{table}

\begin{figure}[tb!]
    \centering
        \subfloat[]{\includegraphics[height=0.3\textwidth]{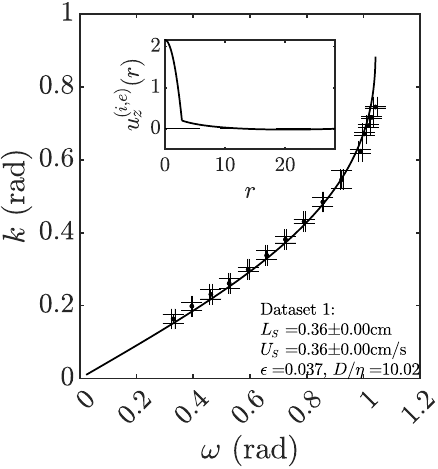} }
        \subfloat[]{\includegraphics[height=0.3\textwidth]{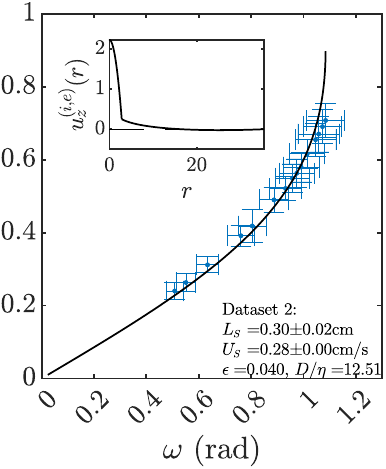} }
        \subfloat[]{\includegraphics[height=0.3\textwidth]{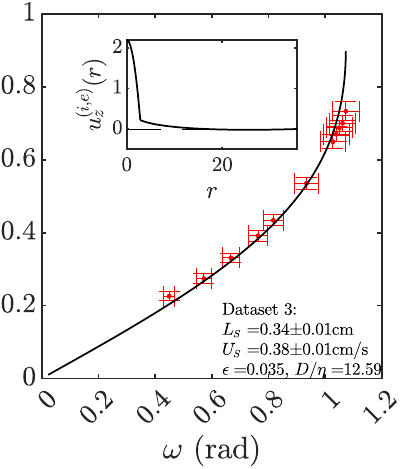} \label{fig:Fig_Stokes_vs_expc} }
    \caption{Experimental results of linear periodic waves $k(\omega)$ (dots) fitted with the two-Stokes dispersion relation assuming (\ref{eq:twoStokes_lambda2}) (black curve). Subfigures report predicted recirculating vertical flow velocities.} 
    \label{fig:Fig_Stokes_vs_exp_2}
\end{figure}

To further interpret the linear periodic waves obtained in experiments and investigate the shorter wave regime in the neighborhood of the critical frequency, we compare the experimental results with the linear dispersion relation for two-Stokes interfacial waves (\ref{eq:two_Stokes_w}) subject to the assumption $Q^{(e)}=0$ in (\ref{eq:twoStokes_Qe}) so that $\lambda$ follows from (\ref{eq:twoStokes_lambda2}). We fit the experimental data $\Tilde{k}(\Tilde{\omega})$ in the subcritical regime with the two-Stokes dispersion relation using effective fitting paramters $\epsilon,D$, as well as the scales $L_S$ and $U_S$ to give an exactly matched critical frequency $\omega_{cr}$. Each dataset was collected independently and therefore possesses distinct fluid parameters although they are all similar. Figure \ref{fig:Fig_Stokes_vs_exp_2} shows the fits and table \ref{tab:linear_3} reports the corresponding $\omega_{cr}$ and $k_{cr}$. It is demonstrated that this fitting procedure results in an upshift in the critical frequency and a downshift in the critical wavenumber. Two-Stokes linear dispersion curves well describe the dispersion relation $k(\omega)$ and provide improved predictions for the critical values. Errors in the critical wavenumber $k_{cr}$ are caused by the step size of the input $\omega$ in experiments, the sensitivity of $k_{cr}$ on fluid parameters, and limitations in the fitting method. As $\omega$ approaches $\omega_{cr}$, $k'(\omega)\to \infty$ so that a small change in $\omega$ will lead to a substantial difference in $k$, leading to difficulties in approximating $k_{cr}$. The predicted interior flow rates from the fits using (\ref{eq:twoStokes_Qi}) for dataset 1 to 3 are $Q^{(i)}=3.1, 1.9, 2.7$ ml/min, reasonably consistent with the input $Q_0=3.0,2.0,3.0$ ml/min, and the predicted interior radius obtained from (\ref{eq:nondim_scale}) are $R_0=0.20, 0.17, 0.18$ cm, respectively.
The two-Stokes dispersion relation follows from the physical scenario that we observed in experiments where the interior fluid flows up and pools at the top over time while the exterior flow recirculates with a zero net flow rate.

We have shown that the two-Stokes recirculating flow that includes an external fluid pressure gradient and exterior flow mass conservation is a slightly more accurate model of propagating linear waves generated by a wavemaker than the long-wave conduit equation for describing relatively short waves between two viscous fluids in the linear regime. It is notable, however, that the conduit equation accurately reproduces the subcritical data.

\begin{table}[tb!] \rule{\textwidth}{0.4pt}
\begin{tabular}{ccccccc}
Dataset & \multicolumn{2}{c}{Measured} & \multicolumn{2}{c}{Two-Stokes} & Relative error  \\
 & $\omega_{cr}$ & $k_{cr}$  & $\omega_{cr}$ & $k_{cr}$ & $\Delta k_{cr}$ \\
01 & $1.04\pm 0.01$ & $0.75\pm0.01$ & 1.04 & 0.88 & 15\%  \\
02 & $1.08\pm 0.07$ & $0.71\pm0.05$ & 1.08 & 0.90 & 21\%  \\
03 & $1.07\pm 0.05$ & $0.73\pm0.03$ & 1.07 & 0.90 & 19\%
\end{tabular} \rule{\textwidth}{0.4pt}
\caption{Comparison of $\omega_{cr}$ and $k_{cr}$ between experimental measurements and the two-Stokes linear dispersion relation assuming (\ref{eq:twoStokes_lambda2}). The fitting method requires an exactly correct $\omega_{cr}$. $\Delta k_{cr}$ reports the relative difference of $k_{cr}$. }
\label{tab:linear_3}
\end{table}

\subsection{Linear wave damping and arresting} \label{sec:experiment_damping}

\begin{figure}
    \centering
        \subfloat[]{\includegraphics[height=0.32\textwidth]{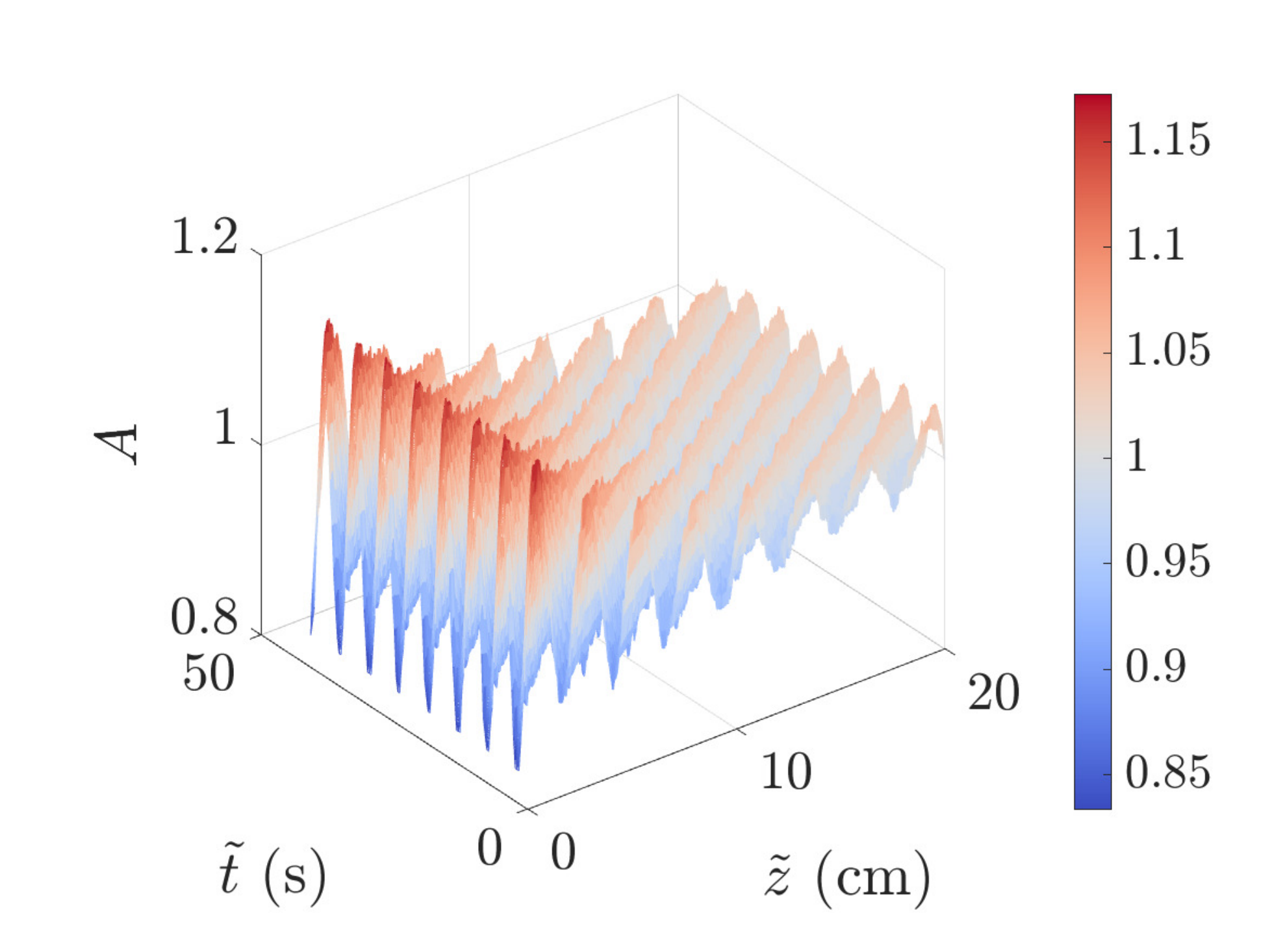} \label{fig:Fig_damping123a}} 
        \subfloat[]{\includegraphics[height=0.32\textwidth]{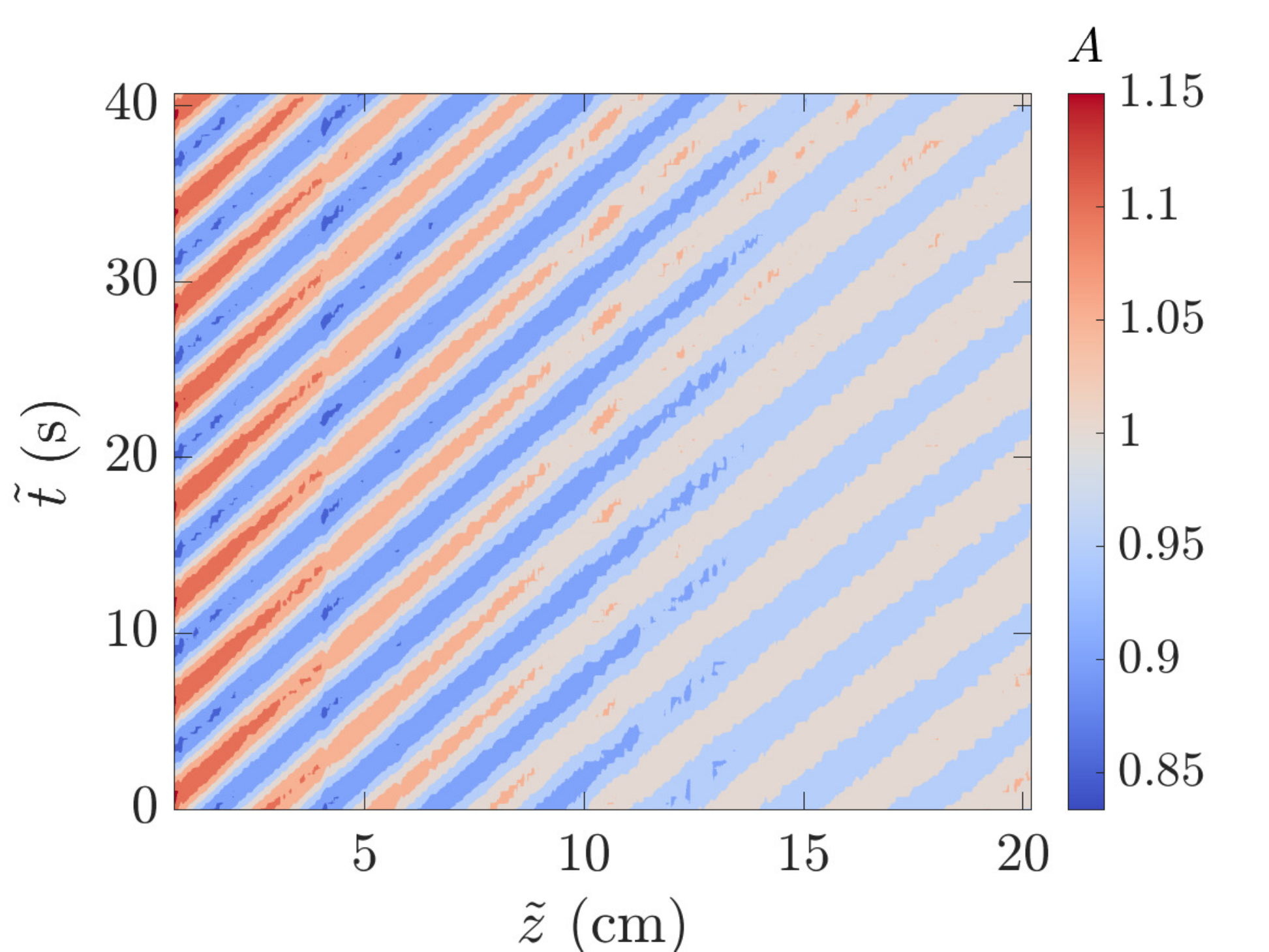} \label{fig:Fig_damping123b}} \\
        \subfloat[]{\includegraphics[width=0.35\textwidth]{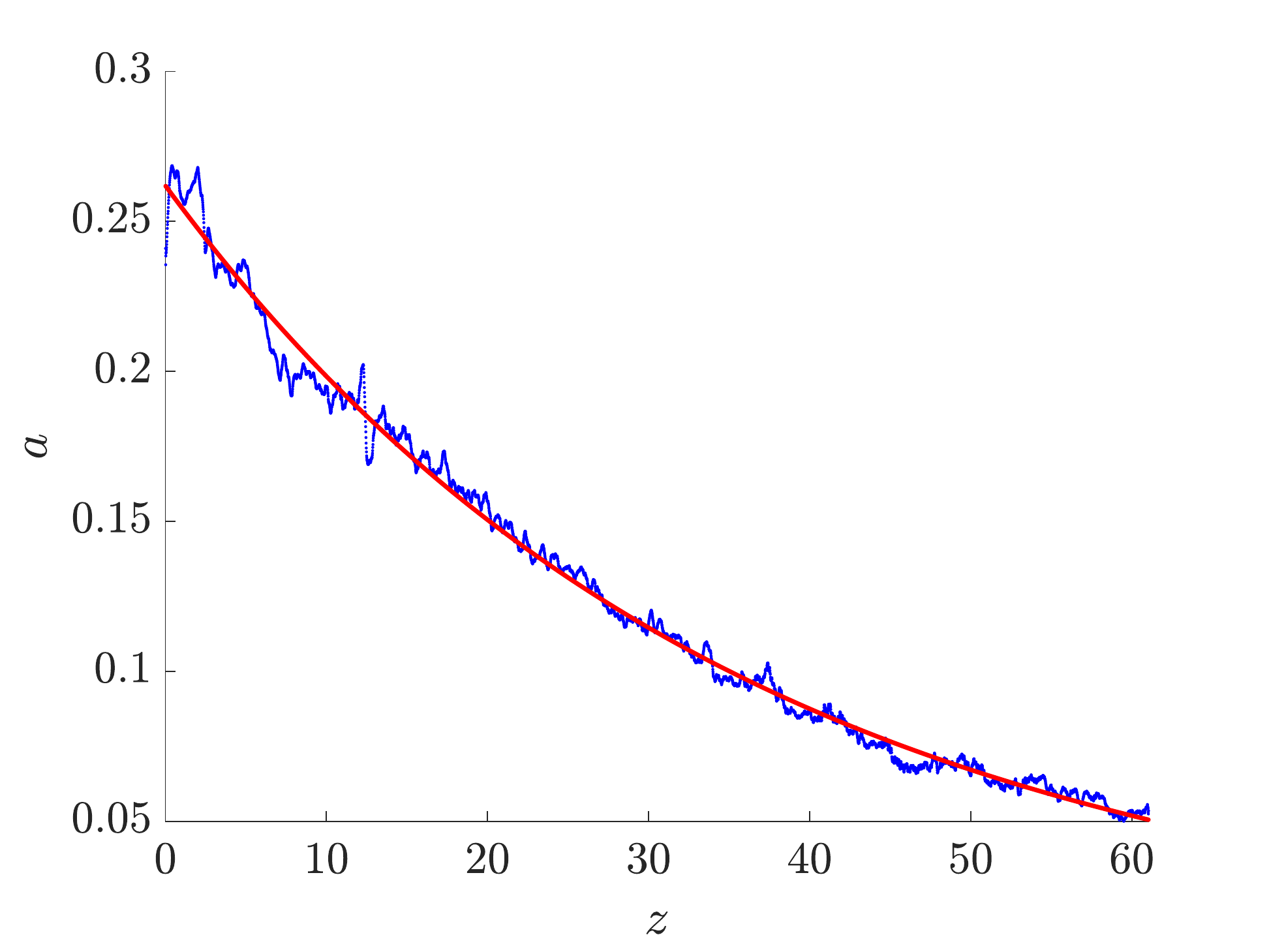} \label{fig:Fig_damping123c}}
    \caption{(a) Surface and (b) contour plots of an experimental small amplitude, damped, non-propagating wave when the injection frequency at the boundary exceeds the critical value. Dimensional wave parameters are $\Tilde{\omega}=1.12\pm0.03$ rad/s and $\Tilde{k}=2.32\pm0.04$ rad/cm with scales $L=0.35$ cm, $U=0.40$ cm/s. (c) Amplitude decay (dotted blue) fitted with $a\exp(-bz)$, $a=0.26\pm0.01$, $b=0.027\pm0.001$ (solid red).} 
    \label{fig:Fig_damping123}
\end{figure}

\begin{figure}
    \centering
        \subfloat[The conduit linear dispersion relation. ] {\includegraphics[height=0.25\textwidth]{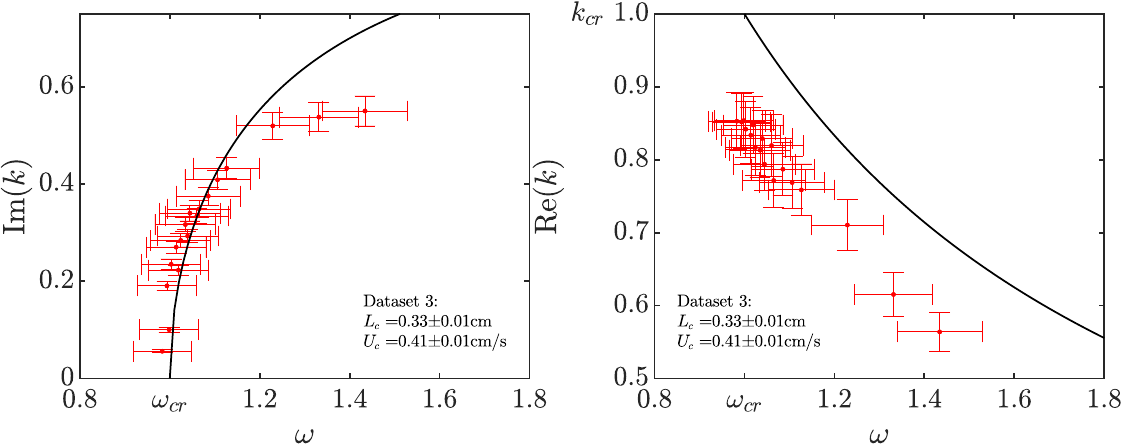} }\\
        \subfloat[The two-Stokes dispersion relation.] {\includegraphics[height=0.25\textwidth]{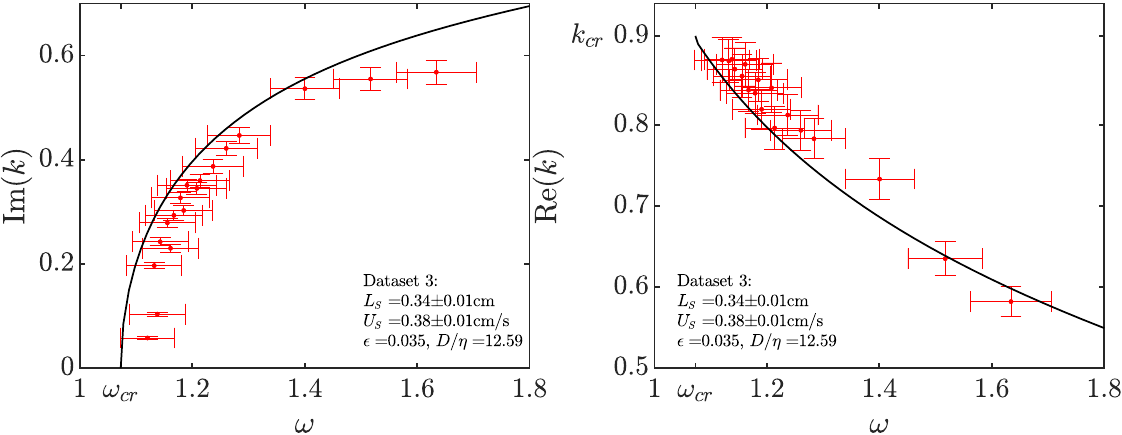}}
    \caption{Comparison between experimental measurements (red dots) of the exponential spatial decay rate (left), spatial frequency (right) and theory (black line) for waves in the supercritical regime $\omega>\omega_{cr}$. } 
    \label{fig:Fig_damping45}
\end{figure}

\begin{figure}
    \centering
        \subfloat[]{\includegraphics[width=0.32\textwidth]{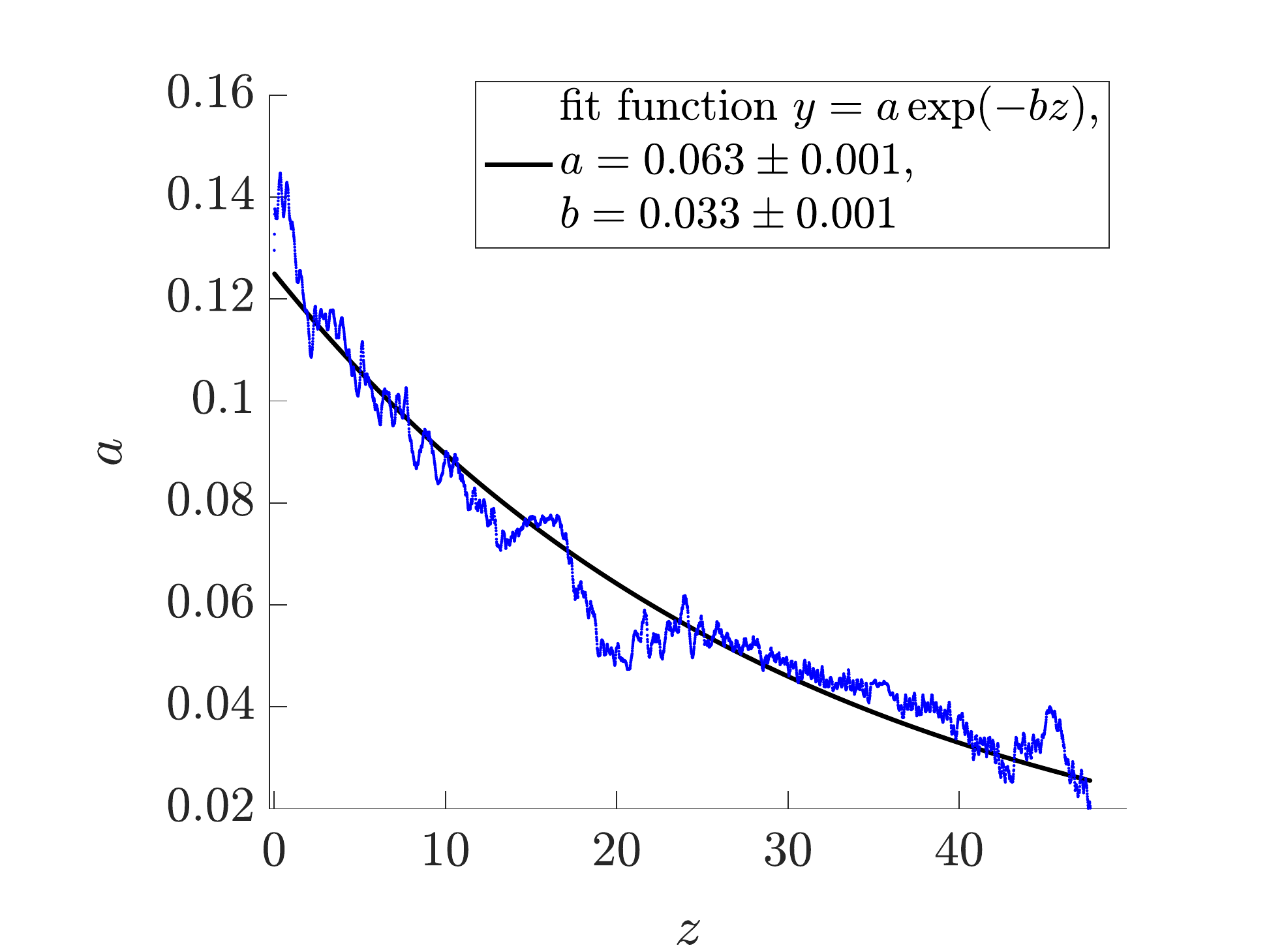}}
        \subfloat[]{\includegraphics[width=0.32\textwidth]{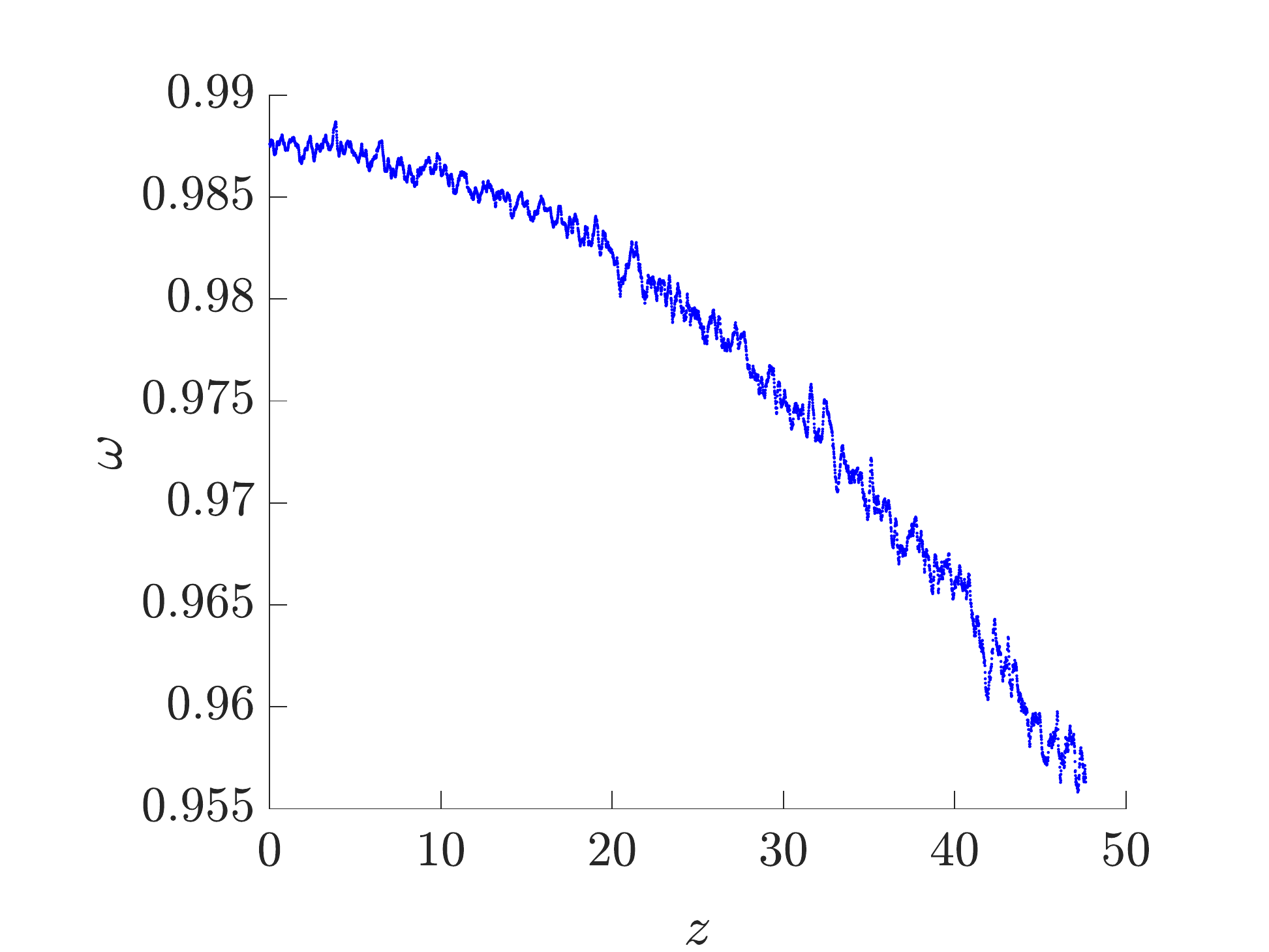}}
        \subfloat[]{\includegraphics[width=0.32\textwidth]{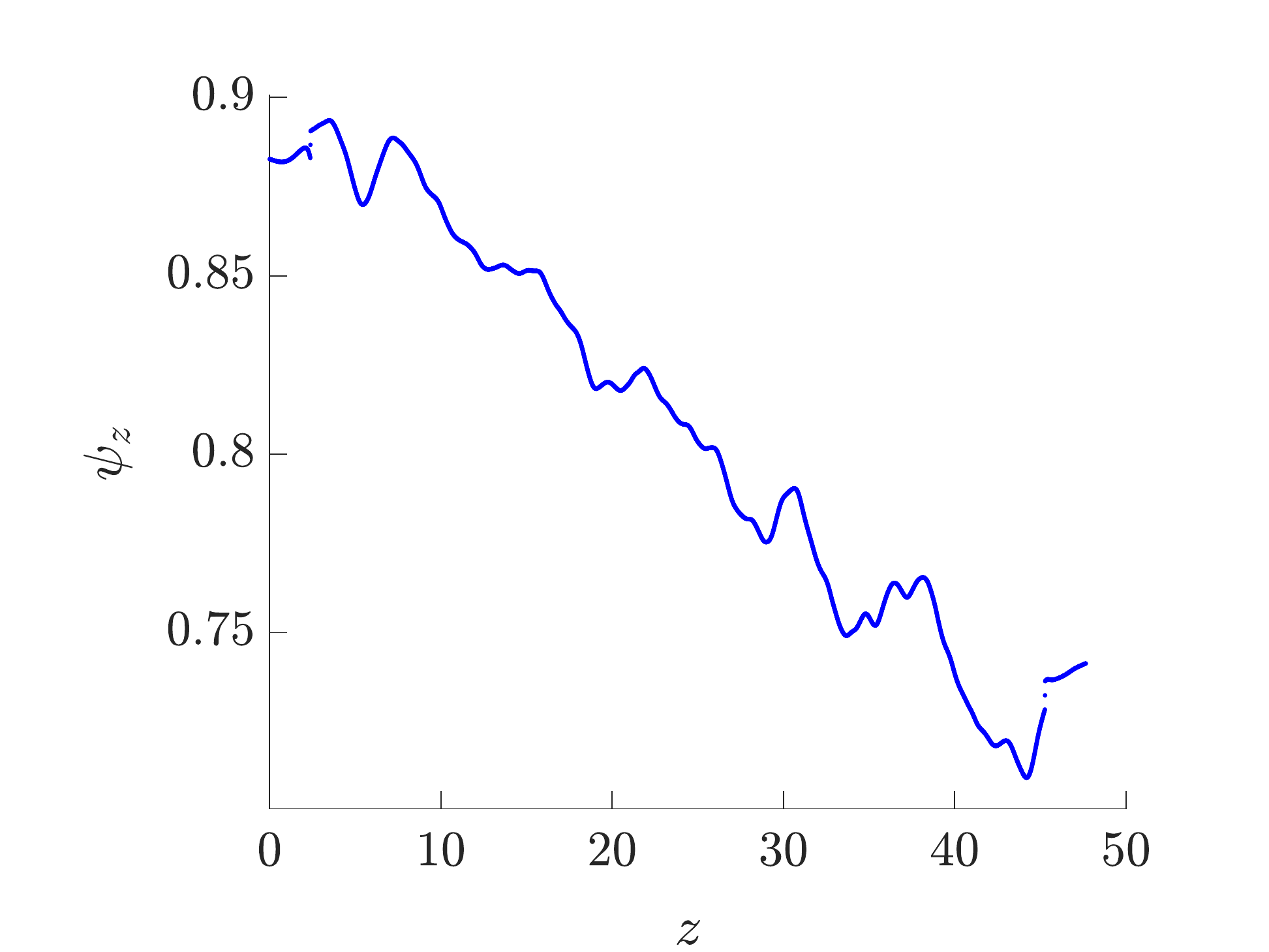}}
        % \subfloat[]{\includegraphics[width=0.45\textwidth]{Fig/Fig_damping9.pdf}}
    \caption{Investigation of a linear damping wave profile during propagation. (a) Decay of amplitude is fitted with an exponential function. (b) Frequency is also found to slightly decrease ($\sim 3\%$) in space. (c) The phase slope is subject to a decay of $\sim 20\%$. Phase is measured from the temporal wave at fixed $z$ and smoothed. 
    % (d) Changes in $k$ and $\omega$ still follow from the full linear dispersion relation at $\epsilon=0.02, \alpha=0.35$. Nondimensionalization scales are $L_S=0.29$ cm and $U=0.35$ cm/s.
    } 
    \label{fig:Fig_damping6-9}
\end{figure}

Linear periodic waves in the short-wave, supercritical regime when the injection frequency $\omega$ exceeds the critical frequency $\omega_{cr}$ exhibit spatial amplitude damping and do not propagate unhindered. Figure \ref{fig:Fig_damping123a} and \ref{fig:Fig_damping123b} demonstrate an example of a damped wave observed in a viscous fluid conduit. The wave generated by a time harmonic injection rate at the nozzle quickly decays spatially while maintaining periodicity in time at each fixed spatial location. By measuring the spatial dependence of the amplitude of the wave in figure \ref{fig:Fig_damping123c}, we find that it is well fitted by an exponential function in space. This agrees with our hypothesis derived from the radiation condition (\ref{eq:conduit_linear_disp_2}) so that for complex dispersion $k=k_\text{Re} + i k_\text{Im}$, the plane wave solution takes the form  
\be 
    A(z,t) = a e^{- k_\text{Im} z}\cos(k_\text{Re} z+ \omega t + \psi),
\ee 
where $k_\text{Im}$ is the spatial damping rate and $k=k(\omega)$ is obtained from the linear dispersion relation in the supercritical regime.  

The exact linear dispersion relation of the conduit equation and the two-Stokes flow system obtained in section \ref{sec:two_Stokes_flow} are used to compare with the experimental wave profiles. Here we report the data points in figure \ref{fig:Fig_damping45} as collected using the same experimental setup for those in figures \ref{fig:Fig_conduit_vs_expc} and \ref{fig:Fig_Stokes_vs_expc}; therefore, identical parameters and nondimensionalization scales are used. No new fitting is applied. For a total of 18 experimental trials, $k_\text{Im}$ is obtained as the spatial amplitude damping rate by fitting an exponential to the wave amplitude, $\omega$ is the averaged frequency over a spatial window size of $z\in[0,2] \to [0,0.7\text{cm}]$ (250 data points) measured through the cosine fitting method described in section \ref{sec:experiment_method}, and $k_\text{Re}$ is the slope of the temporal wave phase over the same spatial window. Figure \ref{fig:Fig_damping45} reports the comparison of the linear dispersion relation in the supercritical regime from experiments and theory. It is shown that the experimental results closely match the two-Stokes theoretical predictions and the first data point of the damping wave is also well approximated by the two-Stokes predicted $\omega_{cr},k_{cr}$. The conduit dispersion relation also performs well at describing $k_\text{Im}$ but significantly overshoots $k_\text{Re}$.

Damped waves are found to exhibit a spatially dependent frequency and wavenumber as shown in figure \ref{fig:Fig_damping6-9}. Through temporal data fitting with the function $\phi(t) = a \cos(-\omega t + \psi)$, we obtain the amplitude $a$, frequency $\omega$ and phase $\psi$, whose slope is the local wavenumber $k$.  With an exponentially decreasing amplitude consistent with linear theory, $\omega$ and $k$ are found to spatially decrease.  We are unable to explain this spatially dependent frequency, and more notably, the decaying wavenumber. For example, nonlinear effects are insufficient to explain these observations.

Nevertheless, the conduit equation again provides a reasonable description of part of the supercritical data. We now focus solely on the conduit equation to model nonlinear waves.

\subsection{Periodic traveling waves in the weakly nonlinear regime} \label{sec:experiment_weakly_nonlinear}

\begin{figure}[tb!]
    \centering
        \subfloat[]{\includegraphics[width=0.4\textwidth]{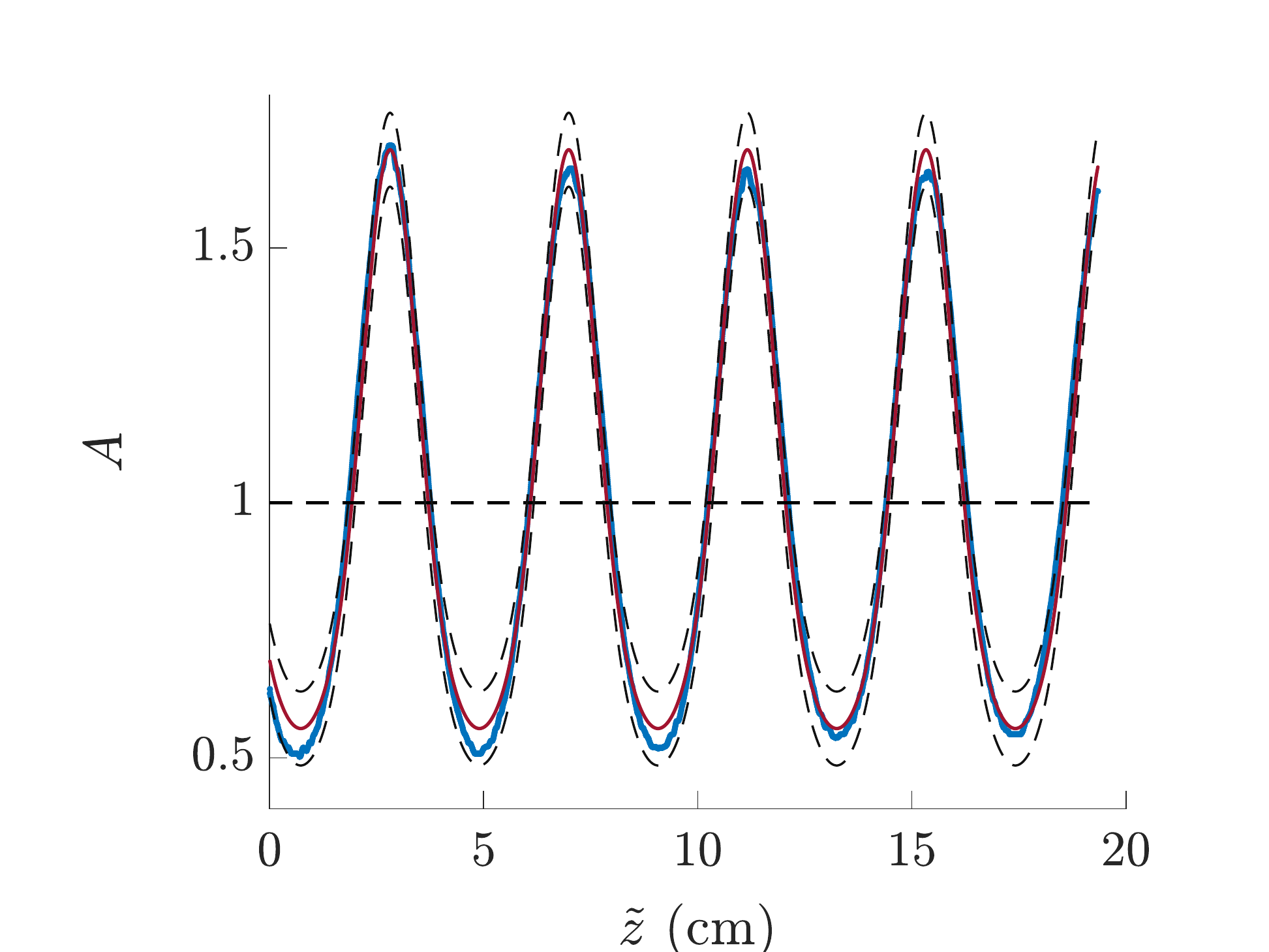} \label{fig:Fig_weakly_nonlinear_1-4a}} 
        \subfloat[]{\includegraphics[width=0.4\textwidth]{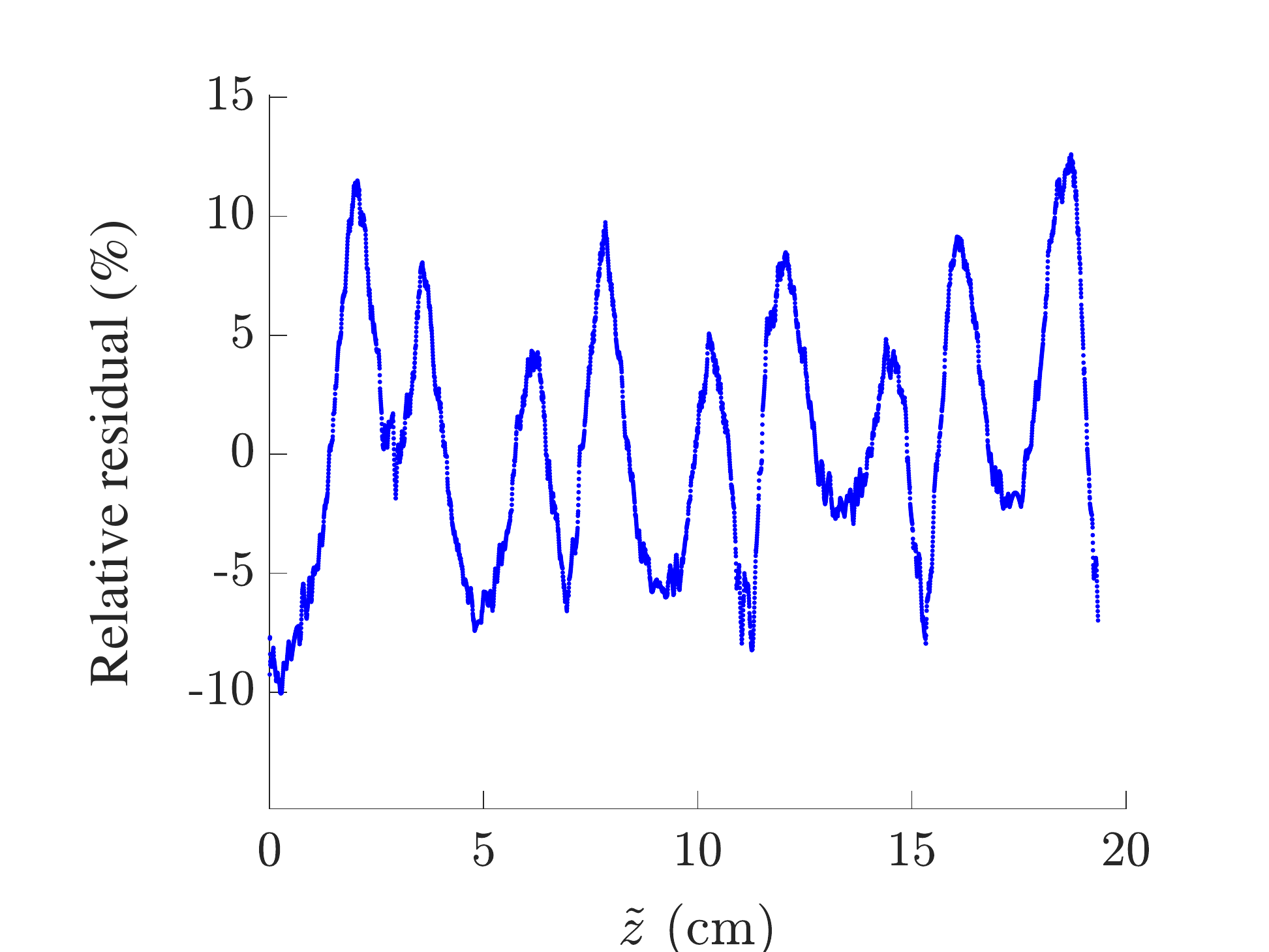}}\\
        \subfloat[]{\includegraphics[width=0.4\textwidth]{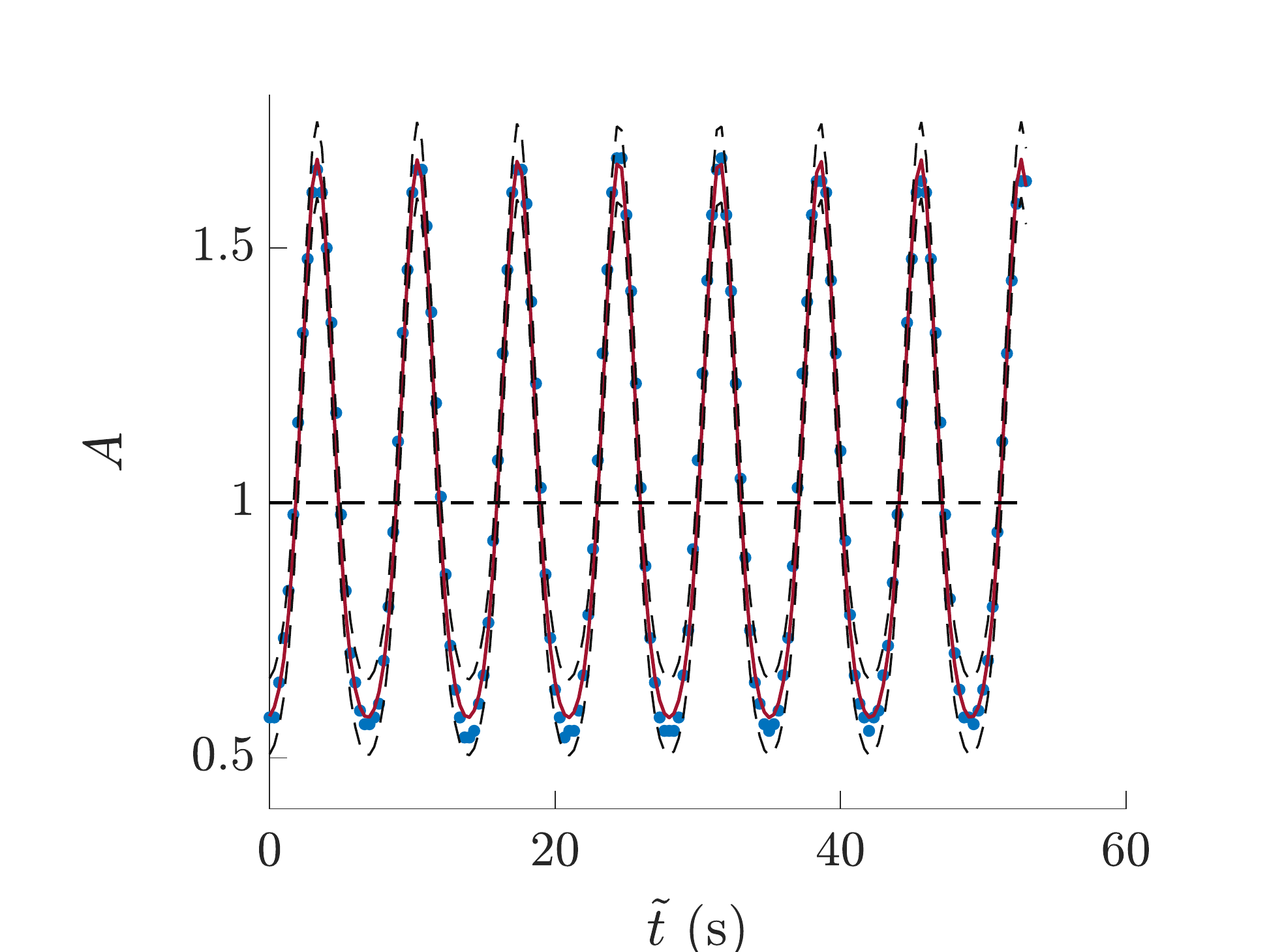} \label{fig:Fig_weakly_nonlinear_1-4c}} 
        \subfloat[]{\includegraphics[width=0.4\textwidth]{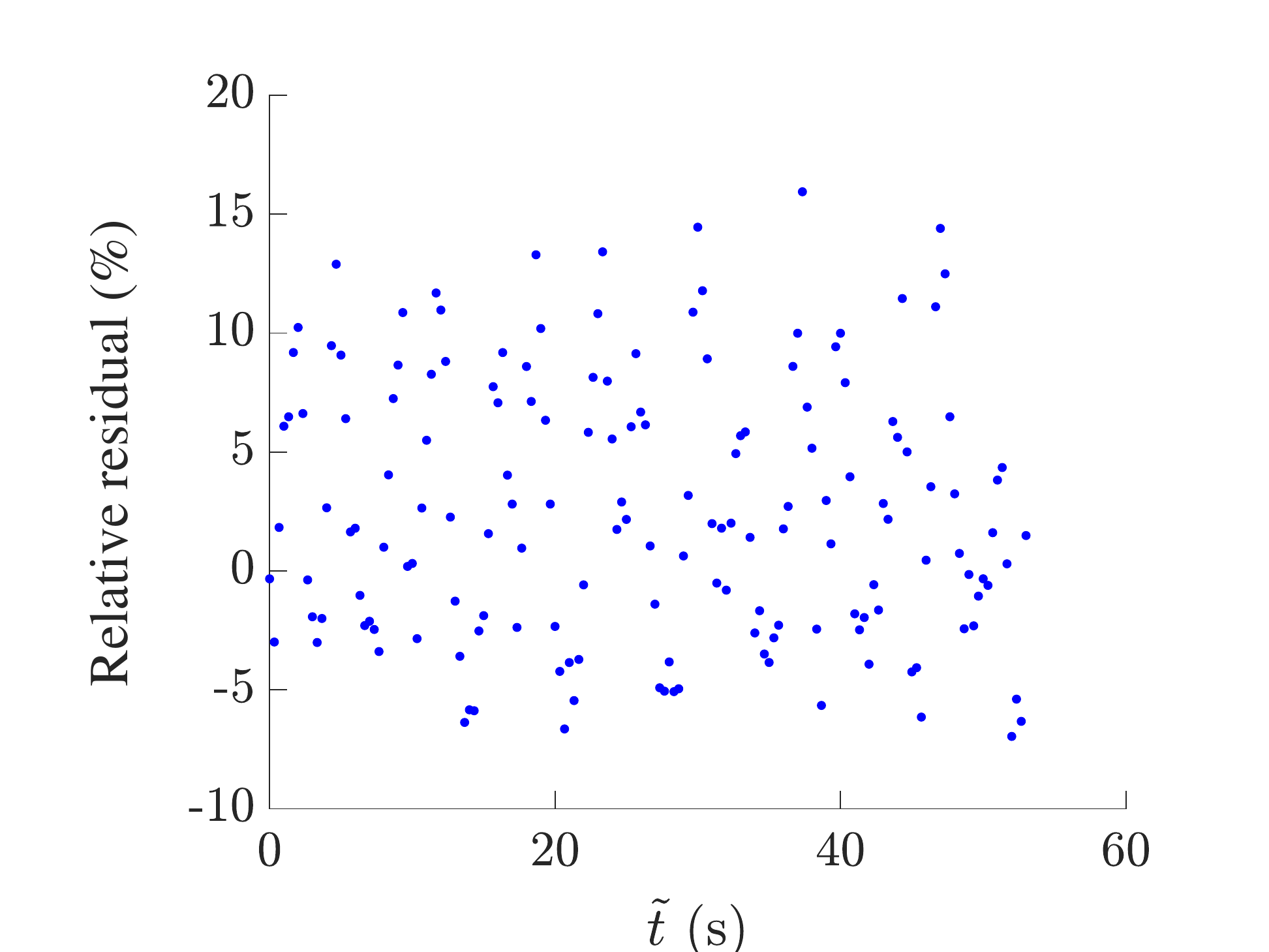}}
    \caption{Measured periodic waves (blue dots) and the weakly nonlinear approximation (red solid lines). Error bars are represented by gray dashed lines and the horizontal black dashed line is the wave mean. (a) The spatial wave at a fixed $t$ fitted with expansion (\ref{eq:weakly_nonlinear_fit1}). (b) Percentage relative error in space. (c) Wave in the time domain at a fixed $z$ fitted with expansion (\ref{eq:weakly_nonlinear_fit2}). (d) Percentage relative error in time.}
    \label{fig:Fig_weakly_nonlinear_1-4}
\end{figure}

Larger-amplitude, periodic traveling waves in the weakly nonlinear regime are distinguished from their linear counterpart by no longer maintaining symmetry about the wave mean and cannot be described by a single trigonometric harmonic. The waves are more accurately modeled by the weakly nonlinear approximation (\ref{eq:intro_weakly_nonlinear_1}, \ref{eq:intro_weakly_nonlinear_2}) of the conduit equation obtained via the amplitude expansions $\phi \sim 1+a\phi_1+a^2\phi_2+a^3\phi_3$ and $k \sim k_0+a^2k_1$, where $0<a\ll1$ is an amplitude scale.
An experimental weakly nonlinear wave is shown in figure \ref{fig:Fig_weakly_nonlinear_1-4}. Length and velocity scales for nondimenionalization are $(L_c,U_c)=(0.33 \pm 0.01 \text{ cm}, 0.38\pm 0.01 \text{ cm/sec})$. The dimensional spatial wave in figure \ref{fig:Fig_weakly_nonlinear_1-4a} is fitted with the three harmonic expansion (\ref{eq:intro_weakly_nonlinear_3}) at $\Tilde{t}=\Tilde{t}_0$
\be 
    \phi(\Tilde{z}) = 1 + \dfrac{a_z}{2} \cos(\Tilde{k} \Tilde{z} + \psi_1) +  \dfrac{1+\Tilde{k}^2}{48\Tilde{k}^2} a_z^2\cos(2(\Tilde{k} \Tilde{z} + \psi_1)) + \dfrac{1+\Tilde{k}^2}{1536\Tilde{k}^4} a_z^3 \cos(3(\Tilde{k} \Tilde{z}+\psi_1)), \label{eq:weakly_nonlinear_fit1}
\ee 
where $a_z=1.10 \pm 0.01$, $k=0.50\pm0.01$ $(\Tilde{k}=1.52\pm 0.01$ rad/cm) and $\psi_1=2.05\pm 0.01$. The approximate frequency is derived from figure \ref{fig:Fig_weakly_nonlinear_1-4c}, where the dimensional temporal wave is approximated by (\ref{eq:intro_weakly_nonlinear_1}) at $\Tilde{z}=\Tilde{z}_0$
\be \begin{split}
    \phi(\Tilde{t}) = 1 &+\frac{a_t}{2}\cos(\Tilde{\omega} \Tilde{t}+\psi_2)+\frac{a_t^2}{24\left(1-\sqrt{1-\Tilde{\omega}^2}\right)}\cos(2(\Tilde{\omega} \Tilde{t} + \psi_2)) \\&+\frac{\left(-1-\sqrt{1-\Tilde{\omega}^2}\right) a_t^3 }{768\left(-2+\Tilde{\omega}^2+2\sqrt{1-\Tilde{\omega}^2}\right)}\cos(3(\Tilde{\omega} \Tilde{t}+\psi_2)), \label{eq:weakly_nonlinear_fit2}
\end{split}\ee 
with $a_t=1.06 \pm 0.02$, $\omega=0.78\pm0.01$ ($\Tilde{\omega}=0.90 \pm 0.01$ rad/s) and $\psi_2=-2.97 \pm 0.03$.  Fits exhibit a maximum absolute error of $0.11$ and have larger discrepancies at the wave peaks. Adding higher order harmonic terms yield an even better approximation. 

\begin{figure}[tb!]
    \centering
        \includegraphics[width=0.4\textwidth]{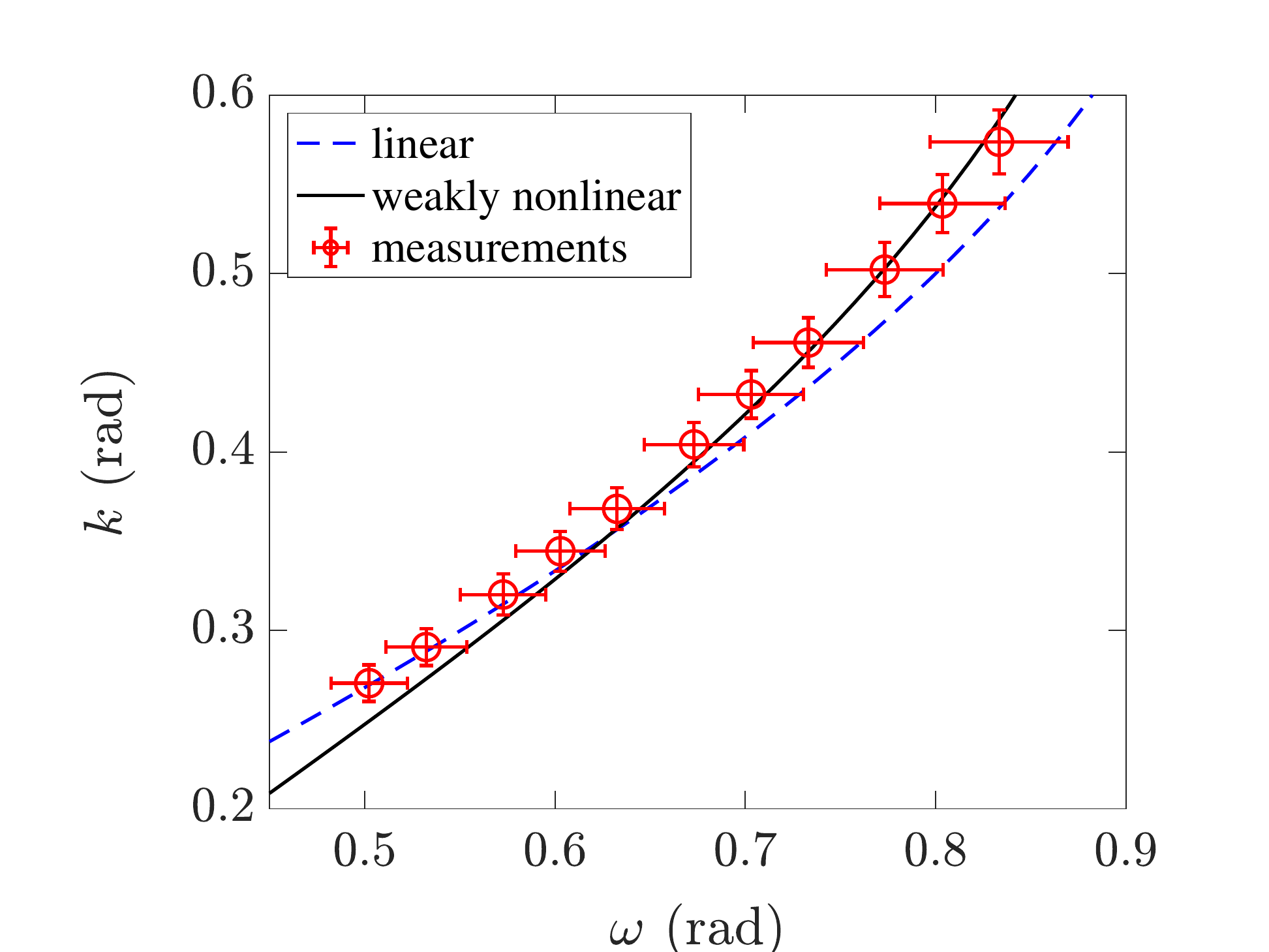}
    \caption{Nondimensional experimental wave profiles $k(\omega)$ (circles) compared with the weakly nonlinear dispersion relation (\ref{eq:intro_weakly_nonlinear_2}) at $a=1.04$ (black solid line) and the linear dispersion relation (blue dashed line).} 
    \label{fig:Fig_weakly_nonlinear_5}
\end{figure}

A total of 11 trials were measured in the weakly nonlinear regime with the same input amplitude and various input frequencies. Wave profiles are measured as follows. Fitting the spatial and temporal waves with the weakly nonlinear expansions (\ref{eq:weakly_nonlinear_fit1}, \ref{eq:weakly_nonlinear_fit2}), we obtain dimensionless $k= \Tilde{k} L,\omega =\Tilde{\omega}\frac{L}{U}$ as well as $a=\frac{a_z+a_t}{2}$ for those 11 trials as the red circles in figure \ref{fig:Fig_weakly_nonlinear_5}. Experimental results are then compared with the weakly nonlinear dispersion relation (\ref{eq:intro_weakly_nonlinear_2}) and the linear dispersion relation (\ref{eq:conduit_linear_disp_1}) in figure \ref{fig:Fig_weakly_nonlinear_5}. Although weak nonlinearity does not lead to a significant deviation from the linear dispersion relation, we have shown that the corresponding wave profiles require amplitude-dependent higher harmonics to quantitatively model the experimental observations.  We have quantitatively validated the weakly nonlinear approximation for measured conduit periodic traveling waves of sufficiently large amplitude and demonstrated the necessity of including higher order harmonic terms in this regime.

\subsection{Cnoidal-like nonlinear periodic waves}

\begin{figure}[tb!]
    \centering
        \subfloat[]{\includegraphics[width=0.4\textwidth]{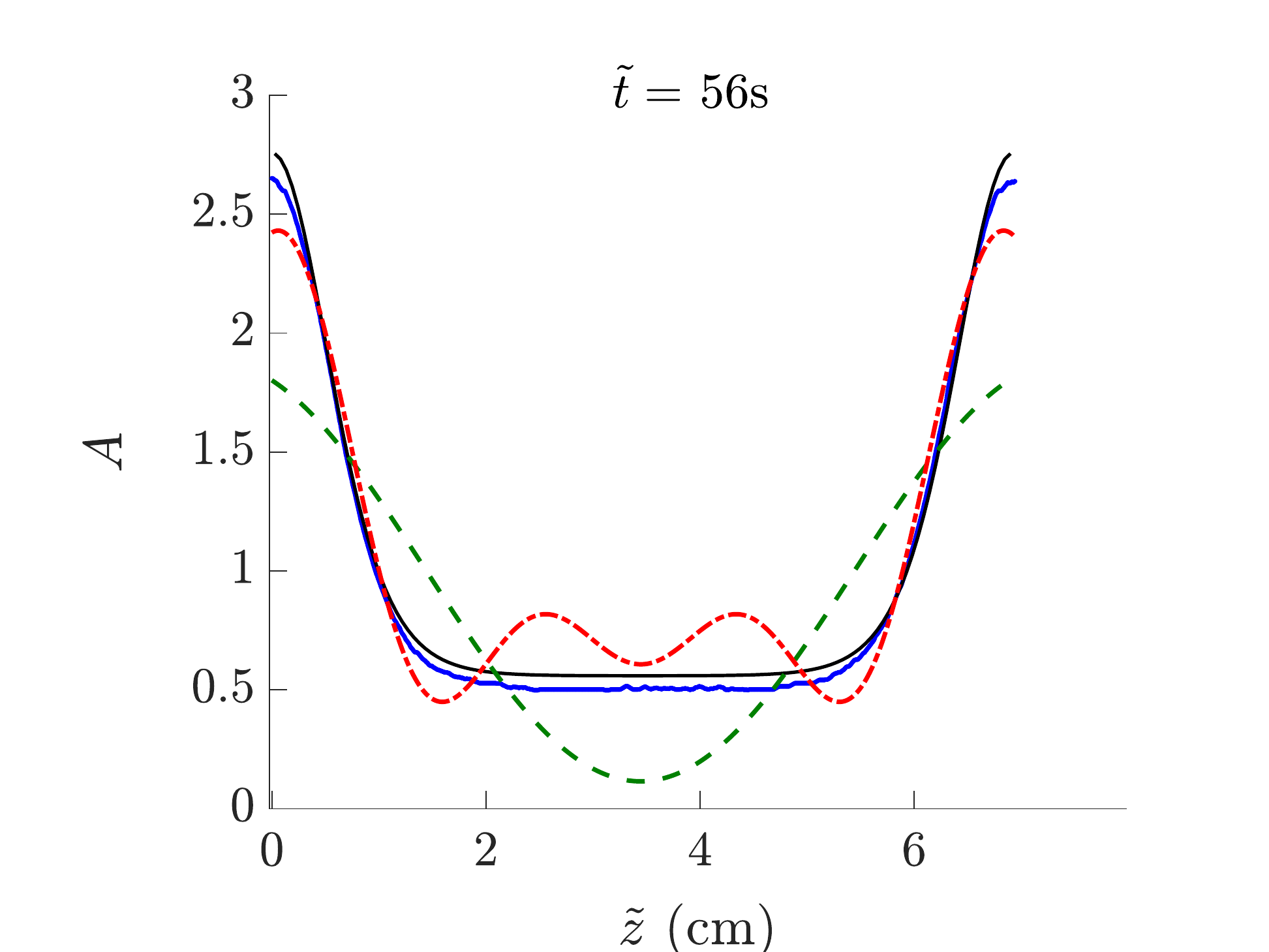}} 
        \subfloat[]{\includegraphics[width=0.4\textwidth]{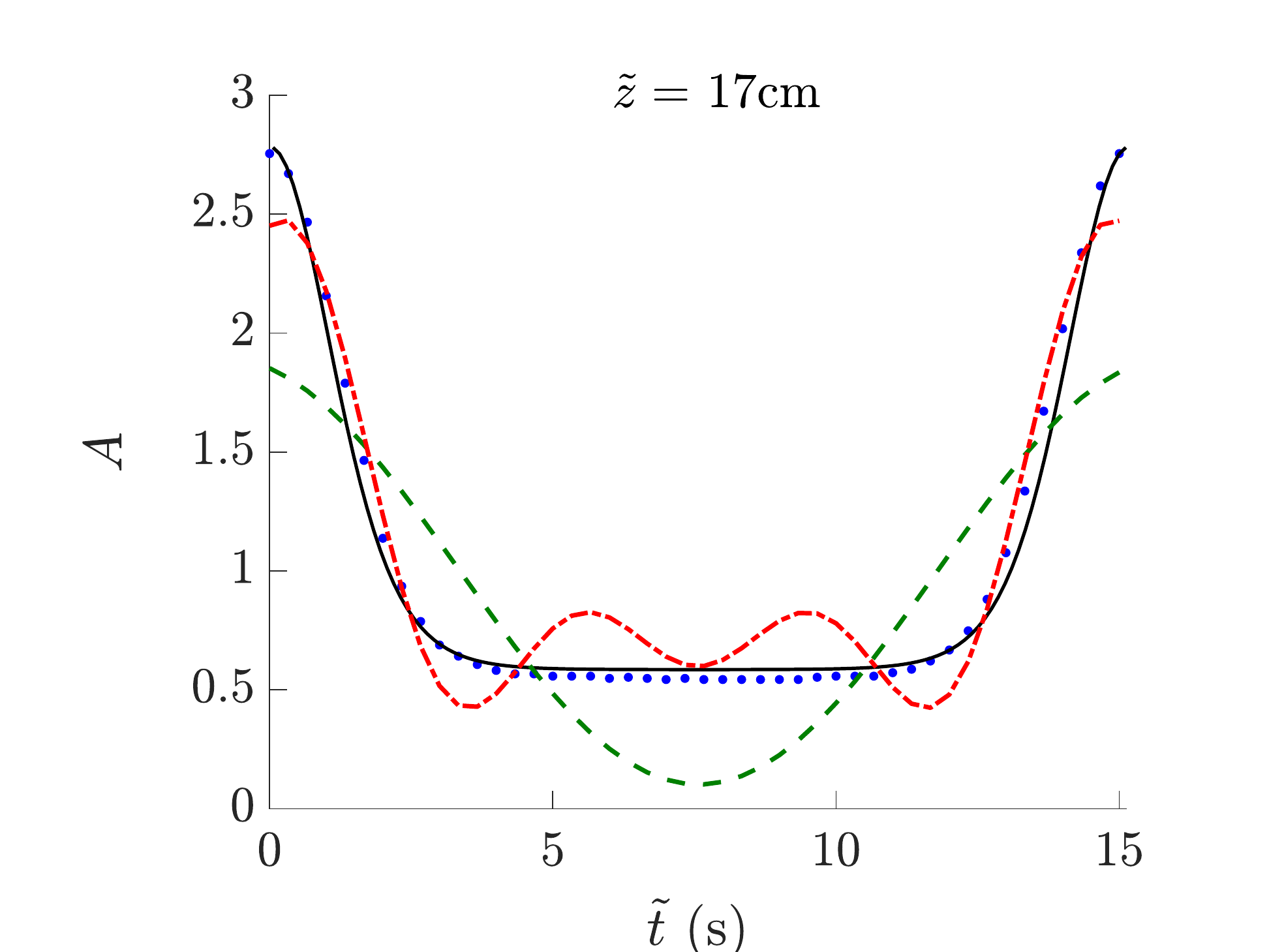}}\\
        \subfloat[]{\includegraphics[width=0.4\textwidth]{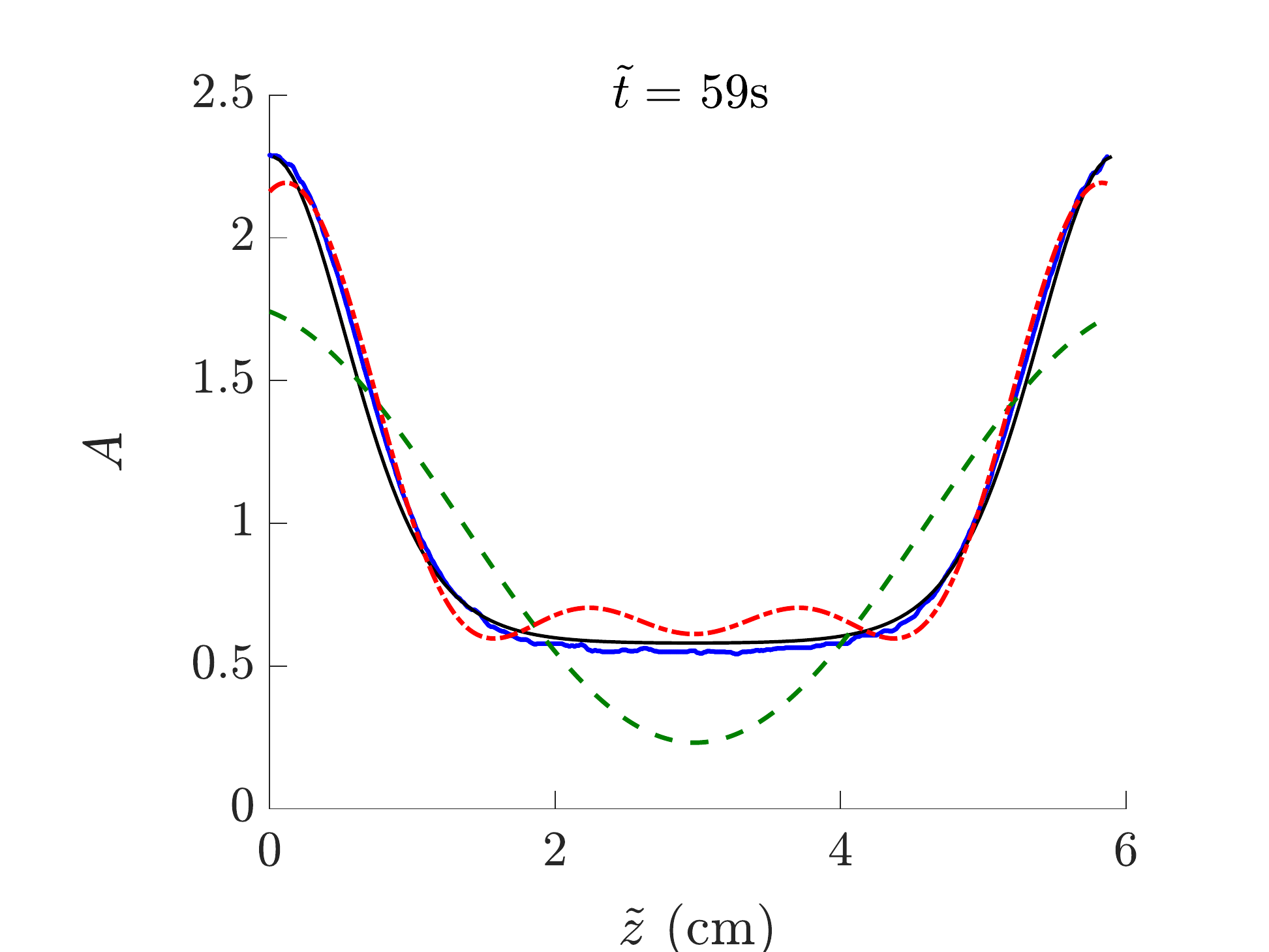}} 
        \subfloat[]{\includegraphics[width=0.4\textwidth]{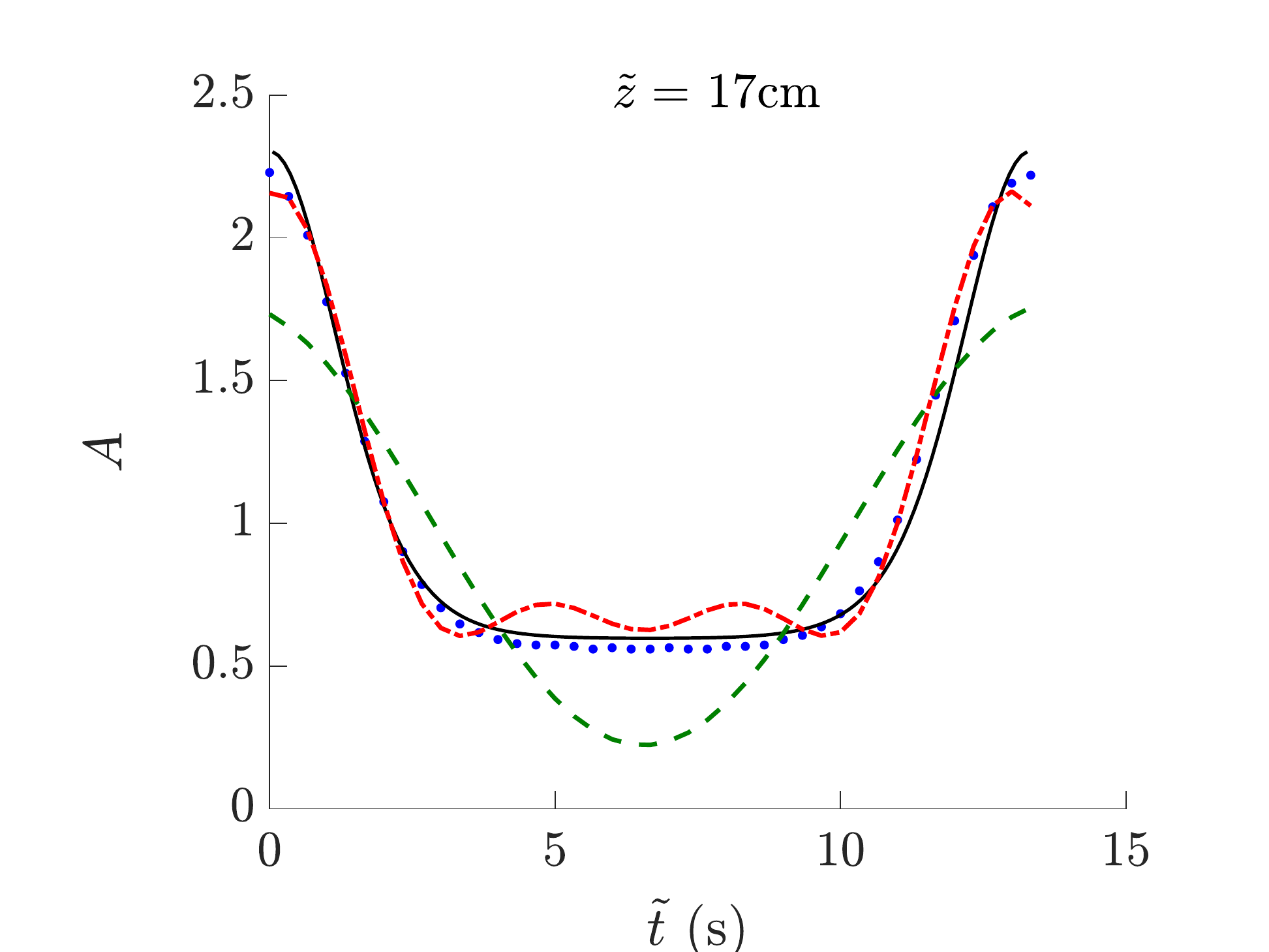}}
    \caption{Averaged experimental wave profiles over one wavelength or period (blue dots) compared with linear (green dashed), weakly nonlinear (red dash-dotted) approximations and cnoidal-like solutions (black solid) in both spatial (a, c) and temporal (b, d) domains. (a-b) The extracted wave parameters are $(a_\text{avg},\omega,k)=(2.20,0.44,0.24)$ (trial 1 in figure \ref{fig:Fig_nonlinear_56}). (c-d) The extracted parameters are $(a_\text{avg},\omega,k)=(1.71,0.50,0.28)$ (trial 3 in figure \ref{fig:Fig_nonlinear_56}). } 
    \label{fig:Fig_nonlinear_1-4}
\end{figure}

\begin{figure}[tb!]
    \centering
        \subfloat[]{\includegraphics[width=0.4\textwidth]{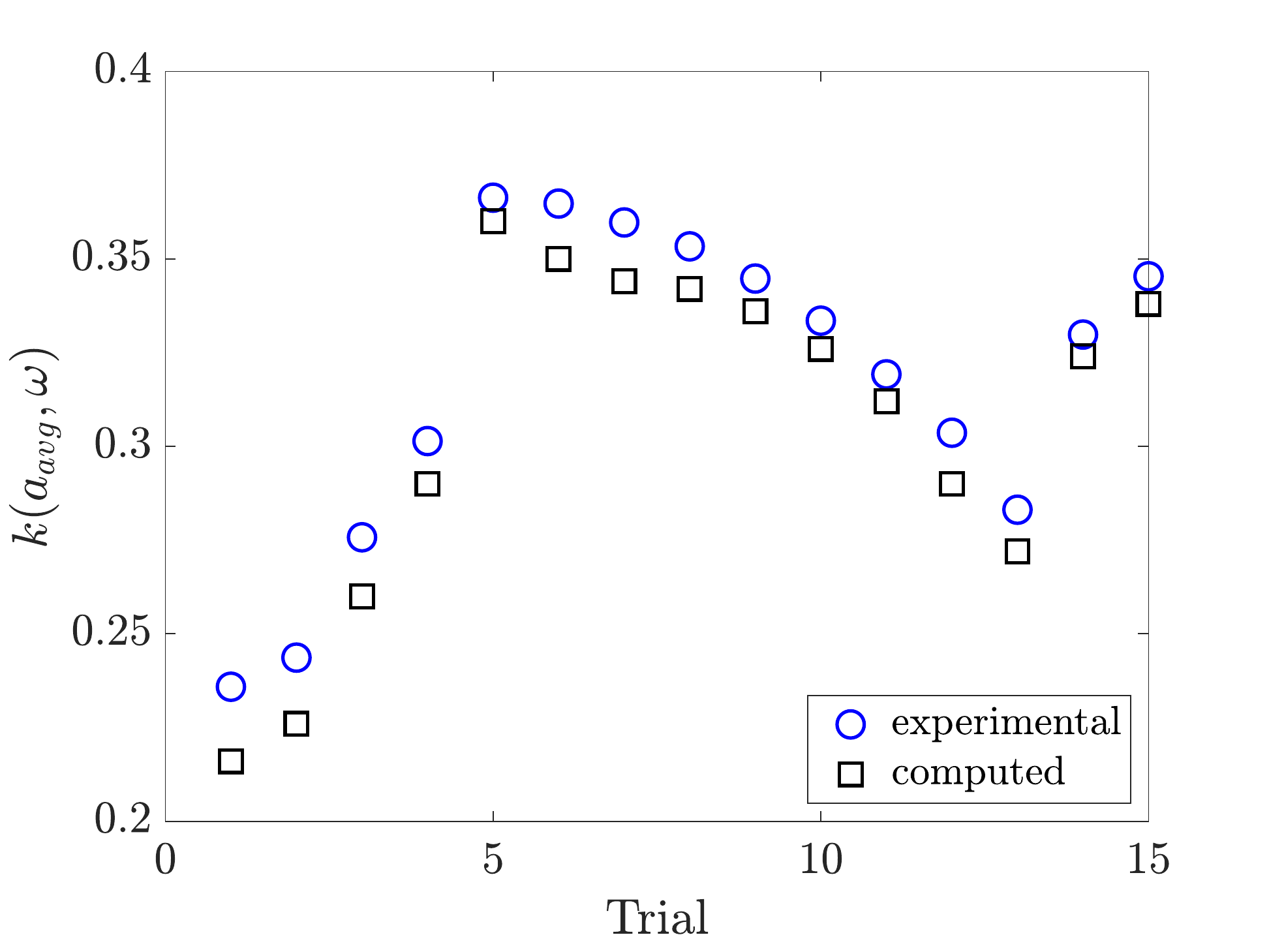}} 
        \subfloat[]{\includegraphics[width=0.4\textwidth]{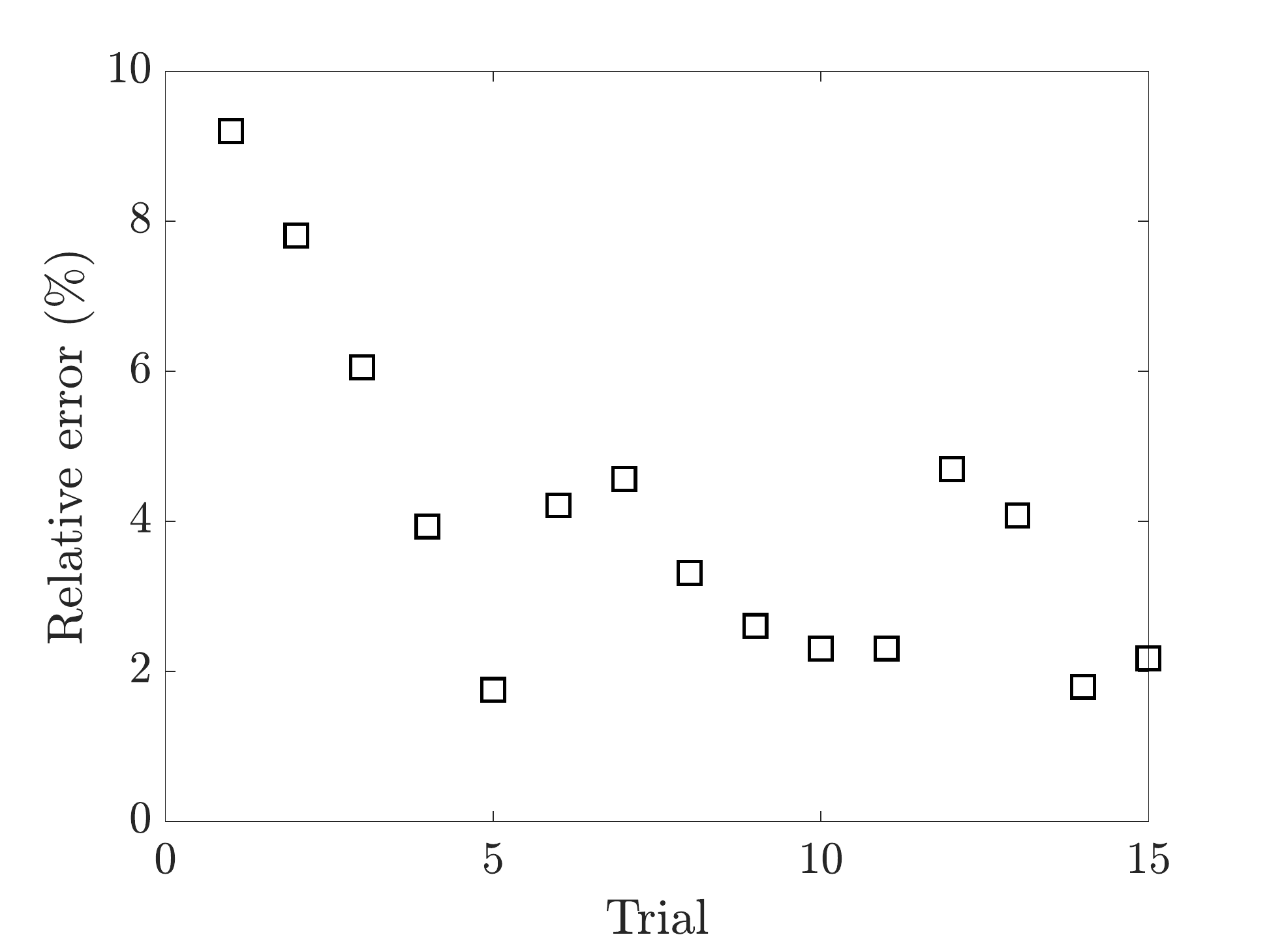}}
    \caption{(a) Comparison of the conduit equation nonlinear dispersion relation for cnoidal-like solutions. Data points are from the trials reported in table \ref{tab:nonlinear}. Utilizing measured $(a_\text{avg}, \omega)$ as the parameters, experimental wavenumbers (blue circles) are compared with the numerically computed nonlinear dispersion relation $k(a_\text{avg},\omega)$ (black squares). (b) Percentage relative error in $k$ for each trial.} 
    \label{fig:Fig_nonlinear_56}
\end{figure}

We now investigate periodic traveling wave solutions with large amplitudes in the fully nonlinear regime. We refer to these waves as cnoidal-like, owing to their similarities to cnoidal shallow water waves and solutions of the Korteweg–de Vries equation (\cite{korteweg1895xli}). First, we generate linear periodic waves for calibrating the scales $L_c=0.26\pm 0.01$ cm and $U_c=0.25\pm 0.01$ cm/sec. We then utilize a numerical database of pre-computed conduit cnoidal-like solutions to experimentally generate and analyze the observed large-amplitude periodic waves. Waves are generated by varying the injection flow rate according to
\be 
    Q(\Tilde{t}) = Q_0\phi^2(- \Tilde{\omega} \Tilde{t}),
\ee 
where $\phi$ is a cnoidal-like solution with unit mean satisfying (\ref{eq:intro_nonlinearODE}). Measured waves are processed by dividing each space and time window into spatial and temporal periods. The average of all processed periods and wave profiles are then reported. Figure \ref{fig:Fig_nonlinear_1-4} displays two example wave profiles, separately in space and time. By direct comparison, it is clear that the observed large amplitude waves cannot be described by linear or weakly nonlinear approximations. Instead, we compare the observed waves with cnoidal-like solutions to the conduit equation. The amplitude is obtained as $a_z=\max(\phi(z))-\min(\phi(z))$ for waves in the spatial domain, and $a_t=\max(\phi(t))-\min(\phi(t))$ for waves in the temporal domain. $\omega$ and $k$ are measured by the relation $\omega=\frac{2\pi}{\overline{T_0}},k=\frac{2\pi}{\overline{\zeta}}$ with nondimensional averaged time period $\overline{T_0}$ and averaged wavelength $\overline{\zeta}$.
The set of unit-mean cnoidal-like solutions is a two-parameter family. Taking $(a_z,k)$ as the parameters for spatial waves and $(a_t,\omega)$ as the parameters for temporal waves, the corresponding cnoidal-like solutions are compared with the experiments in figure \ref{fig:Fig_nonlinear_1-4}. The cnoidal-like wave profiles are barely distinguishable from the observed wave profiles. We stress that we have not performed any fitting in this procedure, other than the independent determination of the scales $L_c$ and $U_c$ from linear wave dispersion measurements. The cnoidal-like solution of the conduit equation is uniquely selected from the two-parameter family using the extracted $(a_z,k)$ in the spatial case or $(a_t,\omega)$ in the temporal case. No profile fitting was performed.
However, discrepancies in spatial and temporal profile measurements lead to $(a_z,k)$ and $(a_t,\omega)$ belonging to two slightly different cnoidal-like solutions for the same trial. We report the average of the two amplitudes $a_\text{avg}=\frac{a_z+a_t}{2}$ in table \ref{tab:nonlinear}. To see how close the solutions are, measurements of wave parameters for 15 trials are reported in table \ref{tab:nonlinear} and figure \ref{fig:Fig_nonlinear_56}.

\begin{table}[tb!] \rule{\textwidth}{0.4pt}
\begin{tabular}{c c c c c c c c c }
\multirow{2}{*}{Trial} & \multirow{2}{*}{$a_\text{avg}$}  & \multirow{2}{*}{$\omega$ (rad)}  & \multirow{2}{*}{$k$ (rad)} & \multirow{2}{*}{$k(a_\text{avg},\omega)$} & \multirow{2}{*}{Relative error}\\
 &  &  &  &  &  &  \\ 
01 & $2.20 \pm 0.03$  & $0.44\pm0.02$  & $0.24\pm0.01$ &  0.22 & $9\%$\\
02 & $2.13 \pm 0.01$  & $0.45\pm0.02$  & $0.24\pm0.01$ &  0.23 & $8\%$\\
03 & $1.71 \pm 0.01$  & $0.50\pm0.01$  & $0.28\pm0.01$ &  0.26 & $6\%$\\
04 & $1.42 \pm 0.01$  & $0.54\pm0.01$  & $0.30\pm0.01$ &  0.29 & $4\%$\\
05 & $1.30 \pm 0.01$  & $0.62\pm0.02$  & $0.37\pm0.01$ &  0.36 & $2\%$\\
06 & $1.27 \pm 0.01$  & $0.61\pm0.03$  & $0.36\pm0.01$ &  0.35 & $4\%$\\
07 & $1.26 \pm 0.01$  & $0.61\pm0.02$  & $0.36\pm0.01$ &  0.34 & $5\%$\\
08 & $1.25 \pm 0.01$  & $0.60\pm0.01$  & $0.35\pm0.01$ &  0.34 & $3\%$\\
09 & $1.24 \pm 0.01$  & $0.60\pm0.03$  & $0.34\pm0.01$ &  0.34 & $3\%$\\
10 & $1.23 \pm 0.01$  & $0.59\pm0.01$  & $0.33\pm0.01$ &  0.33 & $2\%$\\
11 & $1.22 \pm 0.01$  & $0.57\pm0.01$  & $0.32\pm0.01$ &  0.31 & $2\%$\\
12 & $1.18 \pm 0.01$  & $0.54\pm0.02$  & $0.30\pm0.01$ &  0.29 & $5\%$\\
13 & $1.15 \pm 0.01$  & $0.52\pm0.02$  & $0.28\pm0.01$ &  0.27 & $4\%$\\
14 & $1.09 \pm 0.01$  & $0.59\pm0.01$  & $0.33\pm0.01$ &  0.32 & $2\%$\\
15 & $0.89 \pm 0.01$  & $0.60\pm0.01$  & $0.35\pm0.01$ &  0.34 & $2\%$\\
% 16 & 0.77 & 2.22\% & 0.76 & 3.33\% & 0.76 & 0.35 & 1.47\% & 0.62 & 0.36\% \\
% 17 & 0.59 & 2.33\% & 0.58 & 3.83\% & 0.59 & 0.37 & 1.96\% & 0.64 & 1.53\% \\
\end{tabular} \rule{\textwidth}{0.4pt}
\caption{Nonlinear periodic wave measurements. Columns 2-4: measured wave parameters. Column 5: numerically computed solution $k$ at measured $(a_\text{avg}, \omega)$. Column 6: relative error between experimental $k$ and cnoidal-like wave solutions. Trials 1 and 3 are depicted in figure \ref{fig:Fig_nonlinear_1-4}. All trials are shown in figure \ref{fig:Fig_nonlinear_56}. }
\label{tab:nonlinear}
\end{table}

The nonlinear dispersion relation $k(\omega,a_\text{avg})$ is compared between experiments and cnoidal-like solutions in figure \ref{fig:Fig_nonlinear_56}. Agreement is quantitatively achieved for all but three trials within $5\%$ relative error, with at most $10\%$ error otherwise. These measurements are across a range of wave amplitudes $a_\text{avg} \in (0.89,2.2)$ and frequencies $\omega \in (0.44,0.62)$. The trials in figure \ref{fig:Fig_nonlinear_56} are numbered from highest amplitude to lowest, as shown in table \ref{tab:nonlinear}. 
Notably, the experimentally measured wavenumbers exceed the exact values across all 15 trials. Although a different class of waves, we remark that previous measurements of large-amplitude soliton solutions in (\cite{olson1986solitary}) found that the soliton speed was consistently underpredicted by conduit equation soliton solutions. We hypothesize that inclusion of the weakly recirculating flow in the model, could improve upon the already good predictions.

\section{Discussion and conclusion} \label{sec:conclusion}

The main result of this study is the reliable generation of periodic traveling waves in a viscous fluid conduit across a wide range of amplitudes and long to short wavelengths that are well-modeled by the conduit equation. This provides further evidence of the viscous fluid conduit as an accessible and flexible laboratory system for the study of nonlinear wave dynamics with an accurate partial differential equation model in the conduit equation. In order to quantitatively model shorter small amplitude waves, we found it helpful to derive the full two-Stokes linear dispersion relation that accounts for the actual conditions in our experiment. Radial variation in the vertical velocity allows for a non-negligible exterior pressure gradient and recirculating flow. Together with an outer wall, these two main factors give rise to an upshifted critical frequency for the transition from propagating to non-propagating, spatially damped waves. This upshift from the conduit equation's critical frequency in the short-wave regime is observed and verified quantitatively. The two-Stokes recirculating flow system with exterior flow mass conservation is therefore considered to be a more precise model in which we successfully extend the linear dispersive properties of the model to a shorter wave regime. The two-Stokes linear dispersion relation is exactly obtained in terms of Bessel functions. Nevertheless, the conduit equation remains an excellent model of propagating linear waves.

Another key result of this experimental investigation is the existence and stability of periodic traveling waves in the fluid conduit wavemaker problem. Low frequency periodic waves generated at the boundary with a time-periodic injection rate are observed to coherently propagate as traveling waves. Arrested wave propagation for waves generated above the critical injection frequency was observed and found to quantitatively match the two-Stokes dispersion relation for real frequencies and complex wavenumbers. These observations are explained in terms of the radiation condition and linear wave modulation theory.

In addition to presenting a careful analysis of linear periodic waves, our work also contributes to an understanding of conduit periodic traveling wave solutions in the weakly nonlinear and strongly nonlinear regimes. Nonlinear periodic waves are found to require more Fourier harmonics with strongly nonlinear waves well-described by cnoidal-like  solutions of the conduit equation. Quantitative agreement is achieved between experimental results and conduit cnoidal-like solutions for both the nonlinear dispersion relation and the wave profiles. The characterization of fully nonlinear periodic traveling waves and their reliable generation in a viscous fluid conduit experiment hold promise for future studies of more general boundary value problems in dispersive hydrodynamics (\cite{el2016dispersive}), including the generalized Riemann problem and the generation of breathers and breather trains.

% \begin{figure}
%   \centerline{\includegraphics{Fig1}}% Images in 100% size
%   \caption{Trapped-mode wavenumbers, $kd$, plotted against $a/d$ for
%     three ellipses:\protect\\
%     ---$\!$---,
%     $b/a=1$; $\cdots$\,$\cdots$, $b/a=1.5$.}
% \label{fig:ka}
% \end{figure}

% \begin{figure}
%   \centerline{\includegraphics{Fig2}}
%   \caption{The features of the four possible modes corresponding to
%   (\textit{a}) periodic\protect\\ and (\textit{b}) half-periodic solutions.}
% \label{fig:kd}
% \end{figure}

\textbf{Funding.} This work was supported by the National Science Foundation (grant number DMS-1816934).

\textbf{Declaration of interests.} The authors report no conflict of interest.

\appendix

\section{Two-Stokes flow and nondimensionalization} \label{sec:append1}

The governing equations (\ref{eq:twoStokes_mass}-\ref{eq:twoStokes_momentum}) for continuity and momentum conservation will now be nondimensionalized according to (\ref{eq:nondim}). Equations for the interior fluid take the nondimensional form
% \begin{subequations}
% \begin{align}
%     \dfrac{1}{r}\dfrac{\partial}{\partial r}(r u_r^{(i)}) + \epsilon^{1/2}\dfrac{\partial u_z^{(i)}}{\partial z} &= 0, \\
%     Re^{(i)} \left( \frac{\partial u_r^{(i)}}{\partial t} + \epsilon^{-1/2} u_r^{(i)} \frac{\partial u_r^{(i)}}{\partial r} + u_z^{(i)} \frac{\partial u_r^{(i)}}{\partial z} \right) &= -\epsilon^{-3/2}\dfrac{\partial p^{(i)}}{\partial r}+\nabla^2 u_r^{(i)}-\epsilon^{-1}\dfrac{u_r^{(i)}}{r^2}, \\
%     Re^{(i)}\left( \frac{\partial u_z^{(i)}}{\partial t} + \epsilon^{-1/2} u_r^{(i)} \frac{\partial u_z^{(i)}}{\partial r} + u_z^{(i)} \frac{\partial u_z^{(i)}}{\partial z} \right) &= -\epsilon^{-1}\dfrac{\partial p^{(i)}}{\partial z}+\nabla^2 u_z^{(i)},
% \end{align}
% \end{subequations}
\begin{subequations}
\begin{align}
    \dfrac{1}{r}\dfrac{\partial}{\partial r}(r u_r^{(i)}) + \epsilon^{1/2}\dfrac{\partial u_z^{(i)}}{\partial z} &= 0, \\
    -\epsilon^{-3/2}\dfrac{\partial p^{(i)}}{\partial r}+\nabla^2 u_r^{(i)}-\epsilon^{-1}\dfrac{u_r^{(i)}}{r^2} &= 0, \\
    -\epsilon^{-1}\dfrac{\partial p^{(i)}}{\partial z}+\nabla^2 u_z^{(i)} &= 0,
\end{align}
\end{subequations}
where $\nabla^2=\epsilon^{-1}\frac{1}{r}\frac{\partial}{\partial r}(r\frac{\partial}{\partial r})+\frac{\partial^2}{\partial z^2}.$ For the exterior fluid, the equations are 
% \begin{subequations}
% \begin{align}
%     \dfrac{1}{r}\dfrac{\partial}{\partial r}(r u_r^{(e)}) + \epsilon^{1/2}\dfrac{\partial u_z^{(e)}}{\partial z} &= 0, \\
%     Re^{(e)} \left( \frac{\partial u_r^{(e)}}{\partial t} + \epsilon^{-1/2} u_r^{(e)} \frac{\partial u_r^{(e)}}{\partial r} + u_z^{(e)} \frac{\partial u_r^{(e)}}{\partial z} \right) &= -\epsilon^{-1/2}\dfrac{\partial p^{(e)}}{\partial r}+\nabla^2 u_r^{(e)}-\epsilon^{-1}\dfrac{u_r^{(e)}}{r^2} , \\
%     Re^{(e)}\left( \frac{\partial u_z^{(e)}}{\partial t} + \epsilon^{-1/2} u_r^{(e)} \frac{\partial u_z^{(e)}}{\partial r} + u_z^{(e)} \frac{\partial u_z^{(e)}}{\partial z} \right) &= -\dfrac{\partial p^{(e)}}{\partial z}+\nabla^2 u_z^{(e)}.
% \end{align}
% \end{subequations}
\begin{subequations}
\begin{align}
    \dfrac{1}{r}\dfrac{\partial}{\partial r}(r u_r^{(e)}) + \epsilon^{1/2}\dfrac{\partial u_z^{(e)}}{\partial z} &= 0, \\
    -\epsilon^{-1/2}\dfrac{\partial p^{(e)}}{\partial r}+\nabla^2 u_r^{(e)}-\epsilon^{-1}\dfrac{u_r^{(e)}}{r^2} &= 0, \\
    -\dfrac{\partial p^{(e)}}{\partial z}+\nabla^2 u_z^{(e)} &=0.
\end{align}
\end{subequations}
The kinematic boundary condition is 
\be 
    u_r^{(i)}=\epsilon^{1/2}\left(\dfrac{\partial R}{\partial t}+u_z^{(i)}\dfrac{\partial R}{\partial z}\right), \quad r=R(z,t).
\ee 
Equations for continuity of the velocity components are 
\begin{subequations}
\begin{align}
    (u_r^{(i)}-u_r^{(e)}) &=\epsilon^{1/2}\dfrac{\partial R}{\partial z}(u_z^{(i)}-u_z^{(e)}), \quad r=R(z,t),\\
    (u_z^{(e)}-u_z^{(i)}) &=\epsilon^{1/2}\dfrac{\partial R}{\partial z}(u_r^{(i)}-u_r^{(e)}), \quad r=R(z,t).
\end{align}
\end{subequations}
The nondimensional stress balance conditions in the normal and tangential directions are, respectively,
\begin{subequations}
\begin{align}
    \left[ -\|\mathbf{n}_c\|^2 P + \kappa \left(\sigma_{rr}-2\epsilon^{1/2}\dfrac{\partial R}{\partial z}\sigma_{rz}+\epsilon\left(\dfrac{\partial R}{\partial z}\right)^2 \sigma_{zz}\right)\right]_j &= 0, \quad r=R(z,t),\\
    \left[ \kappa \left\{ - \left(1-\epsilon\left(\dfrac{\partial R}{\partial z}\right)^2\right) \sigma_{rz} + \epsilon^{1/2} \dfrac{\partial R}{\partial z} (\sigma_{zz}-\sigma_{rr})\right\}\right]_j &= 0, \quad r=R(z,t),
\end{align}
\end{subequations}
where $\kappa^{(e)}=\epsilon^{-1}, \kappa^{(i)}=1$ are the fluid-specific coefficients, 
$\mathbf{n}_c=\begin{bmatrix}
1 \\ - \epsilon^{1/2} \frac{\partial R}{\partial z}
\end{bmatrix}$ is the normal vector for the conduit, and $P$ is the scaled, absolute pressure with $ P^{(i,e)} = p^{(i,e)}-\frac{\rho^{(i,e)}z}{\rho^{(e)}-\rho^{(i)}}$. The normalized deviatoric stress tensor is 
\be
    \sigma^{(i,e)}= 
    \begin{bmatrix}
        \sigma_{rr}^{(i,e)}  &   \sigma_{rz}^{(i,e)} \\
        \sigma_{zr}^{(i,e)} &   \sigma_{zz}^{(i,e)}
    \end{bmatrix}
    = \epsilon^{1/2} 
    \begin{bmatrix}
        2\dfrac{\partial u_r^{(i,e)}}{\partial r}  &  \dfrac{\partial u_z^{(i,e)}}{\partial r}+\epsilon^{1/2}\dfrac{\partial u_r^{(i,e)}}{\partial z}\\
        \dfrac{\partial u_z^{(i,e)}}{\partial r}+\epsilon^{1/2}\dfrac{\partial u_r^{(i,e)}}{\partial z} & 2\epsilon^{1/2} \dfrac{\partial u_z^{(i,e)}}{\partial z}
    \end{bmatrix}.
\ee 
The above nondimensional governing equations associated with the interfacial boundary conditions, as well as boundary conditions along the symmetry axis $u_r^{(i)} = \frac{\partial u_z^{(i)}}{\partial r} = \frac{\partial p^{(i)}}{\partial r}=0$ at $r=0$ and the outer wall $u_r^{(e)}=u_z^{(e)}=0$ at $r=D$ provide a complete description of the dynamics of two-Stokes interfacial flow.

% \begin{subequations}
% \begin{align}
%     p^{(i)} &=p_0^{(i)}+a p_1^{(i)} + \cdots,\\
%     p^{(e)} &=p_0^{(e)}+a p_1^{(e)} + \cdots,\\
%     u_z^{(i)} &=u_{z,0}^{(i)}+a u_{z,1}^{(i)}+\cdots,\\ 
%     u_z^{(e)}&=\epsilon\left(u_{z,0}^{(e)}+a u_{z,1}^{(e)}+\cdots\right),\\
%     u_r^{(i)}&=\epsilon^{1/2}\left(u_{r,0}^{(i)}+a u_{r,1}^{(i)}+\cdots\right),\\
%     u_r^{(e)}&=\epsilon^{1/2}\left(u_{r,0}^{(e)}+a u_{r,1}^{(e)}+\cdots\right).
% \end{align}
% \end{subequations}

\section{Two-Stokes exact linear dispersion relation} \label{sec:append2}

\begin{figure}
    \centering
        \subfloat[]{\includegraphics[height=0.3\textwidth]{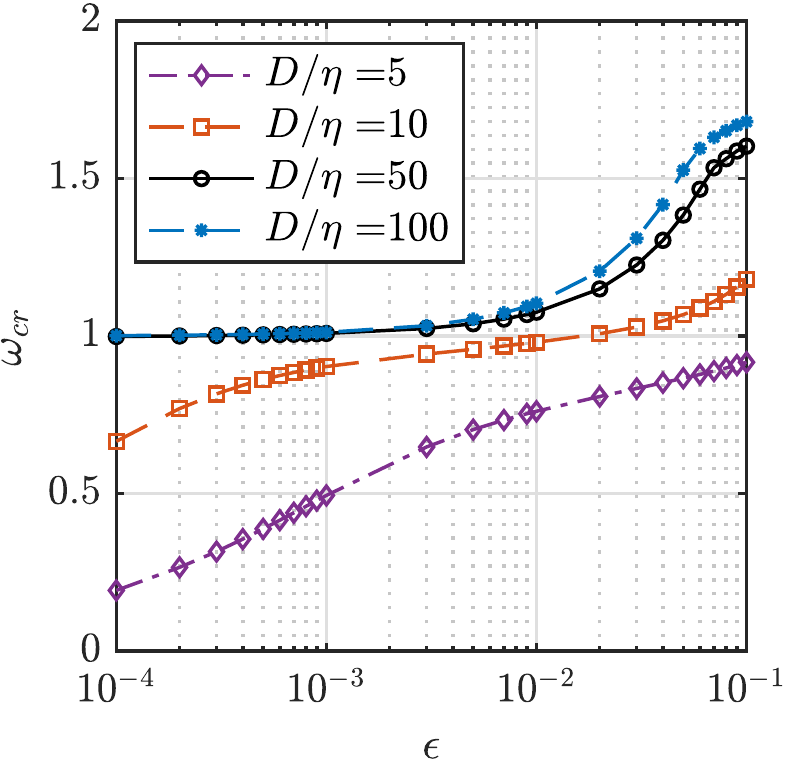}} 
        \subfloat[]{\includegraphics[height=0.3\textwidth]{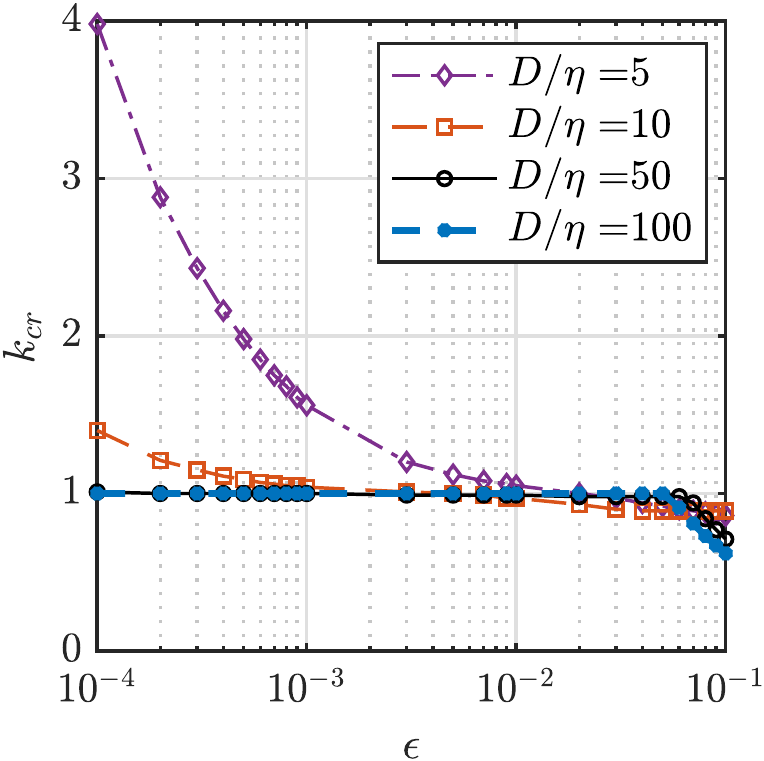}} 
        \subfloat[]{\includegraphics[height=0.3\textwidth]{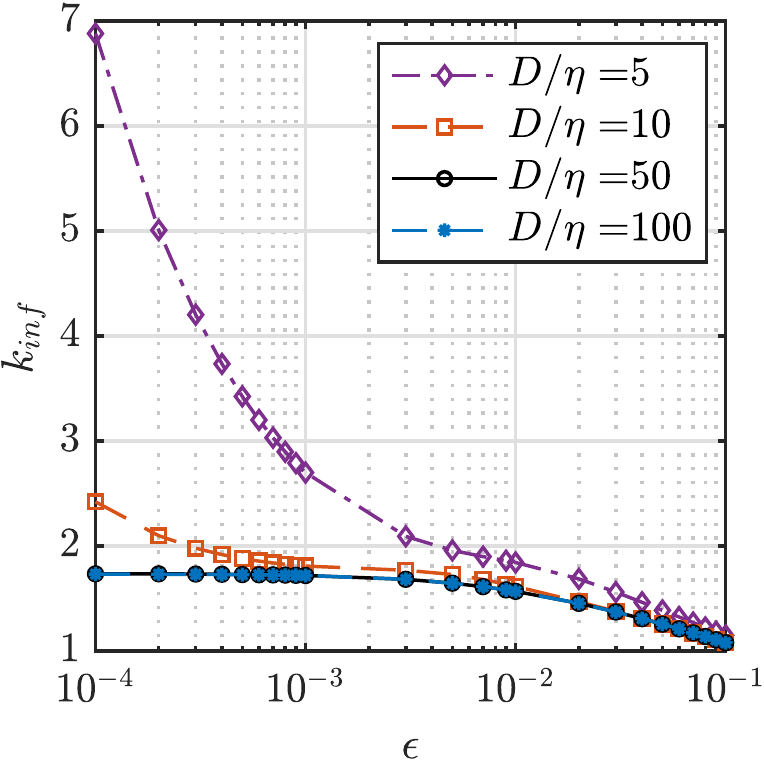}} 
    \caption{Investigation of (a) the critical frequency $\omega_{cr}$, (b) the critical wavenumber $k_{cr}$, and (c) the inflection point $k_{inf}$ at selected parameters $\epsilon,D$ for recirculating flow (\ref{eq:twoStokes_lambda2}).}
    \label{fig:Fig_Stokes_disp2}
\end{figure}

In this section, we derive the solutions for the eigenfunctions in the two-Stokes flow linearization (\ref{eq:twoStokes_linearize1}-\ref{eq:twoStokes_linearize4}) and compute the exact linear dispersion relation. 
The differential equations for two-Stokes flow disturbances are 
% \begin{subequations}
% \begin{align}
%     \dfrac{d g_r(r)}{dr} +\frac{g_r(r)}{r}+ k g_z(r) &=0, \label{eq:twoStokes_linear_int_1} \\
%     \epsilon  g''_r +\frac{ \epsilon  g'_r }{r}- k^2 \epsilon ^2
%     g_r -\frac{ \epsilon  g_r }{r^2}+i h'(r) &= i  Re^{(i)} \epsilon^2 \left( -\omega g_r + k u_{z,0}^{(i)} g_r \right), \label{eq:twoStokes_linear_int_2} \\
%     g''_z +\frac{g'_z }{r}-k^2 \epsilon  g_z -i k h(r) & = i Re^{(i)} \epsilon \left( - \omega g_z + k u_{z,0}^{(i)} g_z + \frac{\partial u_{z,0}^{(i)}}{\partial r} g_r \right)  \label{eq:twoStokes_linear_int_3}
% \end{align} 
% \end{subequations}
\begin{subequations}
\begin{align}
    \dfrac{d g_r(r)}{dr} +\frac{g_r(r)}{r}+ k g_z(r) &=0, \label{eq:twoStokes_linear_int_1} \\
    \epsilon  g''_r +\frac{ \epsilon  g'_r }{r}- k^2 \epsilon ^2
    g_r -\frac{ \epsilon  g_r }{r^2}+i h'(r) &= 0, \label{eq:twoStokes_linear_int_2} \\
    g''_z +\frac{g'_z }{r}-k^2 \epsilon  g_z -i k h(r) & = 0 \label{eq:twoStokes_linear_int_3}
\end{align} 
\end{subequations}
for the interior fluid on the interval $0\leq r \leq \eta$ and
% \begin{subequations}
% \begin{align}
%   \dfrac{d G_r(r)}{d r} +\frac{G_r(r) }{r}+ k \epsilon  G_z(r) &=0, \label{eq:twoStokes_linear_ext_1} \\
%   G_r'' +\frac{G_r'}{r} - k^2 \epsilon  G_r -\frac{G_r }{r^2}+ i  H'(r) &= i  Re^{(e)} \epsilon \left( -\omega G_r + \epsilon k u_{z,0}^{(e)} G_r \right), \label{eq:twoStokes_linear_ext_2} \\
%   G_z''+\frac{G_z'}{r} - k^2 \epsilon  G_z -i k  H(r) &= i Re^{(e)} \epsilon \left( - \omega G_z + \epsilon k u_{z,0}^{(e)} G_z + \frac{\partial u_{z,0}^{(e)}}{\partial r} G_r \right) \label{eq:twoStokes_linear_ext_3}
% \end{align} 
% \end{subequations}
\begin{subequations}
\begin{align}
   \dfrac{d G_r(r)}{d r} +\frac{G_r(r) }{r}+ k \epsilon  G_z(r) &=0, \label{eq:twoStokes_linear_ext_1} \\
   G_r'' +\frac{G_r'}{r} - k^2 \epsilon  G_r -\frac{G_r }{r^2}+ i  H'(r) &= 0, \label{eq:twoStokes_linear_ext_2} \\
   G_z''+\frac{G_z'}{r} - k^2 \epsilon  G_z -i k  H(r) &= 0 \label{eq:twoStokes_linear_ext_3}
\end{align} 
\end{subequations}
for the exterior fluid on the interval $\eta \leq r \leq D$. Along with the interfacial boundary conditions expanded about $r=R(z,t)=\eta+ae^{i(kz-\omega t)}$:
\begin{subequations}
\begin{align}
    k u_{z,0}^{(i)}(\eta) - g_r(\eta) - \omega &=0,  \label{eq:twoStokes_kinematic}  \\
    \epsilon k u_{z,0}^{(e)}(\eta) - k u_{z,0}^{(i)}(\eta) + g_r(\eta) - G_r(\eta) &=0,\\
    \epsilon \dfrac{\partial u_{z,0}^{(e)}}{\partial r}(\eta) - \dfrac{\partial u_{z,0}^{(i)}}{\partial r}(\eta) - g_z(\eta) + \epsilon G_z(\eta)  &=0,\\
    -2i\epsilon k \dfrac{\partial u_{z,0}^{(e)}}{\partial r}(\eta) + 2i\epsilon k \dfrac{\partial u_{z,0}^{(i)}}{\partial r}(\eta) - 2i\epsilon g'_r(\eta) +2iG'_r(\eta) + h(\eta) - \epsilon H(\eta)  &=0,\\
    -\dfrac{\partial^2 u_{z,0}^{(e)}}{\partial r^2}(\eta) + \dfrac{\partial^2 u_{z,0}^{(i)}}{\partial r^2}(\eta) - \epsilon k g_r(\eta) + k G_r(\eta) + g'_z(\eta) -G'_z(\eta) &=0, 
\end{align}  
\end{subequations}
and $g_r(0) = 0, g_z'(0) = 0, G_r(D) = 0, G_z(D) = 0$,  the equations can be solved in terms of modified Bessel functions. The internal flow admits the general solution:
\begin{subequations}
\begin{align}
    g_r(r) &= a_1 K_1\left(\epsilon^{1/2} k r \right) + a_2 I_1\left(\epsilon^{1/2} k r \right) + a_3 r K_0\left(\epsilon^{1/2} k r \right) + a_4 r I_0\left(\epsilon^{1/2} k r \right), \\
    g_z(r) &= a_1 \epsilon^{1/2} K_0\left(\epsilon^{1/2} k r \right) - a_2 \epsilon^{1/2}I_0\left(\epsilon^{1/2} k r \right) - \frac{2a_3}{k}K_0\left(\epsilon^{1/2} k r \right) + a_3 \epsilon^{1/2} r K_1 \left(\epsilon^{1/2} k r \right) \nonumber \\
    & \hphantom{=a_1} - \frac{2a_4}{k} I_0\left(\epsilon^{1/2} k r \right) - a_4 \epsilon^{1/2} r I_1\left(\epsilon^{1/2} k r \right),\\
    ih(r) &= -2\epsilon \left( a_3 K_0\left(\epsilon^{1/2} k r \right) + a_4 I_0\left(\epsilon^{1/2} k r \right) \right).
\end{align}
\end{subequations}
The solution set is linearly independent since the Wronskian $W=-\frac{4k^2 \epsilon}{r^2}$ is nonzero. Symmetry along $r=0$ requires $a_1=a_3=0$.
Similarly, we obtain the general solution for the extrusive velocities and pressure as
\begin{subequations}
\begin{align}
    G_r(r) &= b_1 K_1\left(\epsilon^{1/2} k r \right) + b_2 I_1\left(\epsilon^{1/2} k r \right) + b_3 r K_0\left(\epsilon^{1/2} k r \right) + b_4 r I_0\left(\epsilon^{1/2} k r \right), \\
    G_z(r) &= \frac{1}{\epsilon} \Bigg[ b_1 \epsilon^{1/2} K_0\left(\epsilon^{1/2} k r \right) - b_2 \epsilon^{1/2}I_0\left(\epsilon^{1/2} k r \right) - \frac{2b_3}{k}K_0\left(\epsilon^{1/2} k r \right) + b_3 \epsilon^{1/2} r K_1 \left(\epsilon^{1/2} k r \right) \nonumber \\
    & \hphantom{=a_1} - \frac{2b_4}{k} I_0\left(\epsilon^{1/2} k r \right) - b_4 \epsilon^{1/2} r I_1\left(\epsilon^{1/2} k r \right) \Bigg],\\
    iH(r) &= -2 \left( b_3 K_0\left(\epsilon^{1/2} k r \right) + b_4 I_0\left(\epsilon^{1/2} k r \right) \right).
\end{align}
\end{subequations}
Applying the interfacial boundary conditions and no-slip, no-penetration outer conditions, a linear system is derived for the coefficients $a_{2,4}, b_{1-4}$:
\be 
    A \begin{pmatrix} a_2 \\ a_4 \\ b_1 \\ b_2\\ b_3 \\ b_4 \end{pmatrix} = \begin{pmatrix} 0\\ -\frac{\eta}{2}(1-\epsilon)(1-\lambda) \\ 0 \\ 1 \\ 0 \\ 0 \end{pmatrix}, \label{eq:linear_system}
\ee 
where 
\be 
    A= \begin{pmatrix}
    I_1 & I_0 \eta & -K_1 & -I_1 & -K_0 \eta & -I_0\eta \\
    A_{21} & A_{22} & A_{23} & A_{24} & A_{25} & A_{26} \\
    A_{31} & A_{32} & A_{33} & A_{34} & A_{35} & A_{36} \\
    A_{41} & A_{42} & A_{43} & A_{44} &  A_{45}, & A_{46}\\
    0 & 0 & K_1(\epsilon^{1/2}D k) & I_1(\epsilon^{1/2}D k) & D K_0(\epsilon^{1/2}D k) & D I_0(\epsilon^{1/2}D k) \\
    0 & 0 & K_0(\epsilon^{1/2}D k) & - I_0(\epsilon^{1/2}D k) & A_{65} & A_{66}
    \end{pmatrix} \label{eq:matrixA}
\ee 
with 
\be 
\begin{aligned}
& \begin{alignedat}{3}
    A_{21} &= \epsilon^{1/2} I_0, \quad && A_{22} = 2\frac{I_0}{k}+ \epsilon^{1/2} I_1 \eta, \quad && A_{23} = \epsilon^{1/2}K_0, \\
    A_{24} &= -\epsilon^{1/2} I_0, \quad && A_{25} = -2\frac{K_0}{k} + \epsilon^{1/2} K_1 \eta, \quad && A_{26} = -2\frac{I_0}{k}-\epsilon^{1/2}I_1 \eta, \\
    A_{31} &= \frac{2}{\eta}\left(-\epsilon I_1 + \epsilon^{3/2}I_0 k \eta \right), \quad &&
    A_{32} = 2 \epsilon^{3/2} I_1 k \eta , \quad &&
    A_{33} = \frac{2}{\eta}\left(K_1 + \epsilon^{1/2} K_0 k\eta \right), \\
    A_{34} &= \frac{2}{\eta}\left(I_1 - \epsilon^{1/2} I_0 k\eta \right), \quad &&
    A_{35} = 2 \epsilon^{1/2} K_1 k \eta, \quad &&
    A_{36} = -2\epsilon^{1/2} I_1 k\eta, \\
    A_{41} &= -2\epsilon I_1 k, \quad && A_{42} = -2\epsilon^{1/2} I_1 - 2\epsilon I_0 k\eta, \quad && A_{43} = 2K_1 k , \\
    A_{44} &= 2I_1 k, \quad && 
    A_{45} = -\frac{2}{\epsilon^{1/2}} K_1 + 2K_0 k\eta, \quad && A_{46} = \frac{2I_1}{\epsilon^{1/2}} + 2I_0 k \eta,
\end{alignedat} \\
    & A_{65} = \frac{1}{\epsilon^{1/2} k}\left( -2K_0(\epsilon^{1/2}D k) + \epsilon^{1/2} D k K_1(\epsilon^{1/2}D k) \right),  \\
    & A_{66} = \frac{1}{\epsilon^{1/2} k} \left( -2I_0(\epsilon^{1/2}D k) - \epsilon^{1/2} D k I_1(\epsilon^{1/2}D k) \right).  
\end{aligned}
\ee 
and $I_0 = I_0(\epsilon^{1/2} \eta k)$, 
    $I_1 = I_1(\epsilon^{1/2} \eta k)$, 
    $K_0 = K_0(\epsilon^{1/2} \eta k)$, 
    $K_1 = K_1(\epsilon^{1/2} \eta k)$.
Varying $\epsilon, D$ leads to changes in the dispersion curves from the conduit dispersion relation as well as changes in the critical frequency $\omega_{cr}$, critical wavenumber $k_{cr}$ and inflection point $k_{inf}$ (see figure \ref{fig:Fig_Stokes_disp2}).

Following the asymptotic behavior $D=\epsilon^{-1/2} d$ and $\lambda = \epsilon \frac{16}{d^2} + O(\epsilon^2 \ln(\epsilon))$, where $d=O(1)$, we expand the modified Bessel functions in the linear system (\ref{eq:linear_system}) for $\epsilon \to 0$. We seek solutions to the system (\ref{eq:linear_system}) with the asymptotic expansions
\be 
    v = \begin{pmatrix} a_2 \\ a_4 \\ b_1 \\ b_2\\ b_3 \\ b_4 \end{pmatrix} = \begin{pmatrix} \epsilon^{-3/2} \left(a_{20}+ \epsilon \ln(\epsilon) a_{21} + \epsilon a_{22} + \epsilon^2 \ln(\epsilon) a_{23} + \epsilon^2 a_{24} + \dots \right) \\ 
    \epsilon^{-1}  \left( a_{40}+ \epsilon \ln(\epsilon) a_{41} + \epsilon a_{42} + \epsilon^2 \ln(\epsilon) a_{43} + \epsilon^2 a_{44} + \dots \right) \\ 
    \epsilon^{1/2}  \left( b_{10}+ \epsilon \ln(\epsilon) b_{11} + \epsilon b_{12} + \epsilon^2 \ln(\epsilon) b_{13} + \epsilon^2 b_{14} + \dots \right) \\ 
    \epsilon^{1/2}  \left( b_{20}+ \epsilon \ln(\epsilon) b_{21} + \epsilon b_{22} + \epsilon^2 \ln(\epsilon) b_{23} + \epsilon^2 b_{24} + \dots \right) \\ 
    \epsilon  \left( b_{30}+ \epsilon \ln(\epsilon) b_{31} + \epsilon b_{32} + \epsilon^2 \ln(\epsilon) b_{33} + \epsilon^2 b_{34} + \dots \right) \\ 
    \epsilon  \left( b_{40}+ \epsilon \ln(\epsilon) b_{41} + \epsilon b_{42} + \epsilon^2 \ln(\epsilon) b_{43} + \epsilon^2 b_{44} + \dots \right) 
    \end{pmatrix}. \label{eq:linear_system_pertub}
\ee 
Collecting (\ref{eq:linear_system}) with (\ref{eq:linear_system_pertub}) at each order and applying the kinematic boundary condition (\ref{eq:twoStokes_kinematic}), we achieve the asymptotic expression (\ref{eq:twoStokes_w(k)_asym}) for the linear dispersion relation with 
\begin{subequations}
\begin{align}
    \omega_2(k) &= 4k \left(-1+\ln\left(\frac{d}{2\sqrt{2}}\right) \right) - \frac{\sqrt{2}k(1+2k^2)}{1+k^2} b_{20} - 2\sqrt{2} b_{40} \nonumber \\
    & \hphantom{=} + \frac{2k}{3d^2(1+k^2)^2} \Bigg[ d^2 \left(6 \Gamma \left(k^4-2 k^2-1\right)+k^4 (\ln (8)-19)-6 k^2 (4+\ln
   (2))-6-\ln (8)\right)  \nonumber \\
    & \hphantom{=} + 6 d^2 \left(k^4-2 k^2-1\right) \ln (k)-48
    \left(k^2+1\right) \Bigg], \tag{B 10}
\end{align}
where $\Gamma$ is the Euler–Mascheroni constant (0.5772...) and the coefficients are
\begin{align}
    b_{10} &= -\frac{4\sqrt{2}k^2}{1+k^2},\\
    b_{20} &= \frac{I_0(d k) \left(K_1(d k) \left(d^2 k b_{30}+2 b_{10}\right)+d k b_{10} K_0(d k)\right)+d k I_1(d k) (b_{10} K_1(d k)+d b_{30} K_0(d k))}{d k I_0(d k){}^2-2 I_1(d k) I_0(d k)-d k I_1(d k){}^2}, \label{eq:twoStokes_b20} \\
    b_{30} &= -\frac{\sqrt{2}k(1+3k^2)}{1+k^2}, \\
    b_{40} &= -\frac{k I_0(d k) (b_{10} K_1(d k)+d b_{30} K_0(d k))+I_1(d k) (K_0(d k) (k b_{10}-2 b_{30})+d k b_{30} K_1(d k))}{d k I_0(d k){}^2-2 I_1(d k) I_0(d k)-d k I_1(d k){}^2}. \label{eq:twoStokes_b40}
\end{align}
\end{subequations}

\bibliographystyle{main}
\bibliography{main}

\end{document}